\documentclass{aa}
\usepackage[varg]{txfonts}
\usepackage{natbib}
\usepackage{graphicx}
\usepackage{times}
\usepackage{longtable}
\usepackage{colortbl}
\usepackage{enumitem}

\bibpunct{(}{)}{;}{a}{}{,}
\usepackage{color}

\newcommand{\bl}{$\beta$\,Lyr }
\newcommand{\be}{$\beta$\,Lyr}
\newcommand{\bla}{$\beta$\,Lyr~A }
\newcommand{\blae}{$\beta$\,Lyr~A}
\newcommand{\blb}{$\beta$\,Lyr~B }
\newcommand{\ble}{$\beta$\,Lyr~B}
\newcommand{\spefo}{{\tt SPEFO} }
\newcommand{\spefoe}{{\tt SPEFO}}

\newcommand{\korel}{{\tt KOREL} }

\newcommand{\pyt}{{\tt PYTERPOL} }
\newcommand{\pyte}{{\tt PYTERPOL}}
\newcommand{\shellspec}{{\tt SHELLSPEC}}

\newcommand{\ubv}{\hbox{$U\!B{}V$}}
\newcommand{\ubvr}{\hbox{$U\!B{}V\!R$}}
\newcommand{\bvr}{\hbox{$B{}V\!\left(R\right)_\mathsf{c}$}}
\newcommand{\bv}{\hbox{$B\!-\!V$}}
\newcommand{\ub}{\hbox{$U\!-\!B$}}

\newcommand{\m}{$^{\rm m}\!\!.$}

\newcommand{\kms}{km\,s$^{-1}$ }
\newcommand{\ks}{km\,s$^{-1}$}
\newcommand{\vsin}{$v\sin i$ }
\newcommand{\vsine}{$v\sin i$}
\newcommand{\tef}{$T_\mathsf{eff}$ }
\newcommand{\tefe}{$T_\mathsf{eff}$}
\newcommand{\lgg}{$\log g$ }
\newcommand{\lgge}{$\log g$}
\newcommand{\ms}{M$_{\odot}$}
\newcommand{\rs}{R$_{\odot}$}

\newcommand{\ha}{H$\alpha$}

\newcommand{\hae}{H$\alpha$}

\newcommand{\he}{\ion{He}{i}~6678}

\newcommand{\her}{\ion{He}{i}~7065}

\newcommand{\Am}{\r{A}\,mm$^{-1}$ }
\newcommand{\Ame}{\r{A}\,mm$^{-1}$}

\newcommand{\s}{\hbox{\v{s}}}
\begin{document}

\title{Physical properties of $\beta$ Lyr A and its opaque accretion disk
\thanks{Based on spectro-interferometric observations
obtained with the instruments CHARA/VEGA, CHARA/MIRC, and NPOI,
and \ubvr\ photometric observations from Hvar observatory.},
\thanks{Tables~\ref{tab:if:logobs}, and \ref{tab:lc:data}
are available only in electronic form at the CDS through anonymous
ftp to cdarc.u-strasbg.fr (130.79.128.5) or via
http://cdsweb.u-strasbg.fr/cgi-bin/qcat?J/A+A/}
}
\author{
D.~Mourard\inst{1}\and
M.~Bro\v z\inst{2}\and
J.A.~Nemravov\'a\inst{2}\and
P.~Harmanec\inst{2}\and
J.~Budaj\inst{3}\and                   %SHELLSPEC
F.~Baron\inst{4}\and                   %MIRC interferometry, observation and analysis
J.D.~Monnier\inst{5}\and                %MIRC analysis, support
G.H.~Schaefer\inst{4}\and               %MIRC observations
H.~Schmitt\inst{6}\and                 %NPOI observations and reduction
I.~Tallon-Bosc\inst{7}\and             %VEGA observations
%M.~Zhao\inst{8}\and
J.~T.~Armstrong\inst{6}\and            %NPOI Observations
E.~K.~Baines\inst{6}\and               %NPOI Observations
D.~Bonneau\inst{1}\and                 %planning VEGA strategy
H.~Bo\v{z}i\'c\inst{8}\and             %Hvar photometry and data reduction
J.M.~Clausse\inst{1}\and               %VEGA support
C.~Farrington\inst{4}\and              %CHARA observations
D.~Gies\inst{4}\and                    %CHARA Project Scientist
J.~Jury\v{s}ek\inst{2,9}\and          %BLB observations
D.~Kor\v{c}\'akov\'a\inst{2}\and       %Discussion on accretion disks and RT
%B.~Kloppenborg\inst{4}\and             %SIMTOI tool
H.~McAlister\inst{4}\and             %CHARA Director
A.~Meilland\inst{1}\and                 %early modeling, support
N.~Nardetto\inst{1}\and                 %Support VEGA from 2013-2016, + Support VEGA analysis
P.~Svoboda\inst{10}\and                %BVR photometry
M.~\v{S}lechta\inst{11}\and            %BLB reductions
M.~Wolf\inst{2}\and                    %Hvar photometry
P.~Zasche\inst{2}                      %Hvar photometry and BLB observations
}

   \offprints{\email denis.mourard@oca.eu}

% INSTITUTES
  \institute{
%1
  Universit\'{e} C\^{o}te d'Azur, OCA, CNRS, Lagrange, Parc~Valrose,
  B\^{a}t. Fizeau, 06108~Nice, France \and
%2
  Astronomical Institute of the Charles University, Faculty of Mathematics and
  Physics,\\ V~Hole\v{s}ovi\v{c}k\'ach~2, 180~00~Praha~8, Czech~Republic \and
%3
  Astronomical Institute, Slovak Academy of Sciences,
  059~60 Tatransk\'a Lomnica, Slovak~Republic \and
%4
  The CHARA Array of Georgia State University, Mount Wilson Observatory,
  Mount Wilson, California~91023, USA \and
%5
  Astronomy Department, University of Michigan (Astronomy), 500~Church~St,
  Ann Arbor, MI~48109, USA \and
%6
  Naval Research Laboratory, Remote Sensing Division, Code~7215,
  4555~Overlook~Ave.~SW, Washington, DC~20375, USA \and
%7
 Univ Lyon, Univ Lyon1, Ens de Lyon, CNRS, Centre de Recherche Astrophysique de Lyon UMR5574, F-69230, Saint-Genis-Laval, France \and
%old8
%  Jet Propulsion Lab, 4800~Oak Grove Dr, MS~169-327, Pasadena,
%  California 91109, USA \and
%8
  Hvar Observatory, Faculty of Geodesy, University of Zagreb,
  Ka\v{c}i\'ceva~26, 10000~Zagreb, Croatia \and
%9
  Institute of Physics, The Czech Academy of Sciences, Na~Slovance~1999/2,
  181~21~Praha~8, Czech Republic \and
%10
  Private Observatory, V\'ypustky~5, 614~00, Brno, Czech Republic \and
%11
  Astronomical Institute, Czech Academy of Sciences,
  251~65~Ond\v{r}ejov, Czech~Republic
}

% DATE
\date{Received \today}

%%%%%%%%%%%%%%%%%%%%%%%%%%%%%%%%%%%%%%%%%%%%%%%%%%%%%%%%%%%%%%%%%%%%%%%%

\abstract{
Mass exchange and mass loss in close binaries can significantly affect
their evolution, but a complete self-consistent theory of these processes is
still to be developed. Processes such as radiative shielding due to a~hot-spot region,
or a~hydrodynamical interaction of different parts of the gas stream have been
studied previously. In order to test the respective predictions, it is necessary to carry
out detailed observations of binaries undergoing the largescale mass exchange,
especially for those that are in the rapid transfer phase. \bla is an archetype of such
a system, having a long and rich observational history. Our goal for this first
study is to quantitatively estimate the geometry and physical properties of
the optically thick components, namely the Roche-lobe filling mass-losing
star, and the accretion disk surrounding the mass-gaining star of \blae.
A series of continuum visible and NIR spectro-interferometric observations
by the NPOI, CHARA/MIRC and VEGA instruments covering the whole orbit of
\bla acquired during a~two-week campaign in 2013 were complemented with
\ubvr\ photometric observations acquired during a three-year monitoring of
the system. We included NUV and FUV observations from OAO~A-2, IUE, and
Voyager satellites.

All these observations were compared to a complex model of the system.
It is based on the simple LTE radiative transfer code
\shellspec, which was substantially extended to compute all
interferometric observables and to perform both global and local
optimization of system parameters. Several shapes of the accretion disk were
successfully tested --- slab, wedge, and a~disk with an exponential vertical
profile --- and the following properties were consistently found:
the radius of the outer rim is $30.0\pm 1.0\,R_\odot$, the semithickness of the disk
$6.5\pm1.0\,R_\odot$, and the binary orbital inclination
$i = 93.5\pm1.0\,{\rm \deg}$. The temperature profile is a power-law
or a steady-disk in case of the wedge geometry. The properties of
the accretion disk indicate that it cannot be in a vertical hydrostatic
equilibrium, which is in accord with the ongoing mass transfer. The hot spot
was also detected in the continuum but is interpreted as
a~hotter part of the accretion disk illuminated by the donor. As a by-product, accurate kinematic
and radiative properties of \blb were determined.
}


\authorrunning{Mourard et al.}
\titlerunning{Physical properties of \blae}
\maketitle

%%%%%%%%%%%%%%%%%%%%%%%%%%%%%%%%%%%%%%%%%%%%%%%%%%%%%%%%%%%%%%%%%%%%%%%%
%%%%%%%%%%%%%%%%%%%%%%%%%%%%%%%%%%%%%%%%%%%%%%%%%%%%%%%%%%%%%%%%%%%%%%%%

\section{Introduction}

Mass transfer between close binary components has a~profound impact
on their evolution. Early models of the mass transfer
by~\citet{kippenhahn1967} have explained the~Algol paradox
\citep{crawford1955} and demonstrated that about $80\%$
of mass transferred during the whole process is exchanged
in less than $10\%$ of its total duration
\citep[see also][for recent, more sophisticated modeling]{degreve1986}.
The character and outcome of the process, that is whether
the mass transfer is conservative or whether some matter and angular
momentum escapes from the system and forms a~common envelope around
the whole system \citep{kuiper41,paczynski1976}, depend
strongly on the properties of the binary in question before the beginning
of mass exchange and on the actual mechanism of the mass transfer.

\citet{kippenhahn1967} introduced an initial classification
of systems undergoing mass exchange, depending on whether the mass-losing component overflows the Roche limit during the core hydrogen burning (case~A), or shell hydrogen burning (case~B). Later, the term
"case AB" was suggested to distinguish between cases when the
mass exchange starts as case A close to the exhaustion of
hydrogen in the core and continues as case B later during the process.
All such systems are progenitors of Algols, that is systems in the later
stage of the mass exchange when the mass ratio had already been reversed.
Especially for the more massive Algols, the mass exchange appears to be
non-conservative
as found by~\citet{vanrensbergen2006} from the distribution of the mass
ratios among the observed Algols. The actual mechanisms of the mass and
angular momentum loss from the system are not well established and modeled
as yet.  \citet{bis2000} proposed a~purely hydrodynamical mechanism, which
assumes that the gas flow, after encircling the gainer, hits the original,
denser flow from the mass-losing star and is deviated above and below
the orbital plane in the form of bipolar jets. On the other hand,
\citet{vanrensbergen2008} and \citet{deschamps2013} have proposed that
a radiative interaction (a~hot spot) between the flow, accretion disk,
and the gainer may be responsible for the mass loss in the equatorial plane.

One way to discriminate between different scenarios is to carry out
detailed studies of systems undergoing a phase of rapid mass exchange
and deduce the true distribution and kinematics of the circumstellar gas
for them. The bright, well-known binary \bla (HD~174638, HR~7106,
HIP~92420) with a  steadily increasing orbital
period of 12\fd94, is an~archetype of such systems.
The history of its investigation is more than
two hundred years long, and here we refer only to a~subset of the more recent
studies relevant to the topic of this paper.  Detailed reviews of
previous studies can be found in~\citet{sahade1980} and \citet{hec2002}.
Unless a clear distinction is needed, we shall simply use the name
\bl to denote \bla in the rest of the text.

\bl is currently in a phase of rapid mass exchange, although its initial mass ratio  has already
been reversed and is now $q=m_{\rm g}/m_{\rm d} \simeq 4.50$ \citep[][]{hs93}, with the
mass-losing component (donor) being the less massive of the two
($m_\mathsf{d} \simeq 2.9$\,\ms, and $m_\mathsf{g} \simeq 13.3$\,\ms). Its spectral type is B6-8\,II and the effective temperature $T_\mathsf{eff,d} = 13\,300$\,K \citep[][]{balach86}.
It is losing mass via a~Roche-lobe overflow toward its more massive
partner (gainer). The conservation of angular momentum
(or equivalently, the Coriolis acceleration in the non-inertial corotating frame)
forces the gas flow to encircle the gainer
and to form an~accretion disk around it \citep{huang1963,wilson74,hubeny1991}.
Being composed from hot, and mainly ionized material, it is optically (and also
geometrically) thick in the continuum and because it is observed nearly
edge-on, it obscures the gainer completely. It occupies almost the whole
critical Roche lobe around the gainer in the equatorial plane and was found
to have the temperature of its rim of $7\,000$ to $9\,000$\,K
\citep[e.g.,][]{al2000,mennickent2013}. The geometry of the optically thick bodies
(the donor and the disk) was reconstructed from near-infrared interferometric
observations by~\citet{zhao2008}. The current rate of the mass transfer between the
binary components is high, $\approx 2.1\times10^{-5}\,\mathrm{M_\odot\,yr^{-1}}$
for a conservative transfer \citep{hs93}, and
$\approx 2.9\times10^{-5}\,\mathrm{M_\odot\,yr^{-1}}$
for a~non-conservative one \citep{degreve94,vanrensbergen2016}, as deduced from
the large observed secular change of the
orbital period of $\approx 19\,\mathrm{s\,yr^{-1}}$
\citep{hs93,ak2007}.

The presence of a~hot spot has been advocated by~\citet{lomax2012}
and by~\citet{mennickent2013}. The latter authors also postulated
a~second ``bright spot'' that should arise from a~spiral
arm that is formed within the disk. Some part of the gas flow
is also deflected in the~direction perpendicular to
the accretion disk and forms a~pair of jets, whose existence
was first proposed by \citet{hec96}, and was confirmed
by~\citet{hoffman1998,ak2007,ignace2008,bonneau2011} via
different types of observations. Observational evidence
of the mass loss from the system has been presented
by~\citet{umana2000,umana2002}, who resolved a~circumbinary nebula
surrounding the system in radio emission and found that it
extends along the direction of the bipolar jets. An~attempt to
image the optically thin medium in H$\alpha$ has been carried
out by~\citet{schmitt2009}, but their observations lacked the spatial
resolution needed to separate the individual structures.

The evolution of \bl from the initial to the current evolutionary stage
was modeled by several authors. The early conservative model of mass
exchange by~\citet[][]{ziolkowski1976} was found unrealistic by~\citet{packet1979},
because it would lead to a~contact system.
The latest evolutionary tracks  by~\citet{mennickent2013} and by~\citet{vanrensbergen2016}
are in a good mutual agreement, probably thanks to the fact that
the former is based on evolutionary models
by~\citet{vanrensbergen2008}. The latter is based on
improved evolutionary models including tides, and predicts
that the system undergoes a case~AB mass transfer and that it originated
from a detached binary with initial masses of 10.35\,\ms~(donor)
and 7\,\ms, and a~period of 2.36\,d and its current
age is $2.63\times10^{7}$\,yr.

The aim of the present study is to analyze and use
the very rich series of visible and infrared spectro-interferometric
observations covering the whole orbit of \bl, complemented by
series of standard \ubvr, near-infrared and far-UV photometric observations.
All these observations are compared to the predictions of several
working models of optically thick components of the system,
focusing on the size, shape and physical properties of the opaque accretion
disk surrounding the gainer. To this end, we use
an improved version of the modeling tool \shellspec~by~\citet{budaj2004}
developed for binaries embedded in a 3D moving circumstellar
environment.

   A continuation of this study (to be published later) will use differential
spectro-interferometry and analyses of selected emission-line profiles
to investigate the probable distribution and kinematics of optically thin parts of
the circumstellar matter within the system.

%%%%%%%%%%%%%%%%%%%%%%%%%%%%%%%%%%%%%%%%%%%%%%%%%%%%%%%%%%%%%%%%%%%%%%%%
%%%%%%%%%%%%%%%%%%%%%%%%%%%%%%%%%%%%%%%%%%%%%%%%%%%%%%%%%%%%%%%%%%%%%%%%

\section{Observations}\label{sec:observations}

Throughout this paper, reduced Julian dates
$\mathrm{RJD} = \mathrm{HJD} - 2\,400\,000.0$
are used. The quadratic orbital ephemeris by \citet{ak2007}:
\begin{eqnarray}
T_{\rm min.I}\left(\mathrm{HJD}\right) &=& \,2\,408\,247.968(15) + 12.913779(16) \cdot E \nonumber \\
& + & 3.87265(369)\times10^{-6} \cdot E^2\,,\label{eq:ephemeris}
\end{eqnarray}
is used to compute orbital phases of \blae. Phase $0$ corresponds to superior conjunction of the mass donor.

Our study is based on the following sets of dedicated
spectro-interferometric observations and multicolor photometric observations
(the details on the observations and their reductions
being provided in Appendices~\ref{sec:apa} and~\ref{sec:apb}).

\subsection{\em Interferometry}
Three spectro-interferometric instruments took
part in a~twelve nights long observational campaign
aimed at \bl in 2013. We also have at our disposal
all previous interferometric observations as detailed below.

\begin{itemize}
\item {\em Navy Precision Optical Interferometer (NPOI)\/} \citep{npoi}:
These observations were carried out in 16 spectral
channels spread over wavelength region $\Delta\lambda =
562-861$\,nm with two triplets of telescopes.
\item {\em Michigan InfraRed Combiner (MIRC)\/} \citep{mirc, monnier2006}:
These observations were acquired with six telescopes in $H$-band split into eight channels.
Earlier observations in four-telescope mode have already been analyzed by~\citet{zhao2008}
and qualitatively compared to a~working model. We note also that these observations
were acquired before the instrument was equipped
with photometric channels \citep{che2010}.
\item {\em Visible spEctroGraph and polArimeter (VEGA)\/} \citep{vega, vega2}: These observations
were taken in four spectral regions using medium
spectral resolution $R = 5\,000$. In each of
these regions two channels in continuum $\approx 10-15$\,nm
wide were chosen. Either two or three~telescopes were used.
\end{itemize}

\begin{figure}
\centering
\includegraphics[width=0.4\textwidth]{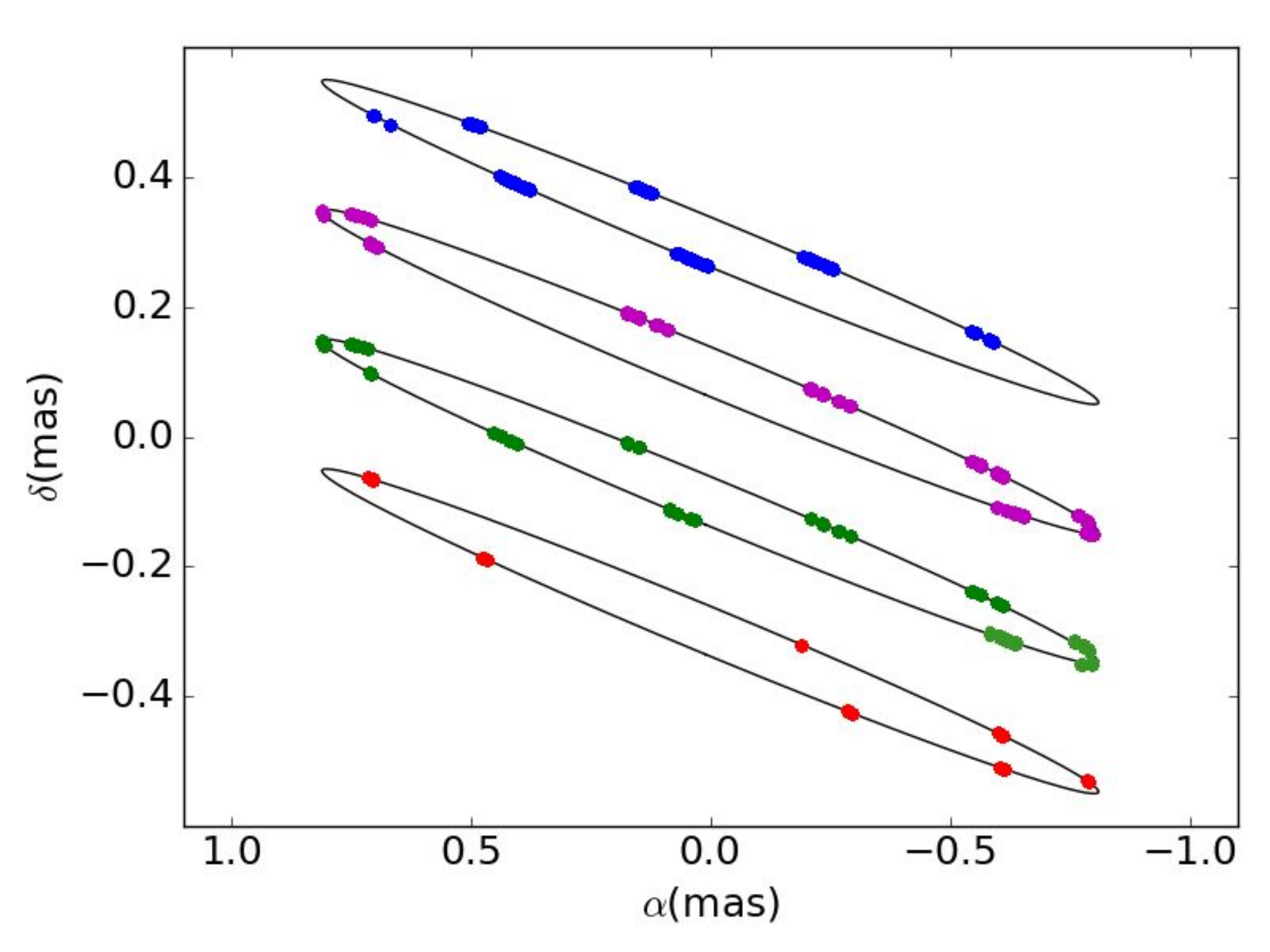}
\caption{Phase coverage of spectro-interferometric
observations of \bl acquired by different instruments.
$\delta$ denotes the relative declination (positive toward the north), and
$\alpha$ the relative right ascension (positive toward the east).
The black line shows the size and orientation of the \bl orbit in the sky,
      the blue dots show orbital phases corresponding to NPOI observations,
the magenta dots to CHARA/VEGA observations,
the green dots to CHARA/MIRC observations acquired in 2013, and
the red dots to CHARA/MIRC observations acquired in 2006/2007. An arbitrary vertical shift of $0.2~mas$ is added to separate the various orbits.
\label{fig:orbit}}
\end{figure}

The VEGA and MIRC instruments are mounted at
Center for High Angular Resolution Astronomy (CHARA) interferometric array \citep{chara}. On nine nights during the campaign, the MIRC and VEGA instruments were co-phased to record fringes simultaneously in the visible and infrared. Basic properties of the spectro-interferometric observations are listed in Table~\ref{tab:if:journal}. The phase coverage of \bla orbit with spectro-interferometric observations from individual instruments is shown in Fig.~\ref{fig:orbit}. A~detailed overview of the observations is listed
in Table~\ref{tab:if:logobs}. The reduced quantities (visibilities, closure phases, and - for MIRC only - triple product amplitudes) are available at CDS in the {\tt OIFITS} format \citep{pauls2005}.

\subsection{\sl Photometry}
A new series of differential Johnson \ubvr\ photometric observations
were acquired at Hvar Observatory in 2013 - 2017,
with earlier infrared photometry acquired by
\citet{jameson1976}, and~\citet{taran2005}.
Apart from that, we included NUV and FUV observations
from OAO~A-2, IUE, and Voyager satellites described in \cite{kondo1994}.

Moreover, there is a~Johnson--Cousins differential \bvr\ photometry
acquired by PS at his private observatory in Brno, Czech~Republic.
The latter observations were not modeled, though, because
of their limited phase coverage and a~relatively simple
standardization procedure.
Therefore these~observations served only as
an~independent check that the Hvar \ubvr\ measurements
do not miss an important light curve feature.

Journal of photometric observations is in
Table~\ref{tab:lc:journal}. The \ubvr\ observations
acquired at Hvar observatory and \bvr\ measurements
collected by~PS are available in Table~\ref{tab:lc:data}
at~CDS.

\begin{table}
\centering
\setlength{\tabcolsep}{5pt}
\caption{Journal of analyzed spectro-interferometry.
\label{tab:if:journal}}
\begin{tabular}{ccrrrrl}
\hline\hline\noalign{\smallskip}
$\Delta T$ & $\Delta \lambda$ & $B_\mathrm{min}$ & $B_\mathrm{max}$ & $N_\mathrm{V^2}$ & $N_\mathrm{CP}$ & \ S.\\
(RJD) & (nm) & (m) & (m) & & \\
\hline\noalign{\smallskip}
$54\,064-54\,293$ & $H$-band & 81 & 330 & 858 & 556 & $2^*$ \\
$56\,465-56\,474$ & $525-840$ & 54 & 246 & 5\,992 & -- & 1 \\
$56\,463-56\,474$ & $H$-band & 33 & 330 & 4\,154 & 4\,978 & 2 \\
$56\,463-56\,471$ & $562-861$ & 12 & 53 & 5\,790 & 1\,892 & 3 \\
% ??? $|T_3|$
\hline
\end{tabular}
\tablefoot{$^*$ These observations were not used in the model fits.\\
$\Delta T$ denotes the time span between the first and the last measurement,
$\Delta\lambda$ the wavelength range of the observations,
$B_\mathrm{min}$ the minimum projected baseline,
$B_\mathrm{max}$ the maximum projected baseline,
$N_\mathrm{V}$ the number of calibrated squared visibility measurements, and
$N_\mathrm{CP}$ the number of closure phase measurements.
In column ``S.'', individual interferometers are distinguished:
1. CHARA/VEGA,
2. CHARA/MIRC,
3. NPOI.
}
\end{table}

\begin{table}
\centering
\caption{Journal of analyzed photometry.
\label{tab:lc:journal}}
\begin{tabular}{cccr}
\hline\hline\noalign{\smallskip}
$\Delta T$ & $N$ & Passband & Source\\
(RJD) & & or $\Delta\lambda$ (nm) & \\
\hline\noalign{\smallskip}
40\,889--40\,904&    783 &   143--332 & 1\\
41\,963--42\,224&    144 &     $JKLM$ & 2\\
45\,207--47\,607&     92 & 125--136.5 & 3 \\
52\,782--53\,311&    158 &    $JHKLM$ & 4\\
56\,488--57\,232& 1\,120 &      \ubvr & 5\\
56\,494--57\,658& 1\,627 &       \bvr & $^*$6\\
\hline
\end{tabular}\\
\tablefoot{
$\Delta T$ denotes the time span between the first and the last measurement,
$N$ denotes the total number of measurements in all passbandes together.
Column ``Passband'' lists photometric filters of Johnson series.
In column ``Source'':
1. OAO A-2 \citet{kondo1994},
2. \citet{jameson1976},
3. IUE \citet{kondo1994}.
4. \citet{taran2005},
5. Hvar observatory,
6. private observatory of PS,
$^*$These observations were not fitted.
}
\end{table}
%

%%%%%%%%%%%%%%%%%%%%%%%%%%%%%%%%%%%%%%%%%%%%%%%%%%%%%%%%%%%%%%%%%%%%%%%%
\section{Choosing the initial physical properties for detailed modeling}

\subsection{Distance estimates}
\subsubsection{Trigonometric parallax}
\cite{esa97} published the \emph{Hipparcos} parallax of
\bl $0\farcs00370\pm0\farcs00052$. \citet{leeuw2007b,leeuw2007a} carried out
a~new reduction of \emph{Hipparcos} data to obtain a~more accurate value of
$0\farcs00339\pm0\farcs00017$.
These values translate to the following distances estimates and $1-\sigma$ ranges:

\centerline{270~pc; range 237 -- 314~pc,}
\centerline{295~pc; range 281 -- 311~pc,}

\noindent for the original and improved \emph{Hipparcos} parallax, respectively.

\subsubsection{Dynamical parallax from the orbital solution and spatial
resolution of the orbit}
\citet{zhao2008} used two different techniques of image reconstruction
and a simple model with two uniformly illuminated ellipsoids and derived
three different distance estimates from the model and two methods of
image reconstruction, respectively:

\centerline{314~pc; range 297 -- 331 pc,}
\centerline{278~pc; range 254 -- 302 pc,}
\centerline{274~pc; range 240 -- 308 pc.}

\noindent It is important to realize that these distances are based on
the value of the projected value of the semimajor axis
$a\sin i = (57.87\pm0.62)$~\rs\, defined by the Kepler's Third Law and the
binary masses, which were estimated from several previous studies.
In this study, we shall use a similar approach to provide an independent
distance estimate based on our new models of the continuum radiation
of \bl.

\subsubsection{$\beta$~Lyr~B as a~distance indicator of $\beta$~Lyr~A}\label{sec:blb}

\blb (HD~174664, BD$+33^\circ3224$, STFA~39B) is the second brightest member
of the \bl visual system. It is a~main-sequence B5\,V star. A~somewhat
puzzling is the fact that it is also an~X-ray source \citep{berg94}.
\citet{abt62} measured its RV on eleven blue photographic spectra
and concluded that the star is a single-line spectroscopic binary
with an orbital period of 4\fd348 and an eccentric orbit.
They also studied available astrometric observations for all visual members
of the \bl~system and concluded that \blb is gravitationally bound to \blae.
\citet{abt76} obtained new photographic RVs of \blb and carried out
an error analysis of newly obtained and earlier RVs to conclude that
there is no evidence for duplicity of \ble. From spectral classification
they concluded that \blb is a normal B7V star close to the zero-age
main sequence.

Here, we critically re-investigate these pieces of information to see whether
\blb can be used to another distance estimate of \blae. We first study the kinematic and radiative properties of \ble.
Radial velocities (RVs) were measured on Ond\v{r}ejov Reticon red spectra
of \blb secured from 1995 to 1996, and on six Ond\v{r}ejov
red CCD spectra from 2003 to 2016. The linear dispersion
of all these spectra was 17.2~\Ame, and their two-pixel
resolution was 12700. Additional details on the reduction
of \blb spectra are presented in Appendix~\ref{sec:apc}.

RVs of \blb were determined using two methods:
\begin{enumerate}
\item An~interactive comparison of direct and flipped line profiles on the
computer screen until the best match is achieved in program {\tt SPEFO}
\citep{sef0,spefo}. RVs were measured independently on
four spectral lines \ion{Si}{ii}{\,6347\,\r{A}}, \ion{Si}{ii}{\,6371\,\r{A}},
\hae, and \ion{He}{i}{\,6678\,\r{A}}.
\item Via an automatic comparison of the observed and
synthetic spectra in the \pyte~program \citep{jn2016}.\footnote{
The program is available at \url{https://github.com/chrysante87/pyterpol}.}
\end{enumerate}
The latter method produced slightly less scattered RVs, hence
only these are presented here.  However, the RVs obtained by both
methods are listed in~Table~\ref{tab:blb:rv}.
At the same time, \pyt also estimated radiative properties of \blae:
\tefe, \lgge, and the projected rotational velocity \vsine.
The red spectra we used ($\Delta\lambda \simeq 6200 - 6800$\,\r{A})
contain numerous telluric lines, which would adversely affect the
results of \pyte. We handle the problem using two methods:
(i)~$1$\,\r{A} intervals centered on each telluric spectral
line were omitted from each spectrum, and
(ii)~telluric and stellar spectra were separated
by spectral disentangling \citep{simon1994} in Fourier space by the \korel
program \citep{hadrava1995,hadrava1997}. \pyt was then used for both, individual
observed spectra with telluric-line regions omitted, and to disentangled
stellar spectrum, free of telluric lines.
The synthetic spectra were taken from the BSTAR \citep{lanz2007} and
AMBRE \citep{palacios2010} grids. The results were similar, though
not exactly the same. We simply adopted their mean as a~realistic estimate
of radiative properties of \ble, noting that both methods may introduce
some systematic errors. The observed spectra still contained
remnants of weaker telluric lines, while the disentangled spectra
were slightly warped and had to be re-normalized.
The formal errors derived from the two solutions are unrealistically small,
as they do not reflect possible systematic errors.
The results are summarized in Table~\ref{tab:blb:spectrafit} and
a~comparison of one observed profile, and disentangled
profiles with the best-fitting synthetic spectra
is shown in Fig.~\ref{fig:blb:spectrafit}.
We assumed the solar composition in these fits guided
by the results of \citet{abt76} from the blue spectra since the red
spectra at our disposal do not contain enough spectral lines to estimate
the metallicity of \blb reliably.

\begin{table}
\begin{center}
\caption{Kinematic and radiative properties of \blb
estimated via a~comparison of its disentangled spectrum and all
observed spectra with synthetic ones using \pyte.
\label{tab:blb:spectrafit}}
\begin{tabular}{lrrrr}
\hline\hline\noalign{\smallskip}
 & \multicolumn{3}{c}{Method} \\
Quantity & \multicolumn{1}{c}{(1)}
& \multicolumn{1}{c}{(2)}
& \multicolumn{1}{c}{(3)} \\
\noalign{\smallskip}\hline\noalign{\smallskip}
\tef (K) & 15\,197(15) &14\,823.9(7.4) &15\,000(200) \\
\lgg\ [cgs]  & 4.3036(25) & 4.1965(23) & 4.25(5)\\
%$L (unit?)$  &  0.9924(8) & 0.9962(1) & 0.994(2) \\
\vsin (\ks)& 95.01(26) & 89.62(34)& 92(3) \\
$\gamma$ (\ks)& $-18.42$(12) & $-17.86$(15) &$-18.1$(0.5) \\
$\chi^2_\mathsf{R}$  & 6.22
& 3.82 & --\\
\noalign{\smallskip}\hline\noalign{\smallskip}
\end{tabular}
\tablefoot{Methods:
(1)~From a fit of the disentangled spectrum,
(2)~from a fit of all observed spectra, and
(3)~the mean values from~(1) and~(2).}
\end{center}
\end{table}

\begin{figure}
\centering
\includegraphics[width=0.5\textwidth]{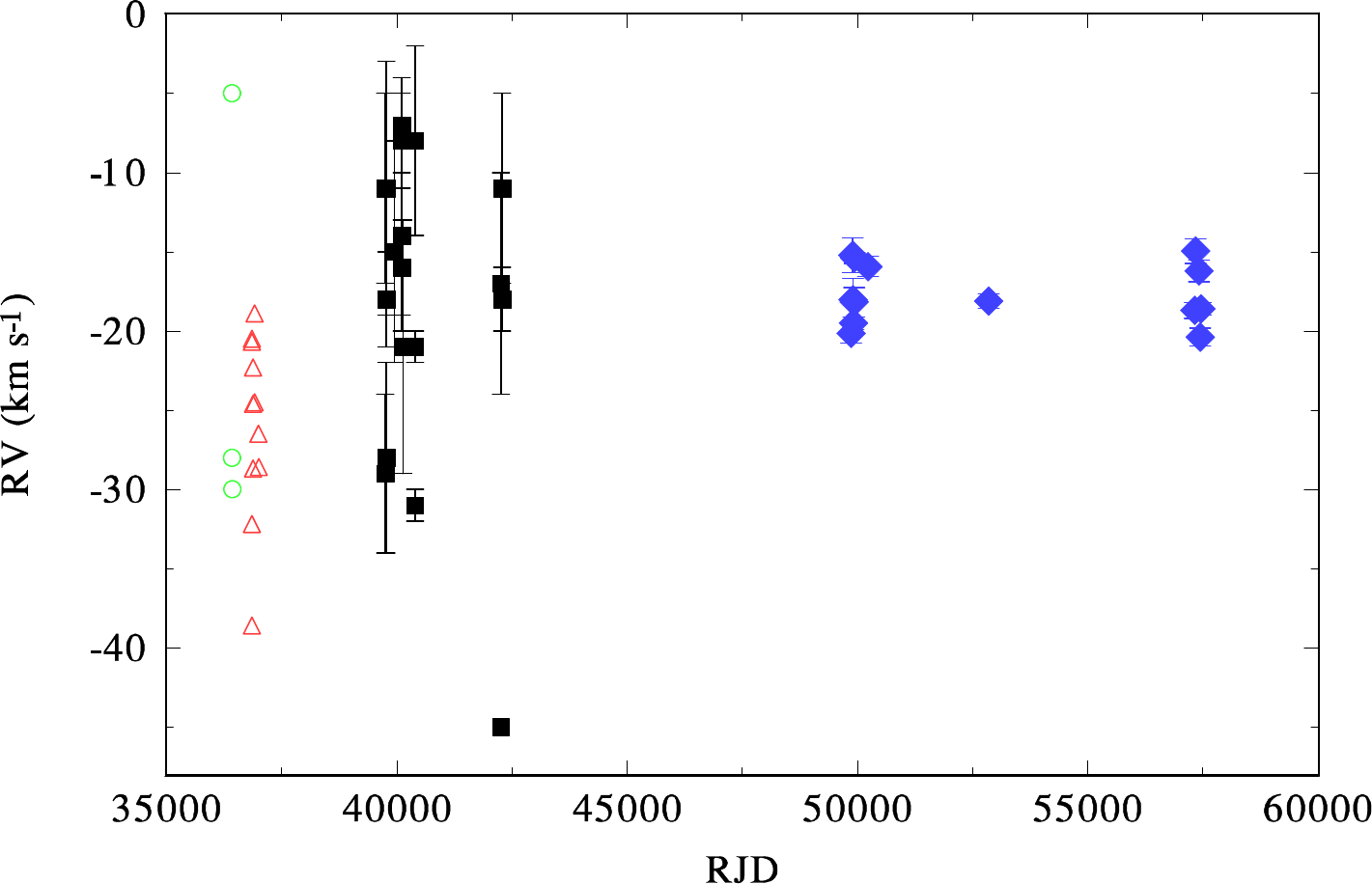}
\caption{Available RVs of \ble. The first three datasets
are RVs measured on photographic spectra of moderate dispersion:
green open circles denote the RVs measured by~\citet{pp61},
the red open triangles come from \citet{abt62}, and the black boxes
are RVs from \citet{abt76}. The filled blue diamonds are the RVs measured
on Ond\v{r}ejov electronic spectra with \pyte. The rms errors, available
only for \citet{abt76} and new RVs are also shown.
We note that the most deviant RV of $-45$~\kms is based on a single line
and comes probably from an underexposed spectrum.}
%The blue lines represent
%the 1-$\sigma$ range of the systemic velocity of \blb,
%$\gamma_\mathsf{B} = -18.1 \pm 0.5$\,\ks. The red lines represent
%the range from $-20$ to $-17$~\ks~of the systemic velocity of \blae.}
\label{fig:blb:rv}
\end{figure}

Do \bla and \blb indeed form a physical bounded system?
All RV measurements of \blb
are plotted in Fig.~\ref{fig:blb:rv}. The rms errors are only available
for \citet{abt76} and for our new RVs and are shown in this plot.
We explain the large scatter of the RVs from the photographic plates
by the combination of four factors: moderate dispersion of the plates,
variable quality of individual spectra, relatively large projected
rotational velocity of \ble, and line blending.
It is our experience that due to blends with fainter spectral lines
the apparent RVs of individual measured lines can differ systematically.
Since different number of spectral lines were measured on different
plates, their average RVs are then affected differently.
This effect should be absent in \pyt RVs, based on the comparison
of whole spectral segments with interpolated synthetic spectra.
Also our manual RV measurements in \spefo should be basically free
from these effects since one sees which part of the profile is measured.
To check on this, we derived robust mean RVs for the four stronger lines
seen in the Ond\v{r}ejov spectra \citep[using the algorithm published
by][]{andrews72}. They are summarized in Table~\ref{robust} and confirm
a very good agreement of all mean RVs of individual lines.
The systemic velocity of \blb based on \pyt RVs
is $\gamma_\mathsf{B}=-(18.1\pm0.5)$\,\ks.
This number is a~weighted mean of two estimates. The first one is
the mean RV derived from RV measurements on individual
spectra, and the latter is measured on the disentangled spectrum
(both are presented in Table~\ref{tab:blb:spectrafit}).
For \spefo RVs from individual lines one gets a 1-$\sigma$ range
from $-17$ to $-21$~\ks. These estimates are to be compared to the
systemic RV of \blae, $\gamma_\mathsf{A}$, which was found
to be in the range from about $-17$ to $-20$~\ks, depending
on the spectrograph used \citep[see for example, solution~5 in Table~10 of][]{hec96}.
From this, one can conclude that the systemic RVs of \bla and \blb are
identical within the range of their measuring errors.
In spite of their large scatter, the RVs of \blb measured on old photographic
spectra by~\citet{pp61} and~\citet{abt62, abt76} do not show
any obvious long-term RV trend.

The range of individual RVs based on our measurements with their corresponding
rms errors could lead to the suspicion that \blb could be a spectroscopic
binary after all. To check on this, we run period searches down to 0\fd5
for \pyt RVs and also \spefo RVs for individual lines. All the periodograms
show a dense forest of comparably deep peaks but the frequencies found
differ mutually, both for the individual lines and for the \pyt RVs.
From this, we reinforce the conclusion of \citet{abt76} that \blb is
a single star.

\begin{table}
\begin{center}
\caption{Robust mean RVs of individual spectral lines measured in \spefo
for Ond\v{r}ejov spectra of \ble.}
\label{robust}
\begin{tabular}{cc}
\hline\hline\noalign{\smallskip}
Line & Robust RV with rms error\\
\noalign{\smallskip}\hline\noalign{\smallskip}
\ion{Si}{ii}~6347 \AA&$-18.6\pm1.3$\\
\ion{Si}{ii}~6371 \AA&$-18.9\pm1.9$\\
\ion{H}{i}~6563 \AA  &$-18.1\pm0.9$\\
\ion{He}{i}~6678 \AA &$-18.4\pm1.2$\\
\noalign{\smallskip}\hline\noalign{\smallskip}
\end{tabular}
\end{center}
\end{table}

\begin{figure}
\centering
\includegraphics[width=0.5\textwidth]{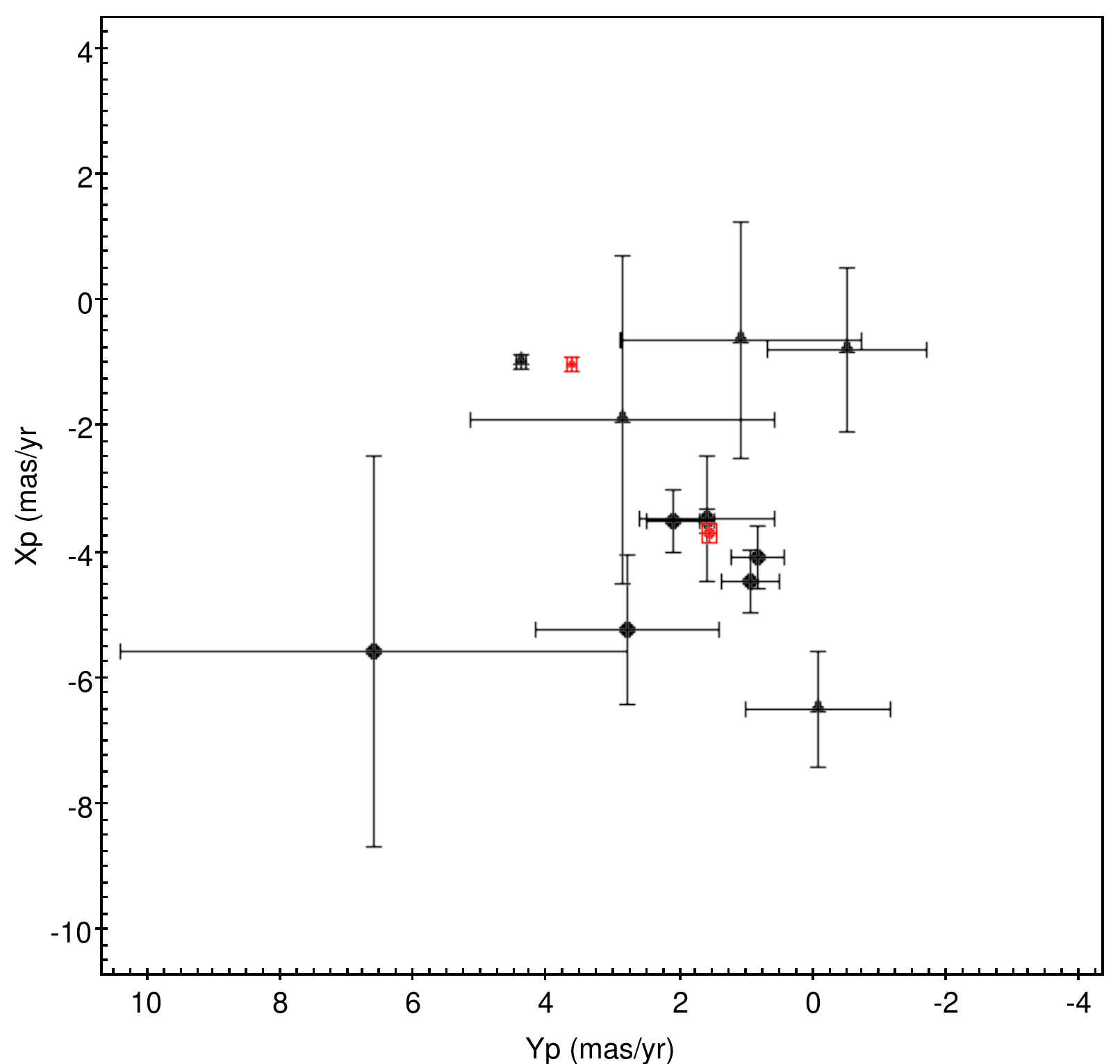}
\caption{Proper motions of \bla and \ble.
The black points denote proper motion measurements of \blae,
the black triangles proper motions of \ble,
the red point the mean weighted proper motion of \blae, and
the red triangle the mean weighted proper motion of \ble.
$X_\mathsf{\mu}$, $Y_\mathsf{\mu}$ are Cartesian
coordinates of the proper motion vector
given by Eqs.~(\ref{eq:pm0}) and~(\ref{eq:pm1}).
The former is north-south oriented, and the latter
east-west oriented.
\label{fig:blb:pm}}
\end{figure}

The proper motions of \bla and \blb were also investigated.
Measurements accessible through VizieR  database
\citep{VizieR} at CDS were downloaded. Only values of
$\mu_\alpha$ and $\mu_\delta$ published after the~Hipparcos
mission were retained.
Proper motions in the right ascension were corrected for the declination of
both systems (see Appendix~\ref{sec:apc:pm} for details). Their weighted mean
is given in Table~\ref{tab:blb:meanpm}.
Individual measurements and the weighted average are shown
in Fig.~\ref{fig:blb:pm}.
\begin{table}[h!]
\centering
\caption{Weighted mean proper motions of \bla and \ble.
\label{tab:blb:meanpm}}
\begin{tabular}{lrr@{$\pm$}lr@{$\pm$}l}
\hline\hline\noalign{\smallskip}
 Component & Unit & \multicolumn{2}{c}{\blae} & \multicolumn{2}{c}{\ble} \\
 \hline\noalign{\smallskip}
 $\mu$ & ($\mathrm{mas\,yr}^{-1}$) & 4.03&0.15 & 4.74&0.25 \\
 $\theta$ & (deg) & 157.3&1.7 & 122.1&2.9 \\
\noalign{\smallskip}\hline
\end{tabular}
\tablefoot{$\mu = \sqrt{\mathrm{X}^2_\mu + \mathrm{Y}^2_\mu}$ denotes
the magnitude of the proper motion vector, and
$\theta = \arctan\left(\mathrm{X}_\mu / \mathrm{Y}_\mu\right)$
the position angle with respect to north, where
$\mathrm{X}_\mu$ and $\mathrm{Y}_\mu$ are given by
Eqs.~(\ref{eq:pm0}) and~(\ref{eq:pm1}).}
\end{table}

The close similarity of systemic RVs of \bla and \blb reinforce
the hypothesis that the two systems formed in the same
association or are even physically linked.
The same cannot be said for the proper motions
or for the respective tangential velocities,
which differ from each other quite significantly.
We conjecture that a~part of this difference can be attributed
to the mutual orbit -- if $\beta$~Lyr~A and~B are bound --
or to the intrinsic velocity dispersion~$\sigma_v$ of the putative stellar
association. If the A-B orbit is circular and seen more or less edge on,
the Keplerian velocity $v_{\rm kepl} = \sqrt{GM_{\rm tot}/(\alpha d)}$
is on the order of $1.2\,{\rm km}\,{\rm s}^{-1}$,
assuming the mass $M_{\rm tot} \simeq 20\,{\rm M}_\odot$,
the angular separation $\alpha \simeq 40\,{\rm arcsec}$,
and the distance $d \simeq 325\,{\rm pc}$.
This should be compared with the relative tangential velocity
$v_{\rm t} = |\vec\mu_\mathsf{A}-\vec\mu_\mathsf{B}| d \simeq 4.2\,{\rm km}\,{\rm s}^{-1}$.
This value seems larger than $v_{\rm kepl}$ and might be on the order of
typical $\sigma_v$.

Nonetheless, the spread of proper motion measurements of \blb is
large and the weighed mean is dominated by the
latest \emph{Gaia} observation. If proper motion measurements of \blb are given equal weight,
the mean proper motion is:
$\mu_\mathsf{B} = 3.4 \pm 2.4$\,$\mathrm{mas\,yr}^{-1}$,
$\theta_\mathsf{B} = 151 \pm 43$\,deg, that is in
agreement with measurements of \blae.
Therefore, we think that the proper motions of both systems do not
contradict the conclusion that the systems formed within
the same association. That alone allows us
to use \blb as a distance indicator for \blae.

\citet{gaia2} published the first parallax derived by the \emph{Gaia}
satellite \citep[see][]{gaia1} for \ble,
$0\farcs00310\pm0\farcs00036$, which translates to a distance

\centerline{$d=323$~pc; range 289 -- 365~pc\,.}

In April 2018, the second DR2 release of the \emph{Gaia} catalog was made
publicly available. The improved value of the \blb parallax is
$0\farcs003006\pm0\farcs000054$, which translates to a rather narrow distance
range

\centerline{$d=333$~pc; range 327 -- 339~pc\,.}

We have also derived a~spectroscopic distance of \blb
using its radiative properties
(see Table~\ref{tab:blb:spectrafit}, third column), and the mean
all-sky~\ubv~magnitudes based on 77 observations acquired at Hvar observatory
during three seasons. The mean all-sky magnitudes:

\centerline{$V=7$\m199, \bv$=-0$\m089, \ub$=-0$\m541}

\noindent were dereddened in a standard way, which resulted in:

\centerline{$V_0=6$\m99, $E$(\bv)=0\m070\,.}

\noindent Assuming that \blb is a~main-sequence star,
its mass and radius can be estimated from relations of \citet{mr88} to be
$M_\mathsf{B} = 4.16^{+0.08}_{-0.09}$\,\ms, and
$R_\mathsf{B} = 2.53^{+0.18}_{-0.16}$\,\rs, respectively.
Adopting bolometric corrections of \citet{popper80}, the absolute magnitude
of \blb is $M_\mathsf{V} = -0.08^{+0.17}_{-0.16}$\,mag, and
its spectroscopic distance is:
\begin{eqnarray}
d_\mathsf{B}^\mathsf{Sp.} = (259 \pm 20)\,\mathrm{pc}. \nonumber
\end{eqnarray}
This is a significantly lower value than what \emph{Gaia} obtained but
it is actually in qualitative agreement with recent findings by
\citet{stassun2016} that \emph{Gaia} DR1 parallaxes are smaller than photometrical ones for well studied eclipsing binaries. Their finding was confirmed also by \citet{graczyk2017}.
Since we note that in many
cases the DR2 parallaxes do agree with the DR1 ones within the quoted errors
(significantly smaller in DR2), the above warning might also be relevant
for the DR2 parallaxes.

As pointed out by the anonymous referee, possible duplicity
of \blb could increase its observed luminosity and, therefore, its distance
from us. Without a direct imaging, one cannot exclude indeed the possibility
that \blb is composed of two similar B7V stars seen pole on. In that case
the spectroscopic distance of such an object could be
as large as $\sim 370$~pc. While it would certainly be desirable to obtain
new, high-dispersion and high-S/N spectra of \blb and derive its more
precise RVs to check on their constancy, we do not find the duplicity of
\blb as too probable. Given the fact that the object was detected as
an X-ray source, it is more probable that any putative secondary
would be a late-type star with a chromospheric emission.
However, no lines of a cool secondary have ever been detected so such
an~object - if present - must be much fainter than the primary. If so,
its presence would not affect our spectroscopic distance
estimate significantly.

\subsection{The mass ratio and orbital solutions}

\begin{table}
\begin{center}
\caption[]{Various estimates of the binary mass ratio of \be.}\label{qest}
\label{jouinter}
\begin{tabular}{cllrrrrrrr}
\hline\hline\noalign{\smallskip}
$m_{\rm g}/m_{\rm d}$& Source & Note\\
\noalign{\smallskip}\hline\noalign{\smallskip}
2.83 -- 4.05 & \citet{hec90}  &1\\
$4.28\pm0.13$& \citet{skul91} &2\\
 $4.68\pm0.1$& \citet{skul92} &2\\
 4.484       & \citet{hs93}   &3\\
 4.432       & \citet{bis2000}&4\\
\noalign{\smallskip}\hline\noalign{\smallskip}
\end{tabular}
\tablefoot{
1. from the spin-orbit corotation of star 2;
2. from a RV solution for Crimean 3~\Ame\ CCD spectra of \ion{Si}{ii} lines;
3. from a RV solution for all at that time available RVs;
4. from a \korel disentangling RV solution for the selection of
     68 red Ond\v{r}ejov Reticon spectra with the highest S/N.
}
\end{center}
\end{table}

Because the RV curve of the star hidden in the disk is only defined by
a pair of fainter \ion{Si}{ii} absorption lines at 6347 and 6371~\AA\,
discovered by \citet{sahade66} and \citet{skul75}, one should also consider
some range of plausible mass ratios. Various relevant estimates are
summarized in Table~\ref{qest}.

For the purpose of our modeling, we shall use the orbital solution
of \citet{hs93} based on \ion{Si}{ii} lines for both stars (their
solution~6 of Table~10), which was basically confirmed by more extended series
of spectra by \citet{hec96}. In particular, we shall adopt
\centerline{$K_1=41.4\pm1.3$, $K_2=186.30\pm0.35$, and $a\sin~i=58.19$~\rs,}
\noindent which implies the mass ratio of 4.500.

%%%%%%%%%%%%%%%%%%%%%%%%%%%%%%%%%%%%%%%%%%%%%%%%%%%%%%%%%%%%%%%%%%%%%%%%

\section{Modeling the continuum radiation}\label{sec:shellspec}

Our modeling of \bla and its application
to spectro-inter\-fero\-met\-ric and photometric observations
is presented in the following section.
Our models are based on the program \shellspec,
developed by \citet{budaj2004}. This program
was equipped with additional features that
simplify modeling of binaries, computations of synthetic interferometric
observables, and a~solution of the inverse task.

First, we briefly outline the program and
its new Python wrapper called {\tt Pyshellspec}. Then, we proceed
to the development of several alternative models of \bl in continuum
and their comparison to observations.

\subsection{Foundations of the model}\label{sec:shellspec:overview}

The model is based on the existing program \shellspec\,
which was designed for the computation of synthetic light curves, spectra and images
of stars and/or binaries surrounded
by a moving 3D circumstellar medium
by means of solving a one-dimensional LTE radiation transfer along some user-specified line-of-sight.

To provide an environment for solving the transfer,
the modeled (non-stellar) objects cannot be represented with
a~\hbox{2-D} mesh covering only the photosphere of each object.
Instead the grid has to sample the whole volume in 3-D.
The only two exceptions are models of stars, which are
always opaque, hence their atmospheres form a~boundary condition.
In \shellspec, each object
is placed into an regularly sampled cuboid
(divided into $n_x \times n_y \times n_z$ cells)
that represents ``the Universe''. All embedded
objects inherit the spatial sampling of the Universe.
Every object has a~simple geometric
shape given by several parameters.
Each cell occupied by an~object is assigned its density (gas, electron, and dust),
temperature and velocity.

We use the following setup:
LTE level populations,
LTE ionisation equilibrium,
the line profile is determined by
thermal,
microturbulent,
natural,
Stark,
Van der Waals broadenings, and
the Doppler shift.
The continuum opacity is caused by
HI~bound-free,
HI~free-free,
${\rm H}^-$~bound-free, and
${\rm H}^-$~free-free transitions.
Moreover, we account for the line opacity of \ha, \he, and \her,
for the future spectral line analysis of the VEGA data.
Abundances are assumed to be solar.
We use a small grid of synthetic spectra for the stars,
generated by Pyterpol \citep{jn2016} from Phoenix, BSTAR, and OSTAR grids \citep{husser2013,lanz2007,lanz2003}.
The stars are subject to the Roche geometry, limb darkening,
gravity darkening (in particular the Roche-filling donor),
and the reflection effect concerning the heat redistribution over the surfaces.

On the other hand, we did {\em not\/} include the Thomson scattering on free electrons,
the Rayleigh scattering on neutral hydrogen,
because these are only implemented as optically thin
(single) scattering in~\shellspec.
It would be a much harder computational problem to account
for multiple scattering in 3D moving optically thick medium,
and would essentially prevent us to converge the model
with many parameters.
There is also neither irradiation nor reflection between the stars.
As the disk is presumably hotter than the silicate condensation temperature,
we account for neither Mie absorption on dust, Mie scattering, nor dust thermal emission.

The velocity field of an~object is given either by a~net velocity or
by rotational velocity, or a combination of both.
The observable quantities (flux and intensity)
are computed for a~line-of-sight, specified by~two angles.
The model is only kinematic, that is the
radiation field has no effect on the state
quantities of the circumstellar medium.

\shellspec~allows modeling of various structures,
but we restrict ourselves to those that are relevant
for the continuum model of \blae. A~very detailed description
of \shellspec~is in~\citet{budaj2004}, and its latest
improvements are described in~\citet{budaj2011a}, \citet{budaj2011b}.

\subsection{Interface for automatic comparison and fitting}\label{sec:shellspec:pyshellspec}

On output, \shellspec~computes the monochromatic flux
$F_\nu$ (in ${\rm erg}\,{\rm s}^{-1}\,{\rm cm}^{-2}\,{\rm Hz}^{-1}$),
and intensity $I_\nu(x, y)$ (in ${\rm erg}\,{\rm s}^{-1}\,{\rm cm}^{-2}\,{\rm sr}^{-1}\,{\rm Hz}^{-1}$) projected onto the
plane perpendicular to the line-of-sight,
where $x$ and $y$ are the Cartesian coordinates in
this plane, but it does not carry out a comparison
of these quantities to observed ones,
or automatic optimization of the model parameters.
Moreover, individual model components implemented in
\shellspec~are not bound by any orbit, although
some model parameters depend on orbital parameters.
The components are almost independent and
do not share parameters that are common. To overcome
these limitations we wrapped \shellspec~in a~Python
interface.%
\footnote{Available at \url{http://sirrah.troja.mff.cuni.cz/~mira/betalyr/}.}

\subsubsection{Computation of synthetic magnitudes, squared visibilities and closure phases}

In our approach, the passband flux (e.g., $F_V$) is computed from
the monochromatic flux $F_\lambda = F_\nu c/\lambda^2$ for a single
(effective) wavelength~$\lambda_{\rm eff}$,
given by the transmission curve of the respective filter.
Its relation to the passband magnitude is simply
$m_V = -2.5\log_{10} {F_\lambda/F_{\rm calib}}$,
where $F_{\rm calib}$ denotes the calibration flux (also monochromatic),
given for example by \cite{johnson1966}, \cite{kondo1994}.
This approximation is required because radiation transfer computations
for many wavelengths would be very time consuming.
Unfortunately, this may lead to a slight offset $\Delta_\mathsf{P}$
between the observed and synthetic light curves.
The offset is thus determined by a minimization of
the following formula:
\begin{eqnarray}
\label{eq:lcoffset}
S(\Delta_\mathsf{P}) = \sum_{i=1}^{N} \left| F_i^\mathsf{obs} - F_i^\mathsf{syn} + \Delta_\mathsf{P}\right|, 	
\end{eqnarray}
where $F^\mathsf{obs}$ is the observed flux,
and $F^\mathsf{syn}$ the synthetic flux.

The FWHM of the passbands~$\Delta\lambda_{\rm eff}$
of interferometric observations were as low as
$20$\,nm in the visible and $50$\,nm in the infrared.
Therefore the passband intensity~$I_{\rm p}$ was calculated
in a similar way from the monochromatic intensity~$I_\nu$ for the effective wavelength.
This approximation is validated by the fact that the continuum does not change significantly through the narrow passbands.
Images are normalized afterward,
rendering any offset with respect to the emergent intensity insignificant.

The images produced by \shellspec~are centered on the primary
component (the gainer), and their nodal line is aligned with the
north-south direction. Hence, the image center has to be
shifted to the system barycenter first and then rotated
to a~given longitude of the ascending node~$\Omega$.
The complex visibility $V(u, v)$ is computed by a two-dimensional
discrete Fourier transform (DFT) of the image at spatial
frequency $\mathbf{b} =  (u,v) = (B_x, B_y) / \lambda$,
where $B_x$ ($B_y$) denotes the~projection of the baseline into east-west
(north-south) direction. The triple product $T_3$ is
then computed as~follows:
\begin{eqnarray}
	\label{eq:tripleproduct}
	T_3(b_1, b_2) & = & V(b_1)V(b_2)V^*(b_1+b_2),
\end{eqnarray}
where $b_1$ and $b_2$ denote
spatial frequencies corresponding to a pair
of baselines in a~closing triangle. Performance of the
two-dimensional fast Fourier transform (FFT) for computation
of interferometric observables was also evaluated.
The need for an~extensive zero-padding of each image
produced by~\shellspec~to obtain a sufficient resolution,
and interpolation within the two-dimensional array made it
actually slower than DFT.

\subsubsection{Interface for the solution of the inverse task}
The optimization of model parameters was carried out through the minimization of the total $\chi^2$ defined as the sum of the $\chi^2$ of the different data sets. MIRC providing triple product quantities that are not totally independent from the $V^2$ estimates, we decided to use a specific weight on the MIRC $V^2$ and $T_3$ data.
\begin{eqnarray}
	\label{eq:chi2total}
	\chi^2 = \chi^2_\mathsf{LC} + \chi^2_\mathsf{IF},
\end{eqnarray}
where the contribution{s} of individual types of
observations are given by the following relations:
\begin{eqnarray}
	\label{eq:chi2lc}
	\chi^2_\mathsf{LC} &=& \sum_{i=1}^{N_\mathsf{P}}\sum_{j=1}^{N_\mathsf{M}}\left(\frac{m^\mathsf{obs}_{i,j} - \tilde{m}^\mathsf{syn}_{i,j}}{\sigma_{i,j}}\right)^2, \\
	\label{eq:chi2if}
    \chi^2_\mathsf{IF} &=& \chi^2_\mathsf{IF_{VEGA}} + \chi^2_\mathsf{IF_{NPOI}} + \chi^2_\mathsf{IF_{MIRC}},
\end{eqnarray}
with
\begin{eqnarray}
\chi^2_\mathsf{IF_{VEGA}} &=& \chi^2_\mathsf{V^2}, \nonumber \\
\chi^2_\mathsf{IF_{NPOI}} &=& \chi^2_\mathsf{V^2} + \chi^2_\mathsf{CP}, \nonumber \\
\chi^2_\mathsf{IF_{MIRC}} &=& \frac{1}{2}\left(\chi^2_\mathsf{V^2} + \chi^2_\mathsf{T_3}\right) + \chi^2_\mathsf{CP}, \nonumber \\
\chi^2_\mathsf{V^2} &=&\sum_{i=1}^{N_\mathsf{V^2}}\left(\frac{\left|V^\mathsf{obs}\right|_i^2 - \left|V^\mathsf{syn}\right|_i^2}{\sigma_i}\right)^2, \nonumber \\
\chi^2_\mathsf{T_3} &=& \sum_{i=1}^{N_\mathsf{T_3}}\left(\frac{\left|T_3^\mathsf{obs}\right|_i - \left|T_3^\mathsf{syn}\right|_i}{\sigma_i}\right)^2, \nonumber \\
\chi^2_\mathsf{CP} &=& \sum_{i=1}^{N_\mathsf{T_3}}\left(\frac{T_3\phi_i^\mathsf{obs} - T_3\phi_i^\mathsf{syn}}{\sigma_i}\right)^2. \nonumber\\
\end{eqnarray}
where
$m$ denotes the magnitude,
$\tilde{m}$ the magnitude corrected
for the offset given by Eq.~(\ref{eq:lcoffset}),
$N_\mathsf{M}$ the number of photometric
observations for $j$-th passband,
$N_\mathsf{P}$ the number of fitted passbands,
$\left|V\right|^2$ the~squared visibility,
$\left|T_3\right|$ the~modulus of the triple product,
$T_3\phi \equiv \arg\lbrace{T_3\rbrace}$ the closure phase,
$N_\mathsf{V^2}$ the number of squared visibility
observations,
$N_\mathsf{T_3}$ the number of triple product
observations, and
$\sigma$'s are the uncertainties of the corresponding
observations.
MIRC provides triple product quantities that are not
entirely independent from the $V^2$ estimates and we thus decided
to use a specific weight on the MIRC $V^2$ and $T_3$ data.
No additional weighting of the data sets is considered
but a detailed analysis of the convergence process is presented in Table~\ref{tab:chi2}.

Available engines for the {\em global\/} minimization of
Eq.~(\ref{eq:chi2total}) are the differential
evolution algorithm by~\citet{storn1997},
implemented within the SciPy library,
and the Simplex algorithm by~\citet{nelder1965},
implemented within the NLOPT library.

%\footnote{Description and the source code are available at \url{https://docs.scipy.org/doc/scipy-0.18.1/reference/generated/scipy.optimize.differential_evolution.html}.}
%\footnote{Steven G. Johnson, The NLopt nonlinear-optimization package, \url{http://ab-initio.mit.edu/nlopt}.}

As the properties of some of the objects are linked to the orbital elements, we implemented an~orbit
binding the objects together for the global optimization. The~orbit
is given by the~following elements:
the period $P$ at a reference epoch,
the epoch of primary minimum $T_\mathsf{min}$,
the rate of the period increase $\dot{P}$,
the eccentricity $e$,
the semimajor axis $a$,
the mass ratio $q$,
the inclination $i$,
the argument of periastron $\omega$, and
the longitude of ascending node $\Omega$.
The orbit binds the two stars and
supplies values of orbit-dependent parameters
(the masses, Roche-lobe radii, distance between the
primary and secondary, radial velocities, and
the orbital phase for a~given epoch). With this in hand, we then use \shellspec~to compute
the model at different phases.

\subsection{A~model for $\beta$~Lyr\ae~A}\label{sec:shellspec:model}

A~model for individual components of \bla
is introduced here. While the model of the
two binary components is straightforward,
properties of the accretion disk surrounding
the gainer remain uncertain. Hence several models
of the accretion disk were constructed and tested.

\subsubsection{A~model of binary components}

The two stellar components of \bla were approximated
with the following models:
\begin{itemize}
\item{\em The donor} has a Roche geometry and is completely
filling the Roche limit and its rotation is synchronized
with the orbital motion. Its shape and size
are driven by the semimajor axis~$a$, and the mass ratio~$q$.
The gravity darkening (von~Zeipel law,
$\alpha_\mathsf{GD} = 0.25$) and a~linear limb darkening
were assumed. Coefficients of the limb-darkening
law were taken from \citet{vanhamme1993} using the
tri-linear interpolation scheme. For the interpolation,
the polar temperature was used instead of the effective temperature,
gravitational acceleration was approximated with that of a~sphere
with radius equal to the polar radius,
and the solar metallicity was used.
Parameters describing the donor are denoted by
the index~``d''.

\item{\em The gainer} is approximated with a~sphere,
even though it likely rotates close to its critical velocity,
$v_{\rm crit} = v_{\rm kepl}(R_\mathsf{g})$,
because of the ongoing accretion, and should thus have
an ellipsoidal shape and significant gravity darkening.
Alternatively, there can be a thin transition layer \citep{pringle1981},
if the gainer is not (yet) rotating critically.
We also checked that the accretion rate~$\dot M$ is low enough and that the associated radial velocities within the disk are much smaller than Keplerian, $v_{\rm r}\ll v_{\rm kepl}$.
Nevertheless, due to the presence of the accretion disk, the only radiation
that may be able to penetrate the disk is that coming
from polar regions, whose radiation and shape do not
depart from a~spherical star so much. The component
is limb-darkened using the same tables and interpolation
scheme as for donor. The radius was set to a~value
typical for B0.5\,IV-V star, $R_\mathsf{g}=6\,$\rs~\citep{mr88},
which is also in agreement with radii adopted in
earlier studies \citep[e.g.,][]{hec92,al2000}.
Parameters describing the gainer are denoted by
the index ``g''.
\end{itemize}

\subsubsection{Models of the accretion disk}

\shellspec~allows the user to set up several models of an accretion disk. They differ in shape, radial temperature, and density profiles as presented below. In Sec.~\ref{disk-properties}, we give more details on the preferred geometries.

\begin{figure}
	\centering
	\includegraphics[width=0.3\textwidth]{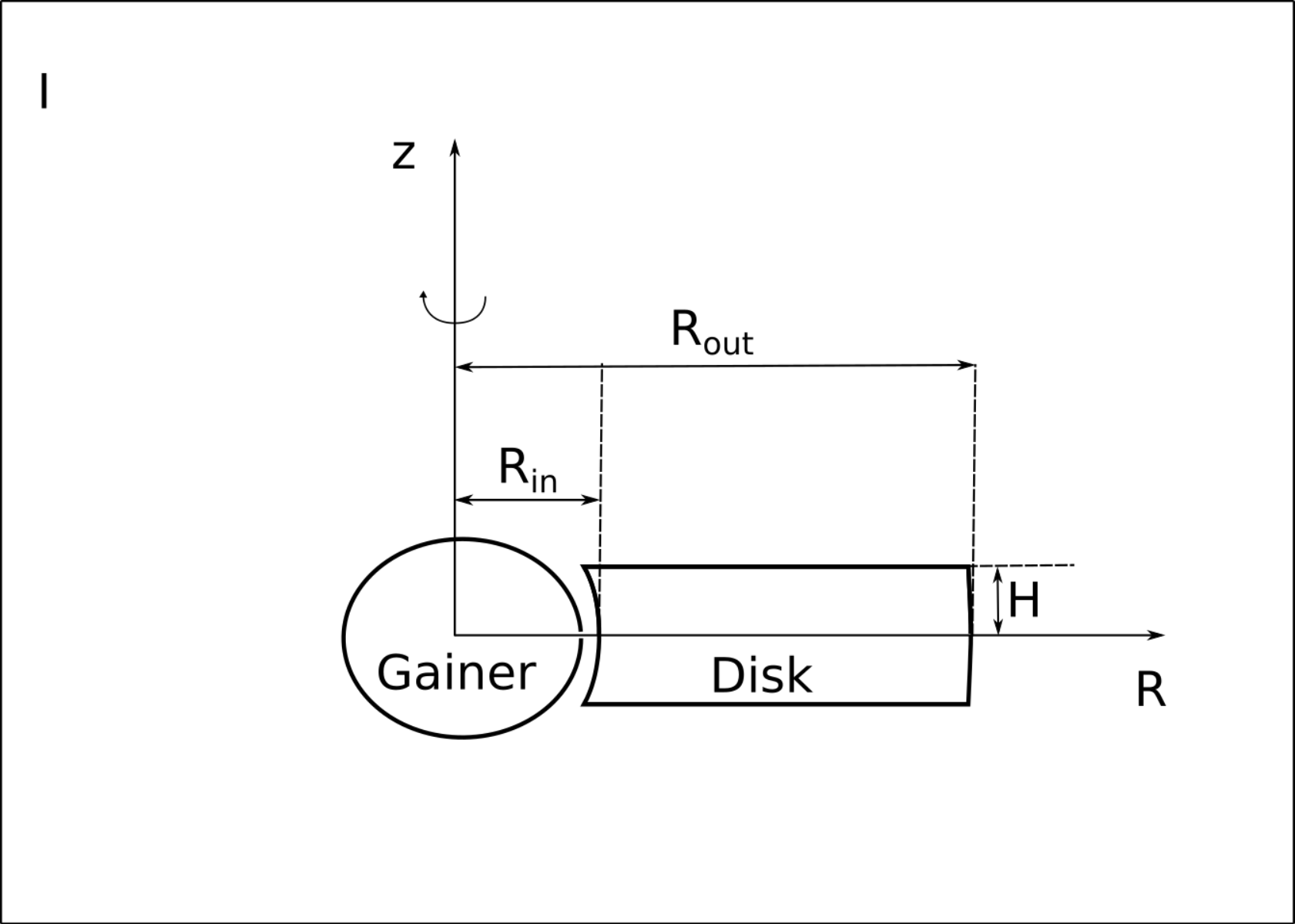}
	\includegraphics[width=0.3\textwidth]{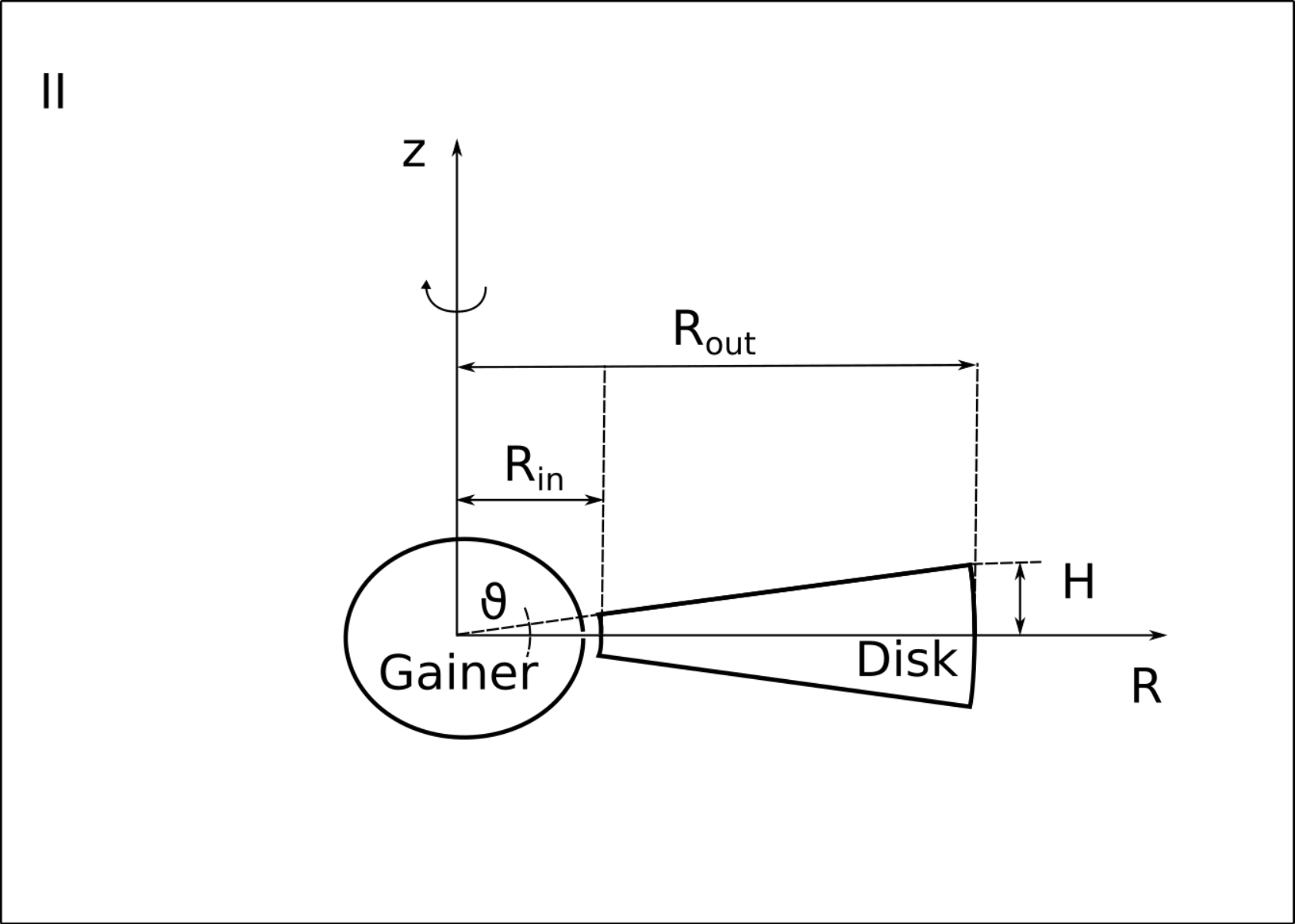}
	\includegraphics[width=0.3\textwidth]{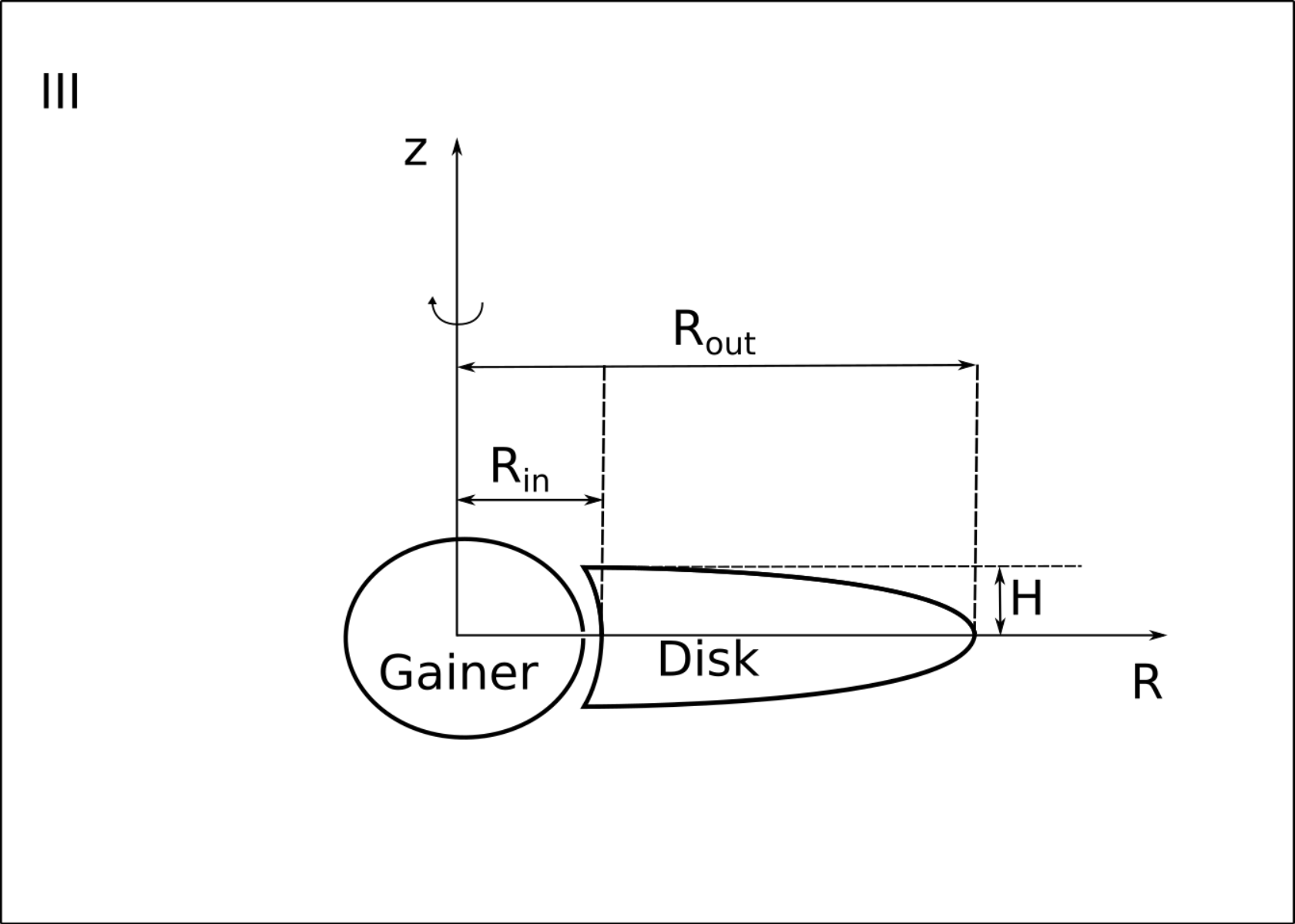}
	\includegraphics[width=0.3\textwidth]{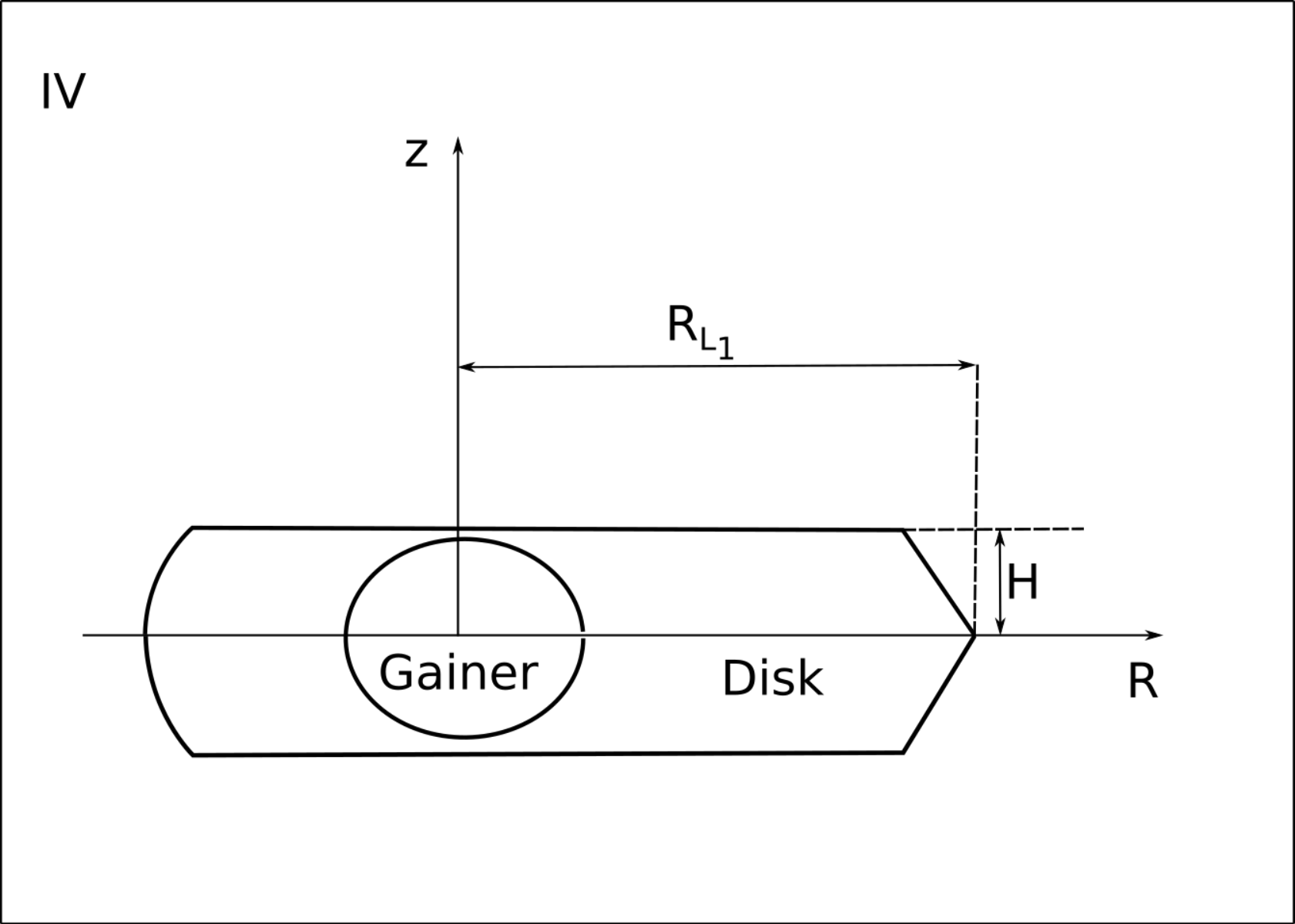}
	\caption{Geometric shapes of models of accretion disks.
		(I)~Denotes the {\em slab},
		(II)~the {\em wedge},
		(III)~the {\em lens}, and
		(IV)~the {\em envelope}.
		The latter is shown as if it was filling the
		Roche limit, that is $R_\mathsf{out} = R_\mathsf{L_1}$,
		where $R_\mathsf{L_1}$ denotes point radius of the
		Roche lobe. Here it is shown this way to emphasize
		the geometry of the disk, but its shape may be more
		symmetric depending on its filling factor.
		See Sect.~\ref{sec:shellspec:model}	for the~description
		of individual disk shapes. The {\em Nebula} is not presented here,
since no simple geometrical sketch can be given. It is illustrated, however,
in Fig.~\ref{fig:shellspec:models}.  	
		\label{fig:model:disk}}
\end{figure}

The disk plane of each model lies in the orbital plane of \be;
$z$-axis is perpendicular to this plane and goes
through the center of each disk. Except for the
envelope-shaped disk, the $z$-axis is also the axis
of symmetry. The radius $R$ is measured in the disk plane
from the center of the disk. The shapes of accretion disks that we tested are (see also Fig.~\ref{fig:model:disk}):
\begin{enumerate}

\item {\em Slab} (panel~I in Fig.~\ref{fig:model:disk})
is limited by two spherical surfaces with radii
$R_\mathsf{in}$, $R_\mathsf{out}$, and two
surfaces $z =\pm H$. The vertical temperature profile~$T(z)$
as well as the density~$\rho(z)$ are constant.
The velocity profile $v(r)$ is Keplerian as in all other cases.

\item {\em Wedge} (panel~II)
is limited by two spherical surfaces with radii
$R_\mathsf{in}$, $R_\mathsf{out}$, and two
conical surfaces $z = \pm R \sin\vartheta$, where $\vartheta$ is
the half-opening angle of the accretion disk.
$H$~denotes the maximal height of the disk
at its outer rim.

\item {\em Lens} (panel~III)
is limited by a~spherical surface with radius $R_\mathsf{in}$
and an~ellipsoidal surface whose semimajor axis is
equal to $R_\mathsf{out}$ and semiminor axis to $H$.

\item {\em Envelope} (panel~IV)
is limited by a~Roche equipotential whose shape is
given by the semimajor axis~$a$ of the orbit, the mass ratio~$q$,
and the filling factor~$f_\mathsf{f}$. A~synchronization between rotation and orbital
motion is assumed. The vertical structure of the envelope
is further limited by two surfaces $z = \pm H$. The
envelope can only be homogeneous and isothermal
($\rho = \mathrm{const.}$, $T = \mathrm{const.}$).

\item {\em Nebula}
is a standard disk with a Gaussian vertical density profile~$\rho(z)$
determined by the hydrostatic equilibrium; in other words, its scale height~$H$
is determined by the local temperature $T(r)$.
Additional parameters ($h_{\rm inv}$, $t_{\rm inv}$, $h_{\rm wind}$)
can be used to account for a temperature inversion in the disk atmosphere,
or a non-zero constant density in the wind region.
However, in order to prevent discretization artifacts
(discussed separately in Appendix~\ref{sec:apd}),
we assume the temperature inversion is not a sudden jump
from $T$ to $t_{\rm inv}T$, and in our \shellspec\
$T(z)$ changes linearly between $h_{\rm inv}H$ and the outer limit.
In order to account also for a non-hydrostatic disks
we modified \shellspec\ to include a~multiplicative factor~$h_{\rm mul}$, which can be treated as a free parameter too.

\end{enumerate}

Two types of radial temperature profiles $T(R)$
were tested. The first one is a power-law:
\begin{equation}
T(R) = T_0\left(\frac{R}{R_\mathsf{in}}\right)^{\alpha_\mathsf{T}}, \label{eq:temperature:powlaw}
\end{equation}
where $T_0$ is the effective temperature at the inner rim
of the accretion disk $R_\mathsf{in}$, and $\alpha_\mathsf{T}$ the exponent of the power-law. The second one corresponds to a~steady accretion disk
heated {\em only\/} by viscous dissipation \citep{shakura1973,pringle1981}:
\begin{equation}
T(R) = T_1 \left(\frac{R_\mathsf{g}}{R}\right)^{\bf\frac{3}{4}}\left(1 - \sqrt{\frac{R_\mathsf{g}}{R}}\right)^{\bf\frac{1}{4}}, \label{eq:temperature:steady}
\end{equation}
where $T_1$ is the characteristic temperature of the disk,
which is actually never attained in the disk.
The maximum temperature $T_{\rm max} = 0.488 T_1$
is at the radius $49/36\,R_\mathsf{g}$.
The temperature along $z$-axis is constant.

The radial density profile of gas $\rho(R)$ is always approximated
by a~power law:
\begin{eqnarray}
\label{eq:density:powlaw}
\rho\left(R\right) &=& \rho_0\left(\frac{R}{R_\mathsf{in}}\right)^{\alpha_\mathsf{D}},
\end{eqnarray}
where $\rho_0$ is the gas density at the inner rim of the
accretion disk. The temperature in the whole accretion disk is
too high $\gtrsim 7\,000\,$K for condensation of dust
grains, so we assumed that the dust density is zero.
The electron density was computed assuming LTE and solar chemical composition.
We simply assume the disk is stable over the time span of our
observations, that is 1970 to 2017.

%%%%%%%%%%%%%%%%%%%%%%%%%%%%%%%%%%%%%%%%%%%%%%%%%%%%%%%%%%%%%%%%%%%%%%%%

\subsection{Realistic uncertainties of observational data}\label{sec:shellspec:datamod}

Our model represents a rather simplified view on \blae, so it
is no surprise that a straight preliminary comparison of the model
and data gave reduced $\chi_\mathsf{R}^2 \gtrsim 20$.
This would mean that our model is wrong,
but our model is not very different from models presented
in earlier studies of the system. We attributed
this mismatch to systematic errors and we tried to compensate
for some of them by taking the following steps:
\begin{itemize}[itemsep=\baselineskip]
	\item The observed light curves of \blae, spread over more than 40~years, exhibit small bumps (e.g., asymmetry near primary and secondary eclipses),
	or flickering (bottom of the primary minimum). They
	represent an intrinsic variability of \bla that
	is beyond the capabilities of our model. The limited
	resolution of the model creates jags in the synthetic
	light curve. They are most pronounced at centers of
	both eclipses. The latter effect is higher, and
	the systematic signal was in general $\lesssim 0.05$\,mag
	(that is $3$ to $5$ times higher than true uncertainty of these measurements).
	Therefore the uncertainty of Hvar \ubvr~light curves
	was set to this value.

    \item The infrared light curves lacked uncertainty estimates.
	They were estimated by inspecting the scatter
	within intervals $0.05$ wide in phase. This uncertainty
	estimate was highly phase-dependent, but their mean
	was $\simeq 0.1$\,mag, hence we adopted this uncertainty
	for all infrared light curves.
	
	\item {\em The squared visibility $|V|^2$ and closure phase $T_3\phi$} are
	commonly affected by systematic effects given by
	atmospheric fluctuation during observations. The CHARA/VEGA
	squared visibility observations obtained with one baseline over a short
	period of time (one block, $\approx 20$\,min) were frequently
	showing much larger spread than uncertainty of individual
	points. This spread is unlikely to be a product of the~slow change
	of the projected baseline caused by the diurnal motion or
	produced by the intrinsic variability of \be. Also our
	experience is that measurements of low squared visibility
	$\lesssim 0.05$ have floor uncertainty $\approx 0.05$.
	The uncertainties of all CHARA/VEGA measurements
	of $|V|^2$ were adjusted according to the following formula:
	\begin{eqnarray}
	\label{eq:if:vega:uncertainty}
	\sigma_{V^2}^\mathsf{new} =
		\begin{cases}
			\max\{\sigma_{V^2}^\mathsf{old}, \sigma_{V^2}^\mathsf{block}\} & \hbox{for } |V|^2 > 0.05\,, \\
                        \noalign{\smallskip}
			\max\{\sigma_{V^2}^\mathsf{old}, \sigma_{V^2}^\mathsf{block}, 0.05\} & \hbox{for } |V|^2 < 0.05\,, \\
		\end{cases}
	\end{eqnarray}
	where $\sigma_{V^2}^\mathsf{old}$ is the uncertainty estimated with
	the reduction pipeline \citep[see][and references therein]{vega},
	and $\sigma^\mathsf{block}_{V^2}$ is the standard deviation of points
	within one block of observations (see Appendix~\ref{sec:apa} for details).

        \item
	For CHARA/MIRC observations, the uncertainty
	of closure phase measurements that satisfy the condition $S/N_\mathsf{|T_3|} \approx 1$
        was adjusted following the formula
	by~\citet{monnier2012}:
	\begin{eqnarray}
	\label{eq:if:mirc:uncertainty}
	\sigma_\mathsf{CP}^\mathsf{new} = \max\{\sigma_\mathsf{CP}^\mathsf{old}, 30\,\mathsf{deg} / \mathrm{S/N}_\mathsf{|T_3|},\frac{1}{5}\left(\Delta T_3\phi\right)_\lambda\}\,,
	\end{eqnarray}
	where $\sigma_\mathsf{CP}^\mathsf{old}$ is the original value
	estimated by the reduction pipeline \citep[see][and references
	therein]{mirc},
	$S/N_\mathsf{|T_3|}$ is the signal-to-noise ratio of
	the corresponding triple amplitude measurement, and
	$\left(\Delta T_3\phi\right)_\lambda$ is the difference
	between the highest and the lowest closure phase
	measurement in a~single passband for one block of
	CHARA/MIRC observations (see Appendix~\ref{sec:apa}
	for details). Similarly to \citet{zhao2011} minimal uncertainty
	$1$\,deg was adopted for {\em all} CHARA/MIRC closure
	phase measurements. The uncertainties on the triple product amplitudes are also corrected to account for systematic effects as described by~\citet{monnier2012}: we use an additive error of $1x10^{-5}$ and a multiplicative factor of $10\%$.

        \item
    A systematic offset was apparent for triple product amplitudes $|T_3|$
    obtained by the NPOI instrument. While the squared visibilities $|V|^2$
    and closure phases $T_3\phi$ can be fitted with our model, there are series of $|T_3|$ observations showing a clear trend with respect to the projected baseline $B/\lambda$, but with sudden increases of the amplitude. These are likely to be of an instrumental origin. Consequently, we decided to fit only the squared visibilities~$|V|^2$ instead. This should not affect the fitting in a negative way, because $|V|^2$ observations should constrain the model anyway and the $|T_3|$ data do not bring important constraints due to the short baselines of NPOI observations.
\end{itemize}

%%%%%%%%%%%%%%%%%%%%%%%%%%%%%%%%%%%%%%%%%%%%%%%%%%%%%%%%%%%%%%%%%%%%%%%%

\subsection{Simplifications reducing the computational time}

The evaluation of $\chi^2$ represented by
Eqs.~(\ref{eq:chi2total})--(\ref{eq:chi2if}) for all
available data turned out to be very demanding.
The total time required for the computation of $\chi^2$
exceeded two hours even for a moderate resolution
$0.6$\,\rs~per pixel, and the grid size
$n_x\times n_y\times n_z = 251\times251\times126$.
Such a long computational time prevents an extensive search of
the parametric space, so three approximations
were introduced:
(i)~low resolution. All models were
computed with the grid resolution $1$\,\rs~per pixel and
grid size $n_x\times n_y\times n_z = 161\times161\times81$;
(ii)~several model parameters were fixed at values
determined by previous investigators of \be; and
(iii)~a ``binning'' of synthetic observable quantities
was introduced. The last approximation differed for the
magnitudes and interferometric observables.

Synthetic magnitudes were not computed for each observation
time. Instead a~synthetic light curve as a~function of orbital phase
was sampled equidistantly with one hundred points for
each passband. Synthetic magnitudes for each observation
time were then obtained by a cubic-spline interpolation
of the synthetic light curve.

The binning in case of the interferometry limited the
number of images $I_\nu(x, y, t, \lambda_\mathsf{eff})$
computed to derive synthetic interferometric observables.
The binning was introduced into the effective wavelength
$\lambda_\mathsf{eff}$ and the orbital phase $\phi(t)$.
The bin size was set to $100$\,nm for $\lambda_\mathsf{eff}$
and $0.001$ for $\phi$. The function
$I_\nu(x, y, t, \lambda_\mathsf{eff})$ was not sampled
equidistantly as it was for the photometry. Instead
the following simplification scheme was adopted:
\begin{enumerate}
\item List pairs $(\lambda_\mathsf{eff}, \phi)$ for all observations.
\item Round the lists to the bin size (precision) given in the preceding paragraph.
\item Remove repeating items.
\item Compute $I_\nu(x, y, t, \lambda_\mathsf{eff})$ only for the remaining items.
\end{enumerate}
The image defined by the pair
$(\lambda_\mathsf{eff}, \phi)$ that
is the nearest to the observation was chosen for
the computation of interferometric observables.
This means that error in the effective wavelength
and orbital phase introduced by this approach cannot
exceed half-width of their respective bin ($\Delta\lambda_\mathsf{eff} = 50$\,nm,
$\Delta\phi = 0.0005$ for this particular application).
We note that for computations of the spatial frequency
($\mathbf{B} / \lambda$) the exact value of effective
wavelength for a~given observation is used.

Using all these steps, the computational time was reduced
by a factor of ten or more, and the evaluation
takes about three\,minutes (on a 8-core processor).
The parallelization was achieved in the Python wrapper
by employing the standard multiprocessing module.

%%%%%%%%%%%%%%%%%%%%%%%%%%%%%%%%%%%%%%%%%%%%%%%%%%%%%%%%%%%%%%%%%%%%%%%%

\subsection{Modeling strategy}

Advances achieved in the previous studies of \bl
allowed us to significantly reduce the number of optimized
parameters and focus more on the properties of accretion
disk. The optimized parameters were:
the inclination $i$,
the longitude of ascending node $\Omega$,
the outer radius~$R_\mathsf{out}$ of the disk,
the semi-thickness~$H$ of the disk,
the radial density $\rho(R)$
and temperature $T(R)$ profiles,
and the distance~$d$ to the system.
An attempt to include the semimajor axis $a$
and the donor temperature $T_\mathsf{eff, d}$ among the
optimized parameters has been done, but the former
turned out to be completely correlated with the
systemic distance, and the latter with the disk
temperature. Therefore both were set to the values reported
in earlier studies, even though they were likely correlated
in these studies, too. Model parameters that
were kept fixed during the optimization
are listed in Table~\ref{tab:shellspec:fixpars}.

The initial set of parameters was based on models developed
by~\citet{al2000} and \citet{ak2007}. As a first step,
photometry,
visible interferometry, and
infrared interferometry
were each fitted independently of the remaining data.
By this procedure, we established that all data are compliant with a similar
model, although not exactly the same. Based on
this information a~generous range
was selected for each optimized parameter
and a search for global minimum of Eq.~(\ref{eq:chi2total})
was run over these ranges. Once the global fit converged
we ran a~local fit using the simplex algorithm
to polish the result of the global method.
This approach was repeated for each eligible
combination of shape and radial temperature profiles.

Uncertainties of the optimal set of parameters
were estimated from the convergence of the global
fit. The resulting $\chi^2$ was scaled down
to the ideal value, that is number of degrees of freedom.
All solutions had their $\chi^2$ scaled down by
the same factor as the optimal set of parameters.
Solutions whose probability $P_\mathsf{\chi^2}(\mathbf{p}) > 0.05$,
where $P_\mathsf{\chi^2}$ is the cumulative
distribution function of the $\chi^2$ probability
density function, and $\mathbf{p}$ is the vector
of optimized parameters, were accepted as possibly
correct ones. The maximal differences between all
accepted results and the optimal one were adopted as
uncertainties.

%%%%%%%%%%%%%%%%%%%%%%%%%%%%%%%%%%%%%%%%%%%%%%%%%%%%%%%%%%%%%%%%%%%%%%%%

\subsection{Results}\label{sec:shellspec:results}

\begin{figure}
\centering
\setlength{\tabcolsep}{0.0pt}
\begin{tabular}{cc}
\includegraphics[width=0.345\textwidth]{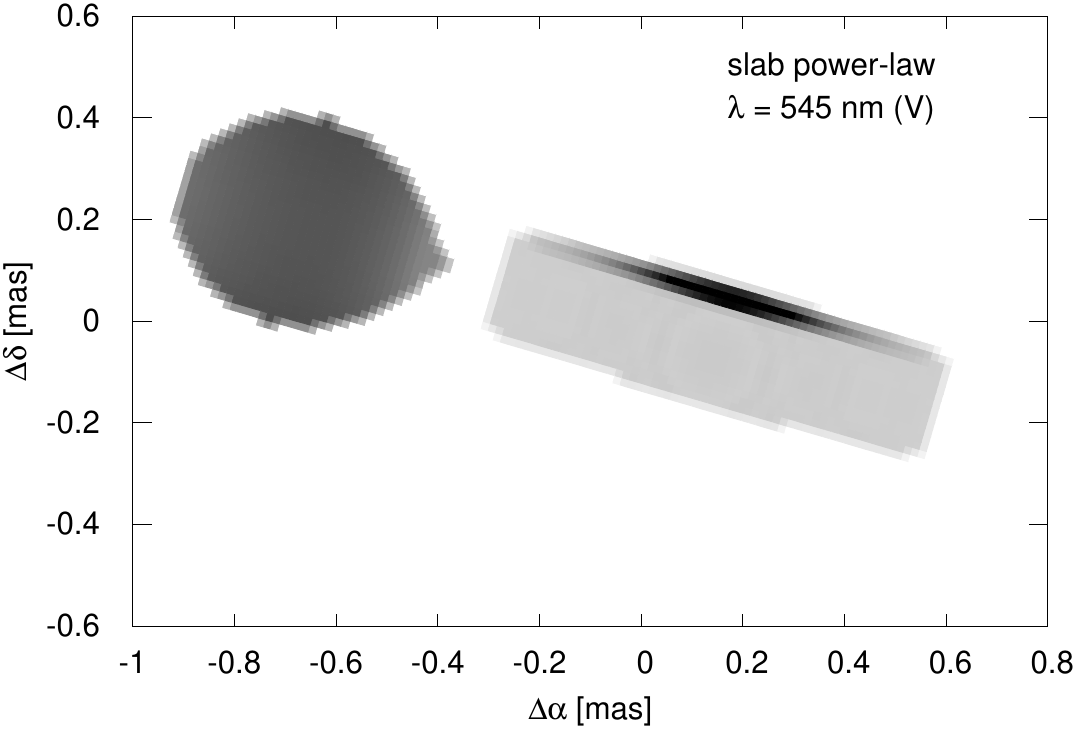} \\
\includegraphics[width=0.345\textwidth]{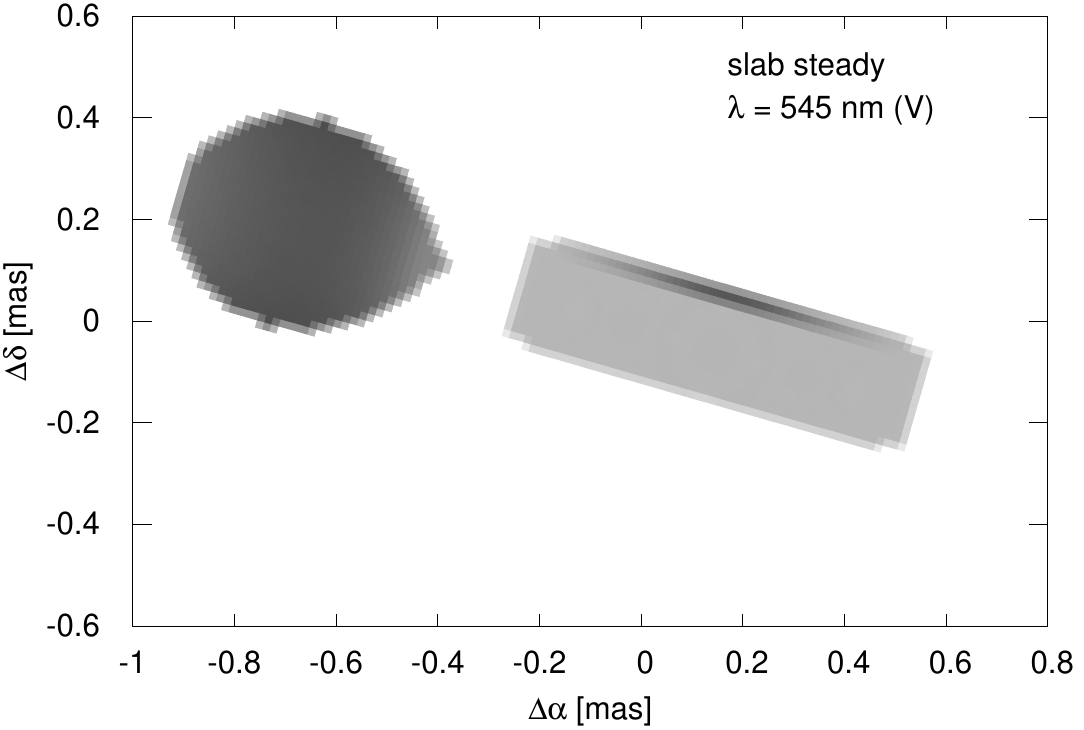} \\
\includegraphics[width=0.345\textwidth]{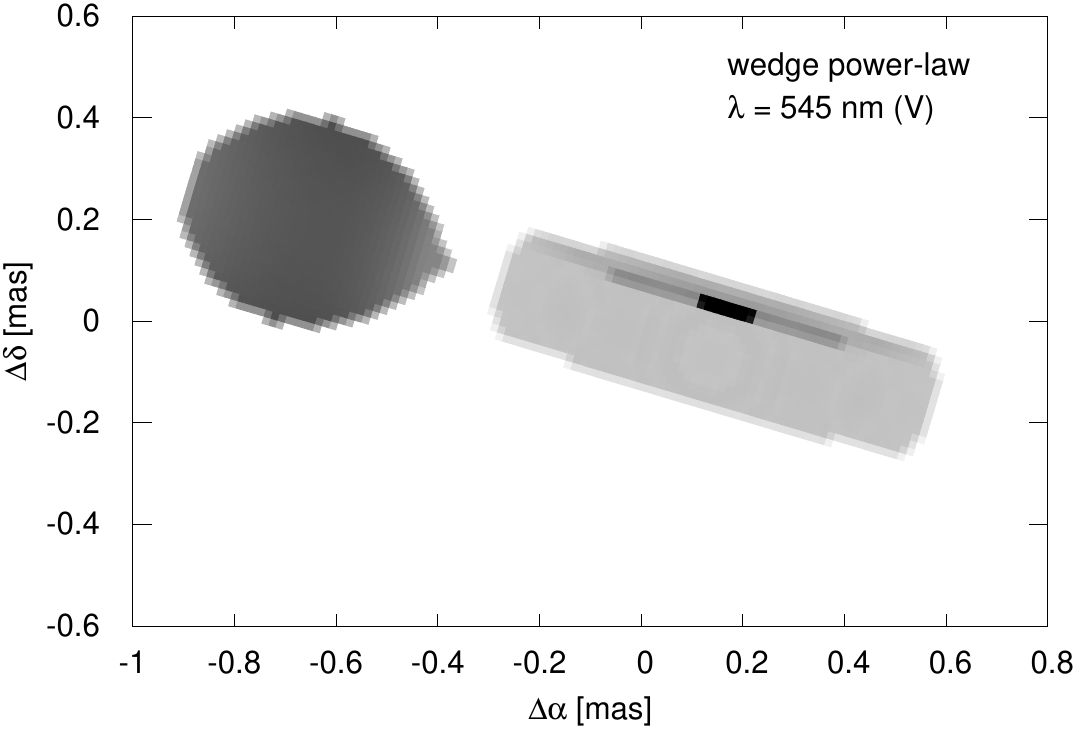} \\
\includegraphics[width=0.345\textwidth]{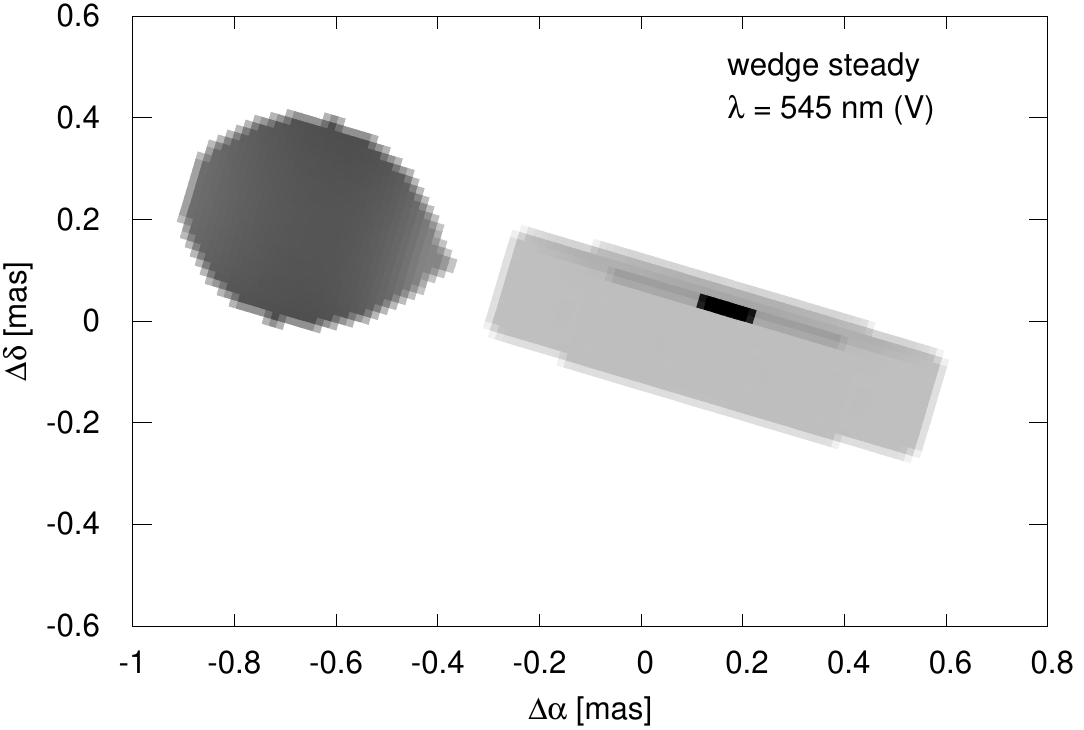} \\
\hspace{0.8cm}\includegraphics[width=0.39\textwidth]{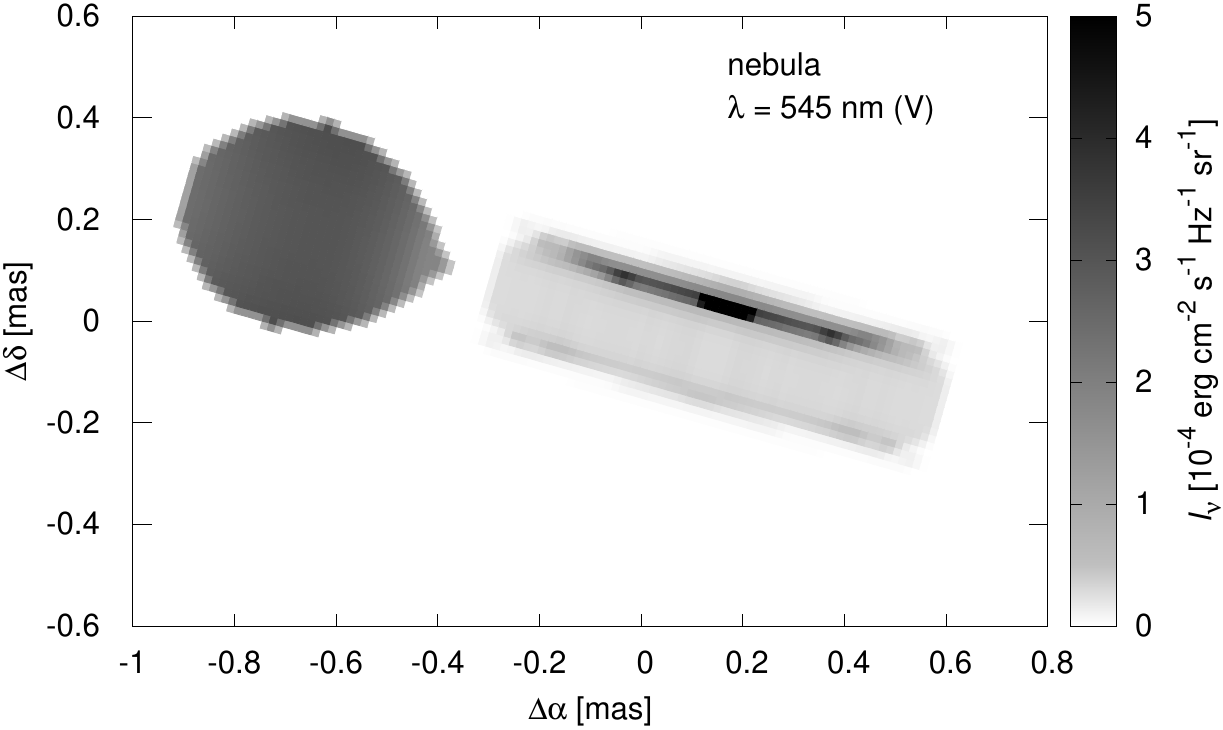} \\
\end{tabular}
\caption{Synthetic images of $\beta$~Lyr~A system for five different models:
slab power-law,
slab steady,
wedge power-law,
wedge steady,
and nebula.
The scale of grays corresponds to the monochromatic intensity~$I_\nu$
(in ${\rm erg}\,{\rm s}^{-1}\,{\rm cm}^{-2}\,{\rm sr}^{-1}\,{\rm Hz}^{-1}$).
The wavelength is always $\lambda = 545$\,mn
and the orbital phase $0.25$;
$\alpha$~goes along east-west direction, and
$\delta$~along north-south direction.
It is worth noting that each model converged independently,
but the outcomes are remarkably similar in terms of geometry.}
\label{fig:shellspec:models}
\end{figure}

The following seven models of the accretion disk were optimized:
  (i)~slab with power-law temperature and density,
 (ii)~slab with steady-disk temperature and power-law density,
(iii)~wedge with power-law temperature and density,
 (iv)~wedge with steady-disk temperature and power-law density,
  (v)~lens with power-law temperature and density,
 (vi)~envelope with homogeneous temperature and density profiles,
(vii)~nebula with power-law radial and exponential vertical density profiles.
Promising optimal models selected out of (i) to (vii)
are shown in Fig.~\ref{fig:shellspec:models}.

\begin{figure}
\centering
\includegraphics[width=7.5cm]{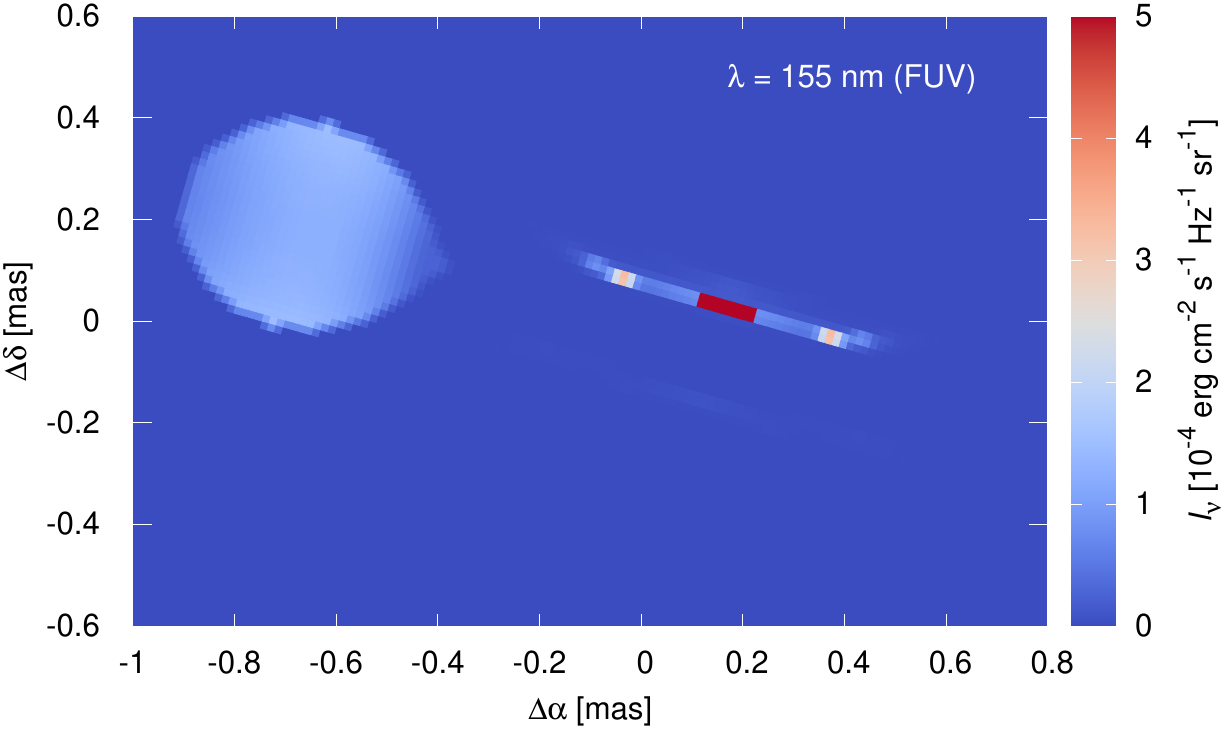}
\includegraphics[width=7.5cm]{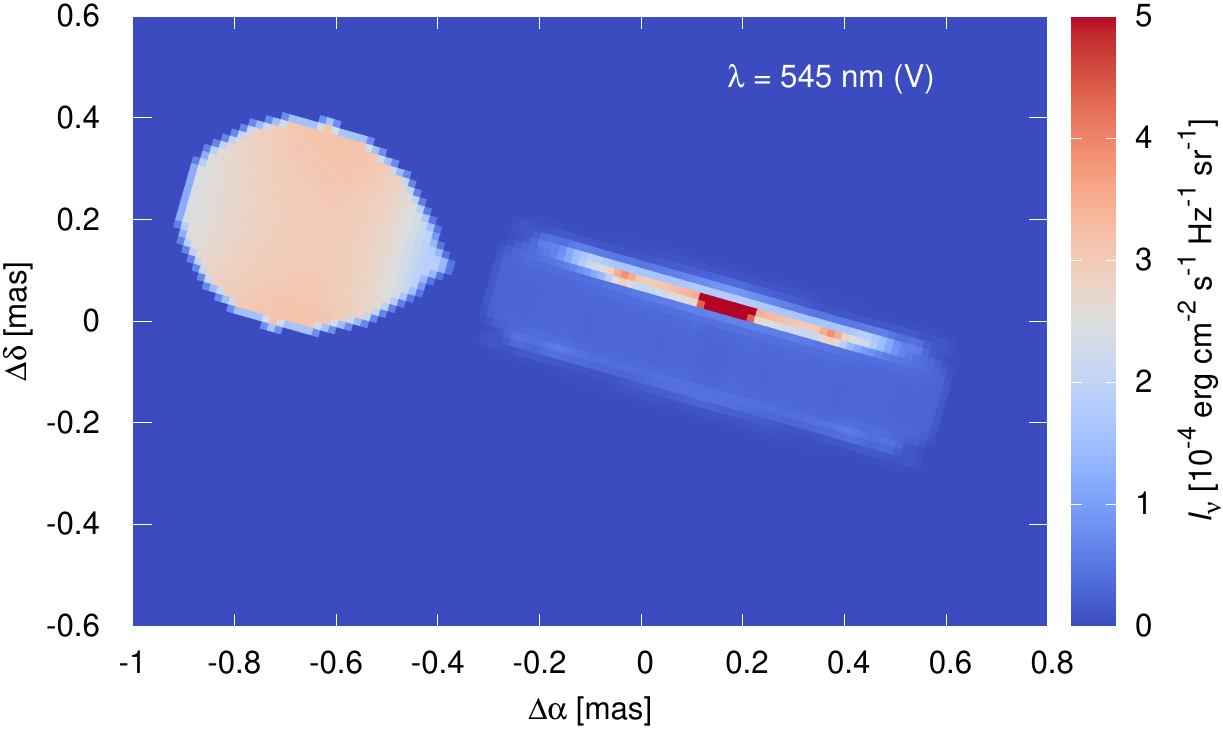}
\includegraphics[width=7.5cm]{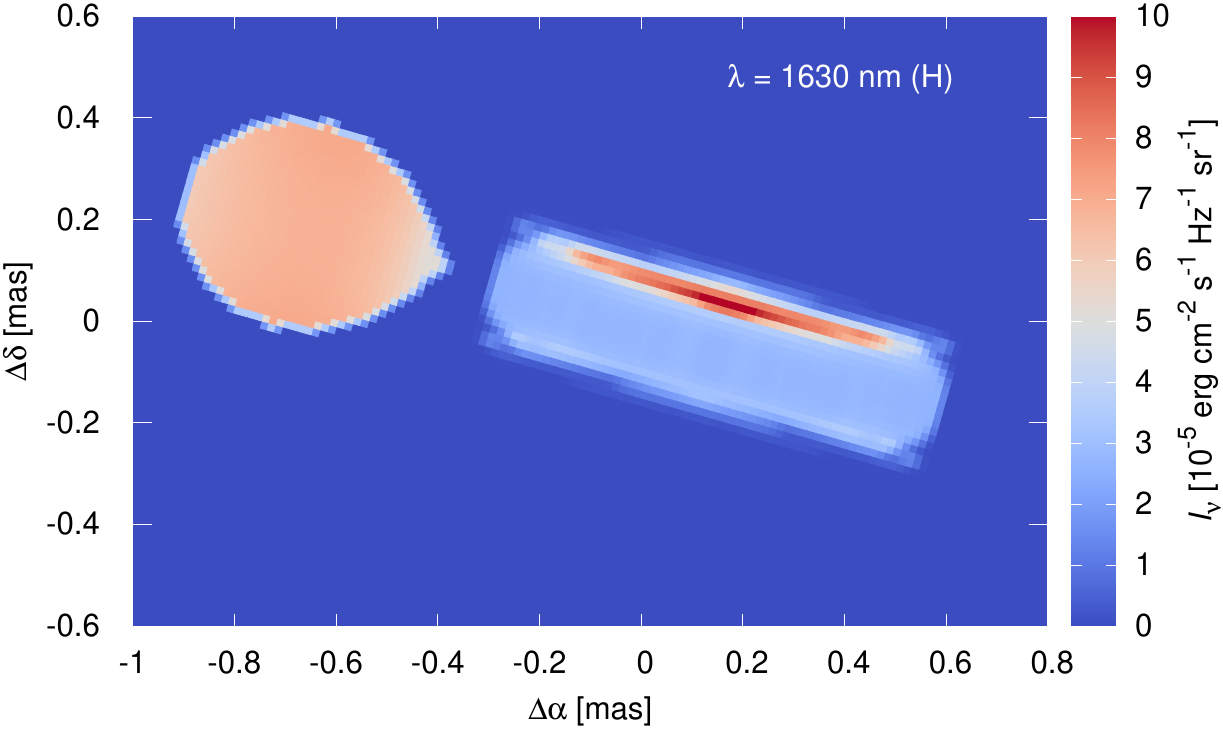}
\includegraphics[width=7.5cm]{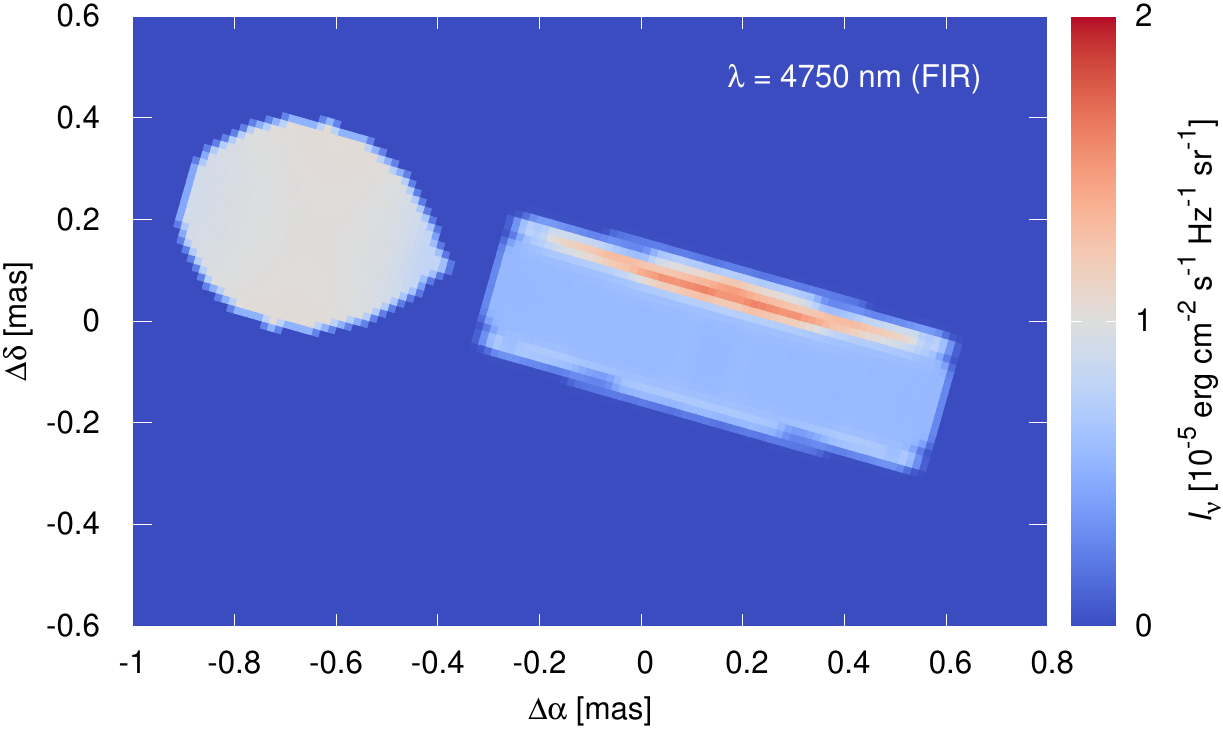}
\caption{Synthetic images of $\beta$~Lyr~A for the best-fit nebula model,
with total (not reduced) $\chi^2 = 103\,233$, shown for four different wavelengths:
$\lambda = 155\,{\rm nm}$ (FUV), 545\,nm (V~band), 1630\,nm (H), and 4750\,nm (M).
The axes correspond to the right ascension~$\alpha$ and declination~$\delta$ (in mas),
while the color scale corresponds to the monochromatic intensity~$I_\nu$
(in ${\rm erg}\,{\rm s}^{-1}\,{\rm cm}^{-2}\,{\rm sr}^{-1}\,{\rm Hz}^{-1}$).
This is a small subset of all 2392 images (per one iteration) used
to derive light curves, interferometric visibilities, closure phases,
and triple product amplitudes.}
\label{nebula_LINES_img_oao2_1550}
\end{figure}

\begin{figure}
\centering
\includegraphics[width=9cm, height=21cm]{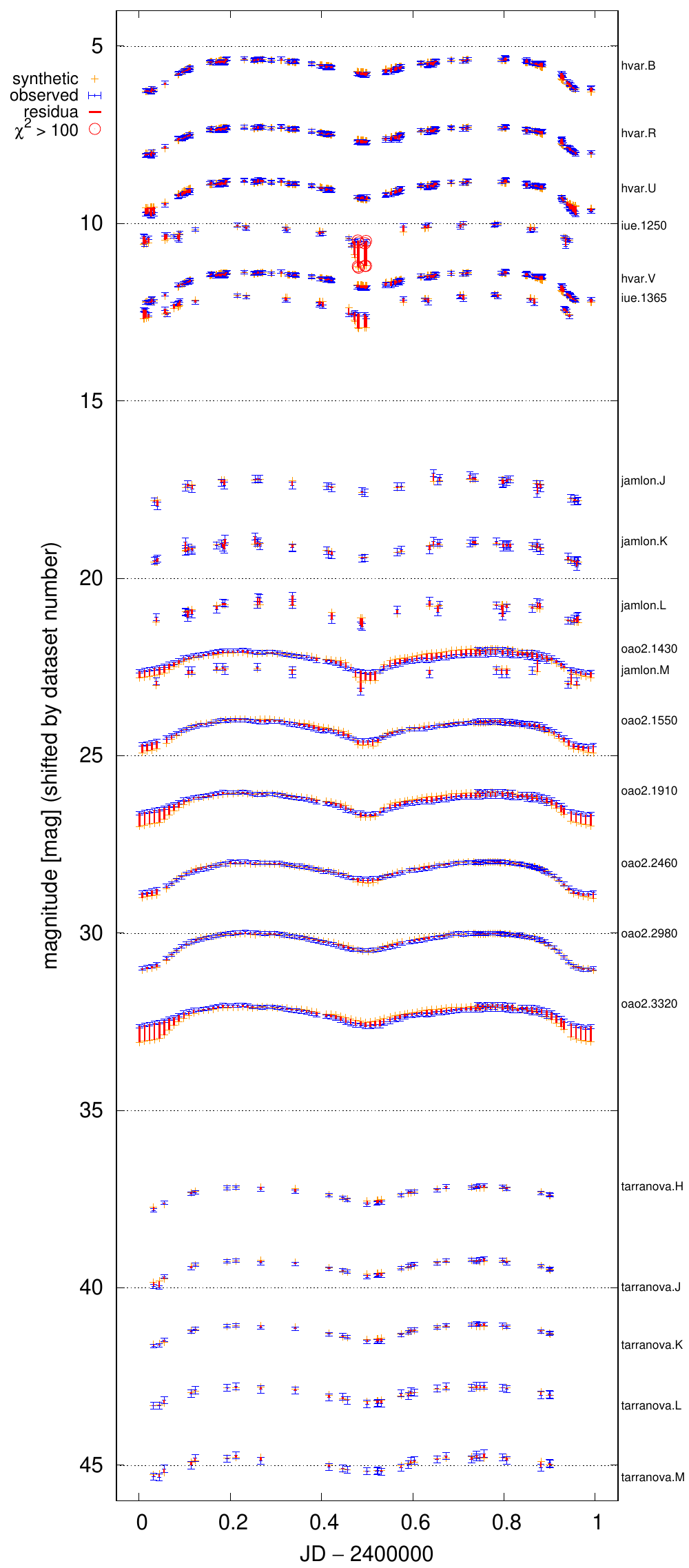}
\caption{Observed and synthetic light curves of $\beta$~Lyr~A,
shown for all 21~datasets (see their names in the right column).
The light curves are phased according to~\ref{eq:ephemeris}.
Vertical offsets are arbitrary.
The best-fit model is again 'nebula' with $\chi^2 = 103\,233$;
the individual contribution arising from light curves comparison
is $\chi^2_{\sf LC} = 6\,918$. Synthetic data are denoted by
yellow crosses, observed data by blue error bars, and residua
by red lines (or circles). There are clear systematic differences
especially for datasets iue.1250, iue.1365, oao2.1910, oao2.3320.
At the same time, there are neighboring datasets matched relatively well.
Sometimes, an intrinsic variability can be also seen (oao2.1430).}
\label{nebula_LINES_chi2_LC_PHASE}
\end{figure}

\begin{figure*}
\centering
\includegraphics[width=16cm]{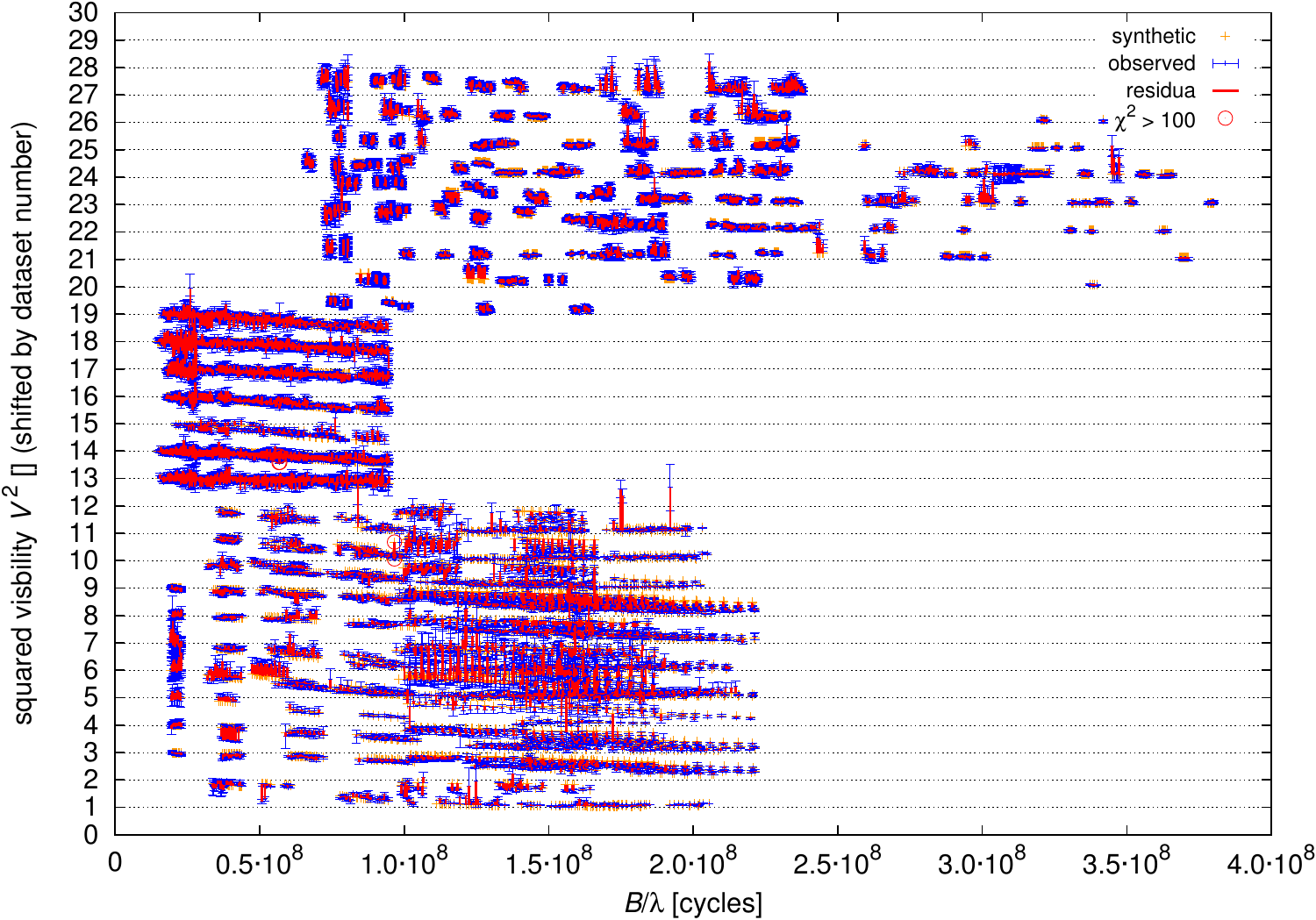}
\caption{Similar comparison as in Figure~\ref{nebula_LINES_chi2_LC_PHASE},
but for squared visibilities $|V|^2$, with a~contribution $\chi^2_{\sf V^2} = 54\,137$.
The $|V|^2$ values are plotted against projected baseline $B/\lambda$ (in cycles),
and shifted vertically according to the dataset number.
The are CHARA/MIRC data at the bottom, NPOI in the middle, and CHARA/VEGA at the top.
Synthetic data are denoted by yellow crosses,
observed data by blue error bars,
and residua by red lines.
A few outliers with large uncertainties, which do not contribute much to $\chi^2$ anyway,
were purposely removed from the plot to prevent clutter.
Even though there are some systematic differences for individual segments of data,
overall trends seem to be correctly matched.}
\label{nebula_LINES_chi2_VIS}
\end{figure*}

\begin{figure*}
\centering
\includegraphics[width=9cm,height=7cm]{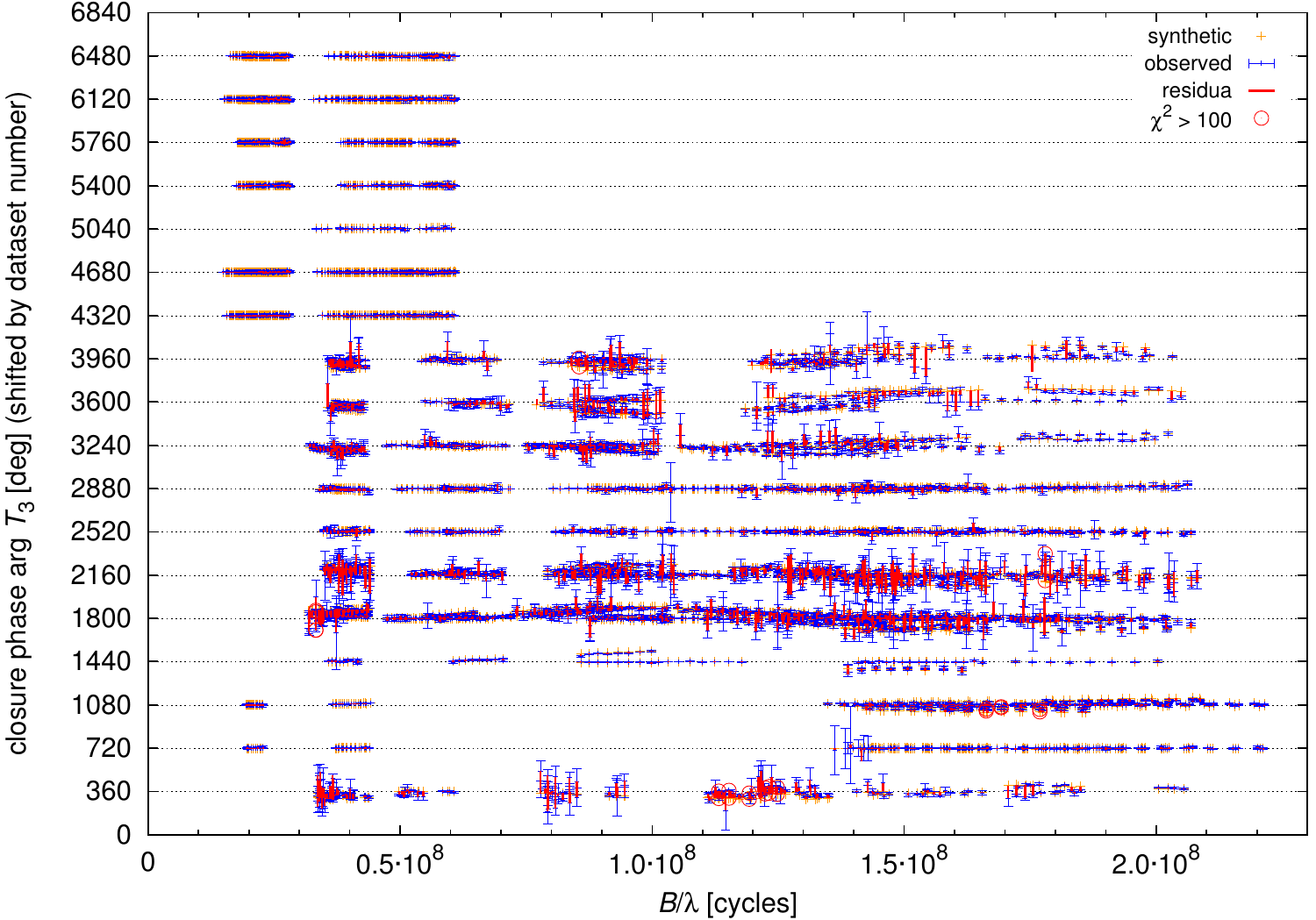}
\includegraphics[width=9cm,height=7cm]{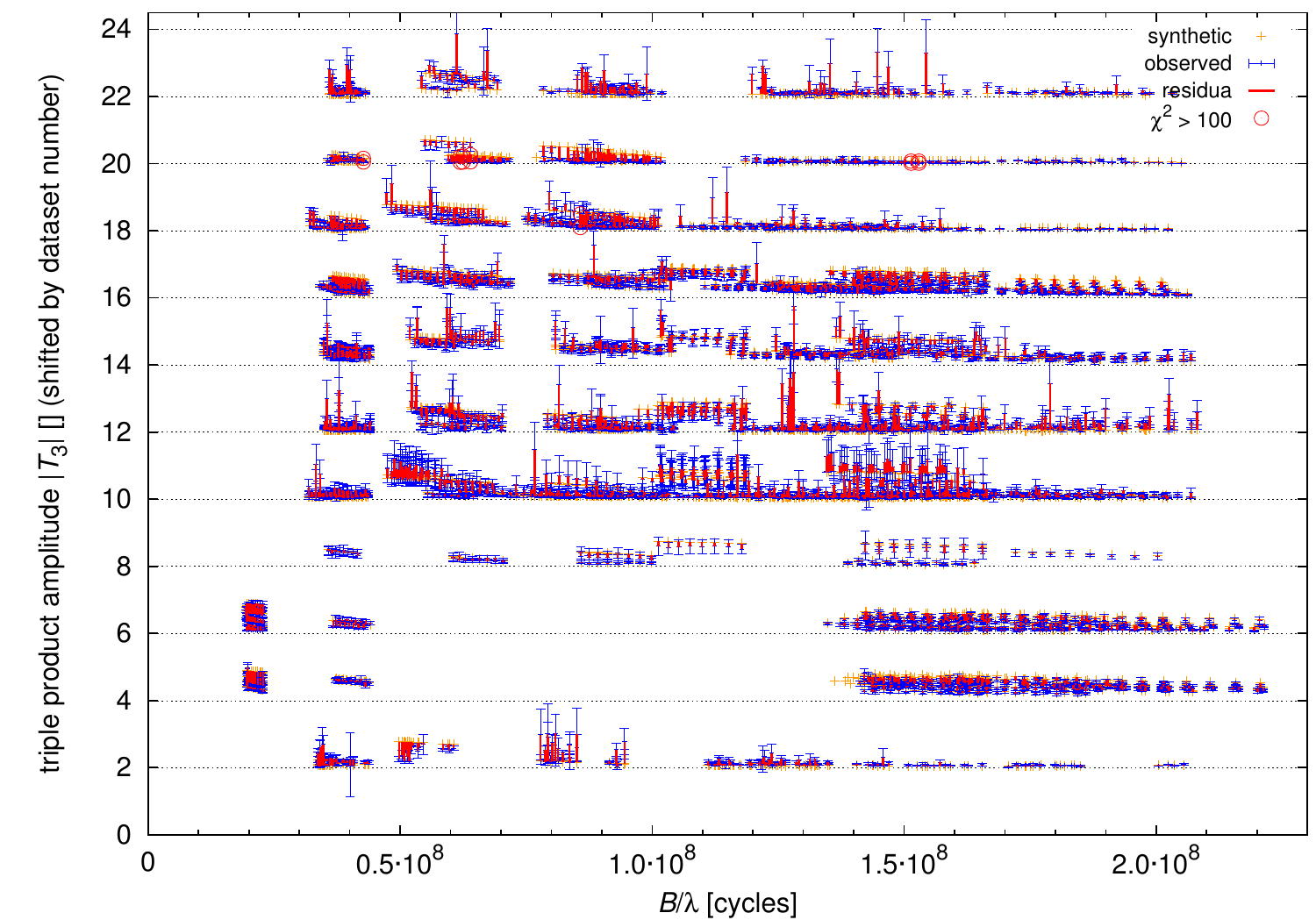}
\caption{Similar comparison as in Figure~\ref{nebula_LINES_chi2_VIS},
but for closure phases $T_3\phi$ (left) and triple product amplitudes $|T_3|$ (right).
Contributions to the total $\chi^2$ are
$\chi^2_{\sf CP} = 29\,153$, and
$\chi^2_{\sf T_3} = 13\,023$.
As before, the values are plotted against projected baseline $B/\lambda$.
$T_3\phi$ measurements were available for NPOI (top half) and CHARA/MIRC (bottom half),
while only $|T_3|$ from MIRC instrument were used.}
\label{nebula_LINES_chi2_CLO}
\end{figure*}

The analysis led to the following findings:
\begin{itemize}
	\item The fit of individual data subsets (photometry, visible
	and infrared interferometry) converged to similar, but
	not identical solutions. A~comparison of these particular
	solutions against all available data have shown clear
	discrepancies for data that were not fitted. For example, a~model fitting the infrared interferometry
	did not reproduce depths of minima of the visible
	light curve. A~detailed uncertainty analysis was
	not carried out though, and the relative importance of these discrepancies
	was not quantitatively evaluated. We report this, because it
	might point to a possible discrepancy in our model
	and/or to a~systematic effect that is still affecting our
	data and that was not suppressed with steps described
	in Sect.~\ref{sec:shellspec:datamod}.
	
	\item The model with homogeneous envelope served primarily as
	a~test whether the accretion disk around gainer
	could be homogeneous in temperature and density.
	This rather non-physical model tested our sensitivity
	to radial temperature and density profiles and it has shown
	that the sensitivity is indeed limited
	because the model with this profile shows only
	slightly higher total $\chi^2$ than the other
        inhomogeneous models (by approximately 5\%,
        which is statistically significant, though).
        The geometric size of the disk is similar to those obtained for the remaining
	models (see~Table~\ref{tab:shellspec:result}) ---
	filling factor $f_\mathsf{f} = 0.910_{-0.073}^{+0.087}$,
	translates into outer (point) radius
	$R_\mathsf{out} = 34.8_{-2.8}^{+3.3}$\,\rs,
	semi-thickness $H = 9.2_{-1.2}^{+1.1}$\,\rs, and
	temperature $T = 7\,230_{-310}^{+480}$\,K that
	is reached all over the surface, because no
	limb-darkening is present.
	It also leads to a higher orbital inclination
	$i = 95.4_{-1.9}^{+0.5}$\,deg. Nevertheless, the
	geometrical configuration and higher $\chi^2$ make this model implausible.
	
	\item The model with a~lens-shaped disk has been
	definitively ruled out. It was tested with both temperature and density
	radial profiles, but the $\chi^2$ was significantly higher, by approximately 35\,\%,
	than that obtained for the remaining shapes.

	\item Both {\em slab\/} and {\em wedge\/} disk shapes turned out
        to be plausible. Nevertheless, if we carefully compare the resulting
        $\chi^2$ values summarized in Table~\ref{tab:chi2}, we can exclude
        the slab with a steady temperature profile, because its (not reduced) $\chi^2 = 114\,981$
        is well above the 3-$\sigma$ level (that is ~105\,669) inferred from the best-fit model.
        The central region of the slab seems too cold, and especially the UV
        light curves do not match the observations (cf.~its $\chi^2_\mathsf{LC}$ contribution).
        In case of the wedge, the steady profile is compensated by the
        central star (gainer), which is partially visible in the opening.
        The resulting $\chi^2$ range from $103\,644$ to $105\,202$,
        that is still within the 3-$\sigma$ level, or in terms of the reduced
        $\chi^2_\mathsf{R} = 3.80$ to $3.86$.

        \item The nebula disk model provides the best-fit, with
        $\chi^2 = 103\,233$, which is equivalent to $\chi^2_\mathsf{R} = 3.78$.
        Even though it is not significantly better on its own,
        if we focus on a {\em subset\/} of observational data,
        namely the light curves ($\chi^2_\mathsf{LC}$ in Table~\ref{tab:chi2}),
        this fit is indeed significantly better than the others
        and it is thus our preferred model.

\end{itemize}
The optimal sets of parameters for the plausible models are listed in
Table~\ref{tab:shellspec:result}. The intervals in which the optimal
solution was searched for with the global-minimization algorithm
are given for each optimized parameter. Although the solutions were
mostly equal from point of the total $\chi^2$, only the {\rm nebula\/}
model is plotted against the data in
Figs.~\ref{nebula_LINES_img_oao2_1550} to~\ref{nebula_LINES_chi2_CLO}.

\begin{table}
\centering
\caption{List of fixed parameters.\label{tab:shellspec:fixpars}}
\begin{tabular}{lrr@{.}lr}
\hline\hline\noalign{\smallskip}
Parameter & Unit & \multicolumn{2}{c}{Value} & References \\
\hline\noalign{\smallskip}
$P_0$ & (d) & 12&913779& 1 \\
$T_\mathsf{min}$ & (HJD) & 2\,408\,254&4248895 & 1 \\
$\dot{P}$ & (d.d$^{-1}$) & 5&9977$\times$10$^{-7}$ & 1\\
$a\sin i$ & (\rs) & 58&19& 2 \\
$q$ & & 4&50& 2\\
$e$ & & 0&0& 2\\
$\omega$ & (deg) & 90&0& 2\\
$R_\mathsf{g}$ & (\rs) & 6&0 & 3\\
$T_\mathsf{eff, d}$ & (K) & 13\,300&0& 4\\
$x_\mathsf{bol, d}$ &     &\multicolumn{2}{c}{$\lambda$-dependent}& 5\\
$x_\mathsf{bol, g}$ &     &\multicolumn{2}{c}{$\lambda$-dependent}& 5\\
$\alpha_\mathsf{GD, d}$ & &0&25& 6\\
$\alpha_\mathsf{GD, g}$ & &0&25& 6\\
\noalign{\smallskip}\hline
\end{tabular}
\tablefoot{
$T_\mathsf{eff, d}$ denotes the polar temperature,
$\alpha_\mathsf{GD}$ the coefficient of gravity darkening,
$x_\mathsf{bol}$ the wavelength-dependent coefficient of linear limb darkening.
References:
1. \citet{ak2007},
2. \citet{hs93},
3. \citet{hec92},
4. \citet{balach86},
5. \citet{vanhamme1993},
6. \citet{vonzeipel1924}.
}
\end{table}

\begin{table*}
\def\x{$\times$}
\def\s{\phantom{$\times$}}
\centering
\caption{Overview of $\beta$~Lyr~A modeling results.}
\label{tab:chi2}
\begin{tabular}{l|rrrrrr}
\hline
\hline
\vrule width 0pt height 11pt depth 5pt
Models
& \multicolumn{1}{c}{$\chi^2$}
& \multicolumn{1}{c}{$\chi^2_{\sf LC}$}
& \multicolumn{1}{c}{$\chi^2_{{\sf V}^2}$}
& \multicolumn{1}{c}{$\chi^2_{\sf CP}$}
& \multicolumn{1}{c}{$\chi^2_{{\sf T}_3}$}
\\
\hline
\vrule width 0pt height 11pt
  slab power-law  & 105\,202 \s &  8\,141 \x & 55\,955 \x & 27\,711 \s  & 13\,394 \s \\
  slab steady     & 114\,981 \x & 15\,879 \x & 56\,229 \x & 28\,820 \s  & 14\,052 \x \\
  wedge power-law & 104\,131 \s &  8\,267 \x & 53\,957 \s & 28\,000 \s  & 13\,906 \s \\
  wedge steady    & 103\,644 \s &  8\,432 \x & 53\,633 \s & 27\,904 \s  & 13\,674 \s \\
  nebula    & 103\,233 \s & 6\,918 \s & 54\,137 \s & 29\,153 \s  & 13\,023 \s \\
\hline
& \multicolumn{1}{c}{$n_{\rm total}$}
& \multicolumn{1}{c}{$n_{\sf LC}$}
& \multicolumn{1}{c}{$n_{\sf VIS}$}
& \multicolumn{1}{c}{$n_{\sf CLO}$}
& \multicolumn{1}{c}{$n_{\sf T3}$}
\\
\hline
\vrule width 0pt height 11pt
  number of observations   & 27289  \s &  2305   \s & 14354  \s & 7717   \s & 2913   \s \\
  reduced $\chi^2_{\rm R}$ & 3.78   \s &  3.00   \s & 3.77   \s & 3.78   \s & 4.47   \s \\
  3-$\sigma$ factor        & 1.0236 \s &  1.0798 \s & 1.0324 \s & 1.0440 \s & 1.0710 \s \\
  3-$\sigma$ level         & 105\,669 \s &  7\,470   \s & 55\,891  \s & 30\,435  \s & 13\,947  \s \\
\hline
\end{tabular}
\tablefoot{ The resulting (not reduced) $\chi^2$ values
and their individual contributions (light curve, squared visibility, closure phase, triple product)
are summarized for five different models. The overall best-fit model is the 'nebula' (bold),
that is a disk with an exponential vertical profile. Below, there are the number of observations,
the respective reduced $\chi^2_{\rm R}$ values,
3-$\sigma$ factors by which the best-fit $\chi^2$ is multiplied
to get corresponding 3-$\sigma$ level.
The crosses ($\times$) denote $\chi^2$ values larger than that.}
\end{table*}

\begin{table*}
\centering
\renewcommand{\arraystretch}{1.3}
\caption{Free parameters of best-fit models of $\beta$~Lyr~A.}
\label{tab:shellspec:result}
\begin{tabular}{l|l|r|r|rrrrr}
\hline
\hline
&
&
&
& \multicolumn{1}{c}{slab}
& \multicolumn{1}{c}{slab}
& \multicolumn{1}{c}{wedge}
& \multicolumn{1}{c}{wedge}
& \multicolumn{1}{c}{nebula}
\\[-3pt]
  \multicolumn{1}{c|}{Par.}
& \multicolumn{1}{c|}{Unit}
& \multicolumn{1}{c|}{Min}
& \multicolumn{1}{c|}{Max}
& \multicolumn{1}{c}{power-law}
& \multicolumn{1}{c}{steady}
& \multicolumn{1}{c}{power-law}
& \multicolumn{1}{c}{steady}
& \multicolumn{1}{c}{power-law}
\\
\hline
$H                $& \rs                               &$ 5.5    $&$ 12    $&$ 6.10   \pm 0.78 $&$ 6.84   \pm 1.34 $&$                 $&$                 $&$                 $\\
$\vartheta        $& deg                               &$ 7.5    $&$ 30    $&$                 $&$                 $&$ 12.75  \pm 0.71 $&$ 12.96  \pm 0.86 $&$                 $\\
$R_{\rm out}      $& \rs                               &$ 26     $&$ 35    $&$ 30.77  \pm 0.75 $&$ 27.71  \pm 1.19 $&$ 30.36  \pm 0.54 $&$ 30.79  \pm 0.55 $&$ 31.11  \pm 0.82 $\\
$T_0              $& K                                 &$ 23000  $&$ 35000 $&$ 33948  \pm 313  $&$                 $&$ 30849  \pm 740  $&$                 $&$ 30292  \pm 1026 $\\
$T_1              $& K                                 &$ 23000  $&$ 35000 $&$                 $&$ 26995  \pm 778  $&$                 $&$ 26424  \pm 619  $&$                 $\\
$\rho_0           $& $10^{-6}\,{\rm g}\,{\rm cm}^{-3}$ &$ 0.001  $&$ 5.00  $&$ 2.90   +   1.07 $&$ 4.82   +   0.17 $&$ 0.99   +   1.21 $&$ 0.07   +   1.24 $&$ 3.46   \pm 1.01 $\\
$\alpha_\mathsf{D}$& 1                                 &$ -4.0   $&$ -0.8  $&$ -0.93  \pm 0.58 $&$ -2.43  \pm 0.67 $&$ -2.44  \pm 0.67 $&$ -3.16  \pm 0.68 $&$ -2.40  \pm 0.25 $\\
$\alpha_\mathsf{T}$& 1                                 &$ -1.2   $&$ -0.7  $&$ -1.03  \pm 0.10 $&$                 $&$ -0.95  \pm 0.10 $&$                 $&$ -1.00  \pm 0.10 $\\
$i                $& deg                               &$ 91     $&$ 95    $&$ 93.26  \pm 0.26 $&$ 92.66  \pm 0.51 $&$ 94.11  \pm 0.45 $&$ 94.15  \pm 0.25 $&$ 93.66  \pm 0.21 $\\
$\Omega           $& deg                               &$ 252    $&$ 255   $&$ 253.65 \pm 0.85 $&$ 253.97 \pm 0.75 $&$ 253.40 \pm 1.26 $&$ 253.42 \pm 0.76 $&$ 254.12 \pm 0.14 $\\
$d                $& pc                                &$ 305    $&$ 330   $&$ 316.7  \pm 5.7  $&$ 315.9  \pm 6.2  $&$ 321.2  \pm 3.7  $&$ 321.4  \pm 6.5  $&$ 319.7  \pm 2.7  $\\
$h_{\rm inv}      $& $H$                               &$ 1      $&$ 9     $&$                 $&$                 $&$                 $&$                 $&$ 3.58   \pm 0.16 $\\
$t_{\rm inv}      $& $T(r)$                            &$ 1      $&$ 9     $&$                 $&$                 $&$                 $&$                 $&$ 8.92   \pm 0.30 $\\
$h_{\rm wind}     $& $H$                               &$ 1      $&$ 15    $&$                 $&$                 $&$                 $&$                 $&$ 5.32   \pm 0.35 $\\
$h_{\rm mul}      $& $H$                               &$ 1      $&$ 15    $&$                 $&$                 $&$                 $&$                 $&$ 4.32   \pm 0.26 $\\
\hline
\end{tabular}
\tablefoot{
$H$ denotes the {\em geometric} semi-thickness,
$\vartheta$ the half opening angle of the wedge,
$R_\mathsf{out}$ the outer {\em geometric} radius of the disk,
$T_0$ the temperature at the inner rim
for a power-law radial temperature profile (given by Eq.~\ref{eq:temperature:powlaw}),
$T_1$ the characteristic temperature of the disk,
with the maximum temperature $T_{\rm max} = 0.488 T_1$ (Eq.~\ref{eq:temperature:steady}),
$\rho_0$ the density at the inner rim,
$\alpha_\mathsf{T}$ the exponent of the temperature profile,
$\alpha_\mathsf{D}$ the exponent of the density profile,
$i$ the orbital inclination,
$\Omega$ the longitude of ascending node, and
$d$ the distance of the system.
``Min'' and ``Max'' denote boundaries of the intervals
that were searched by the global-optimization algorithm.
The lower bound of the density~$\rho$ is not indicated
for the wedge and slab models; it can reach down to
$\sim 10^{-9}\,{\rm g}\,{\rm cm^{-3}}$
while the disk remains optically thick.
}
\end{table*}

%%%%%%%%%%%%%%%%%%%%%%%%%%%%%%%%%%%%%%%%%%%%%%%%%%%%%%%%%%%%%%%%%%%%%%%%
%%%%%%%%%%%%%%%%%%%%%%%%%%%%%%%%%%%%%%%%%%%%%%%%%%%%%%%%%%%%%%%%%%%%%%%%

\section{Discussion}\label{sec:discussion}

This section compares our results from Sec.~\ref{sec:shellspec} with theoretical predictions from earlier studies of \bl.

\subsection{Distance and inclination of \blae}
The distance of $\beta$~Lyr~A inferred from our model
is $d = (319.7\pm2.7)\,{\rm pc}$, which is a bit
larger distance than our preliminary expectation
based on $\beta$~Lyr~B. The final result is essentially based on
$a\sin i = 58.19$\,\rs value (see~Table~\ref{tab:shellspec:fixpars}),
which is $\lesssim 1\%$ larger than
the value accepted by~\citet{zhao2008}
and is well within their uncertainty interval. Hence
the choice of a~slightly different~$a\sin i$
cannot cause the difference between our distance
and that of \citet{zhao2008}. Nevertheless, our estimate
is based on a~more correct physical model
than that by~\citet{zhao2008}, whose estimate
closest to ours relied on a model with two uniform
ellipses.

Our analysis reinforced the hypothesis that \bla and \blb
share a common origin (see Sect.~\ref{sec:blb}).
In particular, \emph{Gaia} distance $d_\mathsf{B} = (333 \pm 6)$\,pc of \blb
is in a fair agreement with our model distance estimate of \blae.
The spectroscopic distance of \blb derived in this study is significantly
lower (259\,pc), but that may result from calibration uncertainties
(see the discussion at the end of sect.~\ref{sec:blb}) or from
possible, still unrecognized duplicity of \ble.

Concerning the inclination of \be, our estimates obtained for the {\em
wedge}- and {\em slab}-shaped disks do not agree with each other. This is not
very surprising, because the wedge shape tends to attenuate the radiation from the
hot central parts of the disk (and the gainer). Because this radiation is in
fact observed, the fit converges to a lower inclination to expose central
parts, and to compensate for the attenuation. The inclination of the {\em nebula} model is just in between.

%%%%%%%%%%%%%%%%%%%%%%%%%%%%%%%%%%%%%%%%%%%%%%%%%%%%%%%%%%%%%%%%%%%%%%%%

\subsection{Properties of the accretion disk}
\label{disk-properties}
\begin{figure}
\centering
\begin{tabular}{cc}
wedge power-law \\
\includegraphics[width=8.5cm]{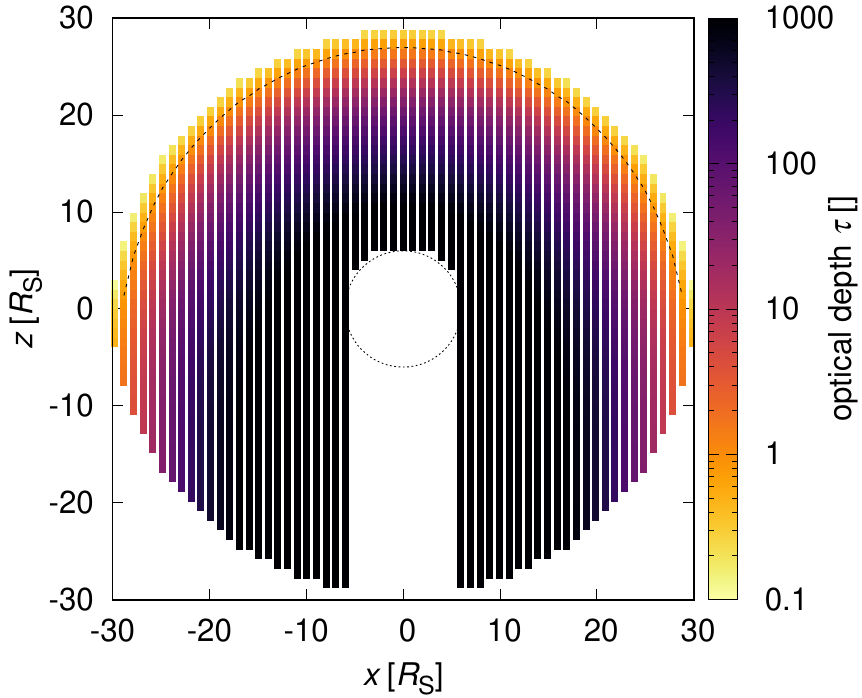} \\
\kern-1.7cm\includegraphics[width=7.3cm]{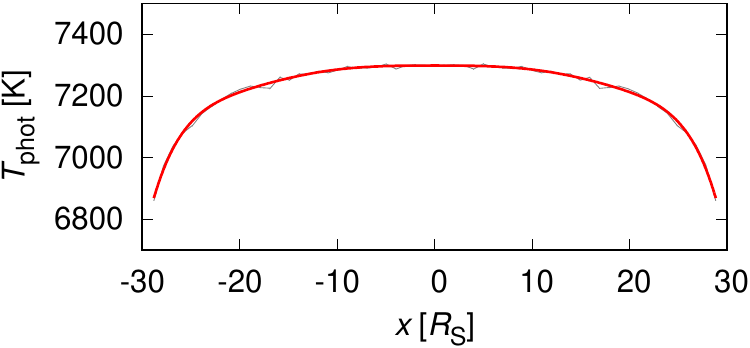} \\
\end{tabular}
\caption{{\em Top panel:}
Optical depth~$\tau$ (computed for the wavelength $\lambda = 1630\,{\rm nm}$) of the \bla disk,
which is observed approximately edge-on (that is from the top).
The coordinate~$z$ thus corresponds to the line of sight,
while $x$ (and $y$) to the sky plane.
The center is empty because the gainer is a~non-transparent object.
In this case, the inner-rim density of the wedge power-law model is set {\em low\/},
$\rho_0 = 10^{-9}\,{\rm g}\,{\rm cm}^{-3}$,
to demonstrate a~separation from the outer rim.
The dashed line shows the physical position of the photosphere ($\tau = 2/3$).
The lines-of-sight grid exhibits 'steps' due to the limited spatial
resolution of our model ($1\,R_\odot$), even though the integration
of the radiation transfer (the contribution function) is internally
performed on a~much finer grid.
{\em Bottom panel:}
Corresponding photospheric temperature $T_{\rm phot}$ of the disk,
which varies along with the radial temperature profile~$T(R)$.
\label{fig:continuum}}
\end{figure}

\begin{figure}
\centering
\begin{tabular}{c}
wedge power-law \\
\includegraphics[width=8.5cm]{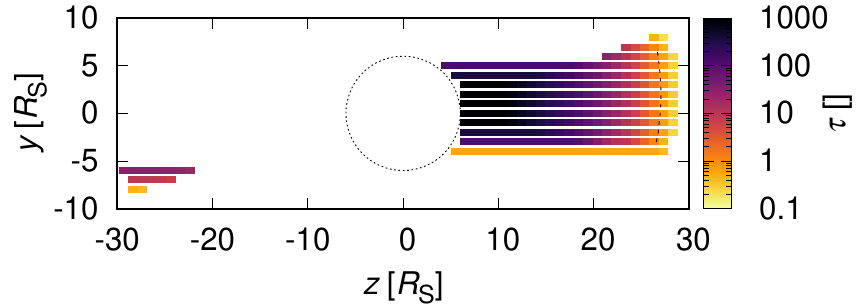} \\
\end{tabular}
\caption{Optical depth~$\tau$ in the vertical (perpendicular) cross-section
of the disk; other parameters are the same as in Fig.~\ref{fig:continuum}.
The lines of sight are seemingly different from the wedge shape,
but this only because they start either in the vacuum, or at the non-transparent object.
\label{fig:continuum:vertical}}
\end{figure}

A critical discussion of disk parameters is presented here:
\begin{itemize}[itemsep=\baselineskip]
	\item {\em Radius of the accretion disk:}
	Dense parts of the accretion disk actually fill the corresponding Roche lobe. The front, back, and side radii are $R_\mathsf{limit, g} = 37.4$\,\rs, $31.7$\,\rs, and
$30.3$\,\rs respectively. The disk that was obtained reaches up to
	the Roche limit, although the 'hard' upper limit of our optimisation procedure was as high as 35\,\rs; the value is thus constrained by our observations.
The tidal cutoff radius is as low as $26.3$\,\rs\
for the given mass ratio~$q$, but this is not necessarily
the edge of a viscous disk \citep{papaloizou1977}.
	In earlier studies \citep[e.g.,][]{al2000}
	the disk was modeled by a~solid body with prescribed
	radiative properties. Outer radii of our disk
	models $R_\mathsf{out}$ listed in
	Table~\ref{tab:shellspec:result} cannot be thus directly
	compared to radii obtained in earlier studies,
	because our disks are not optically thick
	starting from their rim. To obtain a comparable
	radius, a pseudo-photosphere approximated by
	the optical depth $\tau = 2/3$ has to be found.
	It was searched along lines of sight perpendicular to the disk rim
	$\left(x, y=0, z=0\right);~x\in\left[0, R_\mathsf{out}\right]$\,\rs.
	It was realized that the photosphere forms almost up to the geometric
	$R_\mathsf{out}$ for the {\em slab}-shaped disk,
	$R^\mathsf{slab}_\mathsf{out, ph.} \doteq 30\,$\rs, and
	slightly less for the {\em wedge}-shaped disk
	$R^\mathsf{wedge}_\mathsf{out, ph.} \doteq 29\,$\rs,
	where index ``ph'' stands for the photosphere.	
	Hence, our photospheric disk radius is in excellent
	agreement with that obtained by \citet{al2000}, $R = 30\,$\rs, and
	also with \citet{mennickent2013}, $R = (28.3\pm0.3)\,$\rs
	(if uncertainty $\approx 1.5$\,\rs~is taken into account).
	This is demonstrated in Fig.~\ref{fig:continuum}, where
	the physical position of the photosphere within the accretion disk is shown
	for the wavelength $1630$\,nm.

	\item {\em Semi-thickness of the disk:} For the slab-shaped model,
	the value of $H$ is the same as semi-thickness of the disk photosphere.
	For the {\em wedge}-shaped model, we searched for the physical
	position of photosphere along the following lines of sight
	perpendicular to the disk rim:
	$\left(x=0, y=0, z\right); z\in\left[0, H\right]$.
	This is qualitatively demonstrated in Fig.~\ref{fig:continuum:vertical}.
	Up to a certain~$z$ the {\em wedge} is opaque; for intermediate~$z$ the
	ray pierces through the first lobe and the optical depth $\tau = 2/3$
	is reached in the second lobe. For high~$z$ the disk is
	optically thin. Hence the semi-thickness of the ``opaque'' wedge
	is about 30\,\% smaller than the geometric semi-thickness
        $H = R_\mathsf{out}\sin\vartheta \doteq 7$\,\rs\ derived from Table~\ref{tab:shellspec:result}.
	Our semi-thickness of the opaque disk $\simeq 5$\,\rs\
	is in agreement with the result obtained by~\citet{mennickent2013}, $H = (5.50\pm0.10)\,$\rs,
        and substantially lower than the value by~\citet{al2000}, $H = 8.0\,$\rs.
	
\item {\em Shape of the disk:}
Unfortunately, it seems almost impossible to distinguish between
slab, wedge and nebula shapes. Nevertheless, it is clear that their
semi-thickness is so large that the disk {\em cannot\/} be in a vertical
hydrostatic equilibrium. A hydrostatic disk with a constant vertical
temperature profile, $T(z) = {\rm const.}$, would have an exponential
density profile \citep[e.g.,][]{pringle1981}:
\begin{equation}
\rho(r,z) = \rho_0(r)\exp\left({-\frac{z^2}{2H_{\rm eq}^2}}\right)\,,
\end{equation}
with the characteristic scale~$H_{\rm eq}$ given by the temperature profile~$T(r)$:
\begin{equation}
H_{\rm eq}(r) = \sqrt{\frac{{\cal R}T}{\mu}} \frac{1}{\Omega_{\rm k}}\,,
\end{equation}
where
${\cal R}$ denotes the ideal-gas constant,
$\mu$ the mean molecular weight, and
$\Omega_{\rm k}$ the keplerian angular velocity.
As we also verified with \shellspec\ (using the nebula model with $h_{\rm mul} = 1$),
the resulting $H_{\rm eq}$ for our range of temperatures (30,000 to 7,000\,K) is always low,
0.2 to 1.2\,\rs, and the gainer would be always visible.
As a consequence, $H\gg H_{\rm eq}$ is a proof that the
disk is non-equilibrium, and the flow starting from the donor must
have a non-negligible vertical velocities within the accretion disk.
It is a matter of dynamical models (or spectro-interferometry in
individual lines) to constrain the velocity field.

	\item {\em Radial density profile and disk mass:}
	The continuum data do not allow us to see below
	the pseudo-photosphere. This is evident from the
	high correlation between $\rho_0$ and $\alpha_\mathsf{D}$
	resulting from individual models, $\mathrm{corr}\left(\rho_0,
	\alpha_\mathsf{D}\right) \simeq 0.5 - 0.8$.
	Hence our model does not provide the true density
	radial profile, but rather the minimal profile
	required to ``produce continuum'' at the correct radius and height.
	The disk mass given by our radial profiles
	is then $\simeq 10^{-4}$ to $10^{-3}$\,\ms.
	This disk mass estimate is essentially the same as that obtained by~\citet{hubeny1991,hubeny1994}.
	If it was real, it would suggest a very efficient accretion,
	because the accretion time scale
	$\tau_{\rm acc} = M/\dot M \simeq 5\hbox{ to }50\,{\rm yr}$
	is much shorter than the expected duration of \bla mass-transfer phase.
	For a steady disk \citep{pringle1981}, it is simply assumed the viscosity always adapts
	to the constant accretion rate and the viscous time scale is then equal
	to the accretion one, $\tau_{\rm visc} = \tau_{\rm acc}$.
	
	\item {\em Disk-rim temperature:}
	The rim temperature given by Eqs.~(\ref{eq:temperature:powlaw})
	and~(\ref{eq:temperature:steady}) is not directly comparable
	to those obtained with solid surface models, because the
	photosphere does not coincide with the geometric rim.
	To overcome this contradiction, one can adopt the photospheric temperature,
	that is shown in Figs.~\ref{fig:continuum} and~\ref{fig:continuum:vertical}.
	Another approach is to take the mean value of the intensity distribution
	over the photospheric surface
	$\left\langle I_\mathrm{disk}\left(x, y, \lambda\right)\right\rangle$
	for a~given wavelength and find a~temperature corresponding
	to the Planck law $B_\nu\left(T, \lambda\right)$
	for the mean intensity. Using this
	approach, we determined the following rim temperatures
	at three wavelengths (given by the upper index in nm):
	\begin{center}
	\begin{tabular}{cccc}
		\hline\hline\noalign{\smallskip}
		Model & $T^{500}$ & $T^{1000}$ & $T^{2000}$ \\
		\noalign{\smallskip}\hline\noalign{\smallskip}
		slab power-law  & 8\,764 & 8\,323 & 5\,859 \\
		slab steady     & 7\,822 & 7\,434 & 5\,183 \\
%		wedge power-law & \\
		wedge steady    & 6\,847 & 6\,284 & 4\,941 \\
		\noalign{\smallskip}\hline
	\end{tabular}
	\end{center}
	The rim temperature of slab-shaped models
	is comparable to that obtained by~\cite{mennickent2013}, $T_\mathsf{rim} = (8\,200\pm400)$\,K.
	It is difficult to tell which radial temperature profile is correct.
	The power-law seems to give a temperature slightly above,
	and steady-disk gives a temperature that is slightly below the estimate
	by~\citet{mennickent2013}. \citet{al2000} generally gives
	larger rim temperatures. Most of the rim of his
	accretion disk has $T_\mathsf{rim} = 9\,000\,\mathrm{K}$,
	with two strips having twice higher temperature at the
	top and bottom of the disk. Spectroscopic studies \citep[e.g.,][]{hs93,ak2007} do not provide an accurate estimate,
	because spectral types from an~early F-type to late A-type were attributed
	to rim of the accretion disk.

	\item {\em Radial temperature profile:}
	It is almost impossible to determine the whole temperature
	profile, because the disk is opaque (the photosphere
	forms at most a few solar radii below the disk rim), and
	the orbital inclination is very close to $90^\circ$, which
	prevents us from seeing the disk face-on.
	The first issue manifests in itself for the
	power-law radial temperature profile by~the extreme
	correlation between the inner rim temperature
	$T_0$ and the exponent of the power-law
	$\rho\left(T_0, \alpha_\mathsf{T}\right) = -0.92$
	for both models with slab and wedge. A~comparison
	of models with slab and different radial temperature
	profiles (given by Eqs.~\ref{eq:temperature:powlaw},
	and~\ref{eq:temperature:steady}), and the temperature
	profile obtained by~\citet{mennickent2013} is shown
	in Fig.~\ref{fig:shellspec:comptemp}. The radial profiles
	agree with each other up to $\simeq 10$\,\rs~below
	the disk rim, that is below the pseudo-photosphere.
	The radial profile given by Eq.~(\ref{eq:temperature:steady}) derived
	by~\cite{shakura1973} is heated {\em only} by the viscous dissipation,
	but the disk in \bl enshrouds a~B0.5\,V star
	that must considerably heat the disk, too. A~problem is that
	asymptotically a~passive irradiated disk has the same
	radial temperature dependence $T \sim R^{-3/4}$
	as a steady disk \citep{friedjung1985,hubeny1990}.
	Hence it is difficult to discern the two radial temperature
	models from their behavior in outer part of the disk.
	\citet{calvet1991} modeled proto-stellar disks
	and found that the central regions of their disks are significantly heated
	by~the embedded star. It is interesting to note that the temperature $T_0$ of the power law behavior is quite close with the temperature of the central star. This is a good indication for the
the power law temperature behavior in the disk. From~afar the proto-stellar
	accretion disks are very similar to that surrounding
	\bl --- the accretion rate is high, and there
	is a~star in its center. Hence there are reasons to believe that the
	{\em power-law provides a better description of the
	radial temperature profile}. Finally the best-fitting
	power-law $\alpha_\mathsf{T} = -0.95$ is steeper
	than the canonical value ($-3/4$). This may suggest
	a presence of a~transition layer, where the temperature
	falls more steeply at the outer disk rim.

\end{itemize}

\begin{figure}
	\centering
	\includegraphics[width=0.45\textwidth]{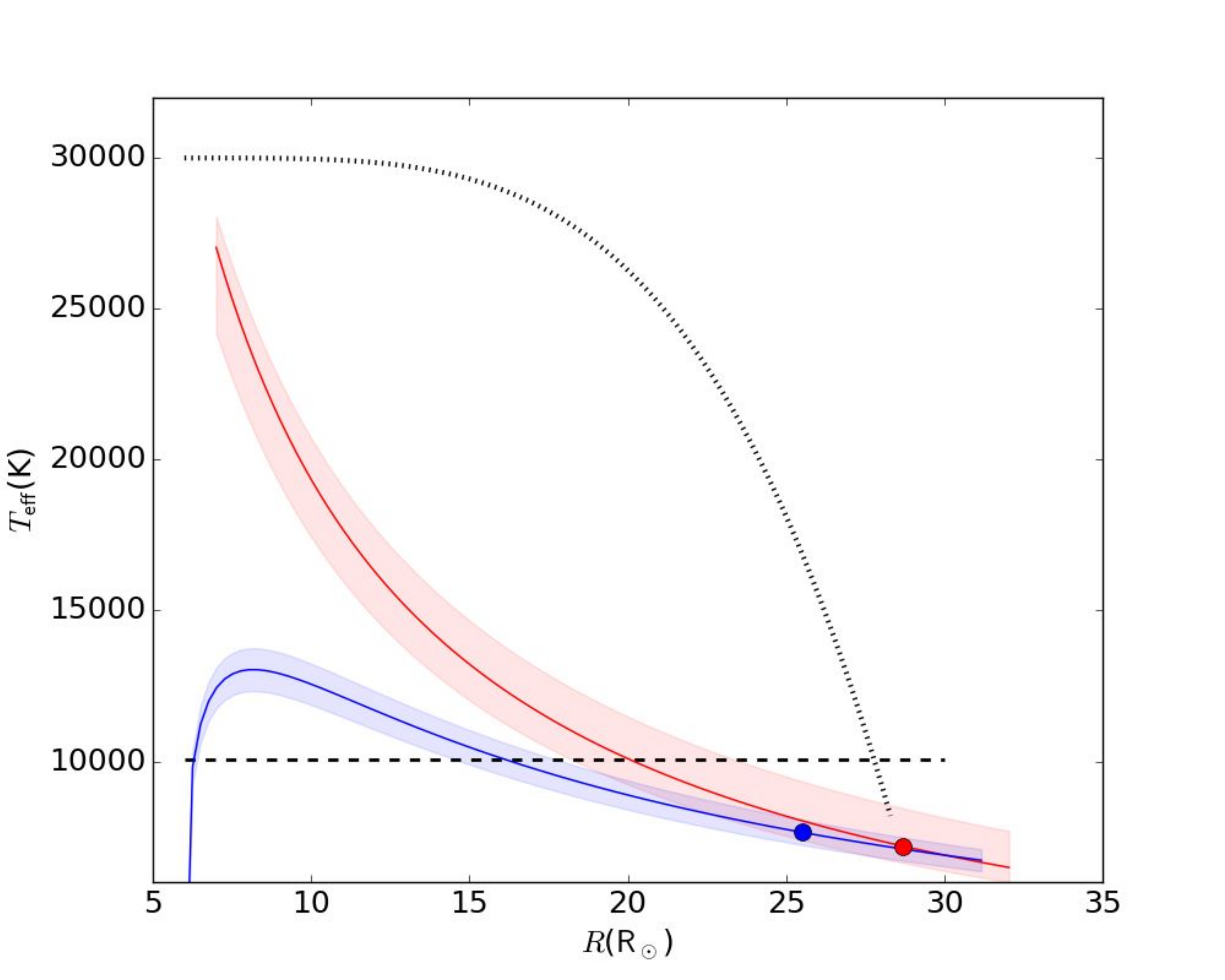}
	\caption{Comparison of radial temperature profiles of the
		accretion disk surrounding the gainer. {\em The dotted line}
		is the profile obtained by~\citet{mennickent2013}, and
		{\em the dashed line} is mean temperature of the two-temperature
		model developed by~\citet{al2000}. {\em The red line} represents
		the best-fitting power-law radial temperature profile
		(solution {\em slab/pl} in Table~\ref{tab:shellspec:result}), and
		{\em the red belt} all plausible solutions given by
		the uncertainty of the inner rim temperature and the
		exponent of the power-law, and
		{\em the red point} position of photosphere ($\tau = 2/3$),
		where temperature $T_\mathsf{ph.} = 7\,192$\,K is reached.
		{\em The blue line} represents
		the best-fitting steady-disk radial temperature profile
		(solution {\em slab/sd} in Table~\ref{tab:shellspec:result}),
		{\em the blue belt} all plausible solutions given
		by uncertainty in the inner rim temperature, and
		{\em the blue point} position of photosphere ($\tau = 2/3$),
		where temperature $T_\mathsf{ph.} = 7\,652$\,K is reached.
		The photospheric temperatures were computed for a~line
		of sight in the plane $z = 0$ and piercing through center
		of the accretion disk
		\label{fig:shellspec:comptemp}}
\end{figure}

%%%%%%%%%%%%%%%%%%%%%%%%%%%%%%%%%%%%%%%%%%%%%%%%%%%%%%%%%%%%%%%%%%%%%%%%

\subsection{Presence of ``a~hot~spot"}

\begin{figure}
\centering
\includegraphics[width=7.5cm]{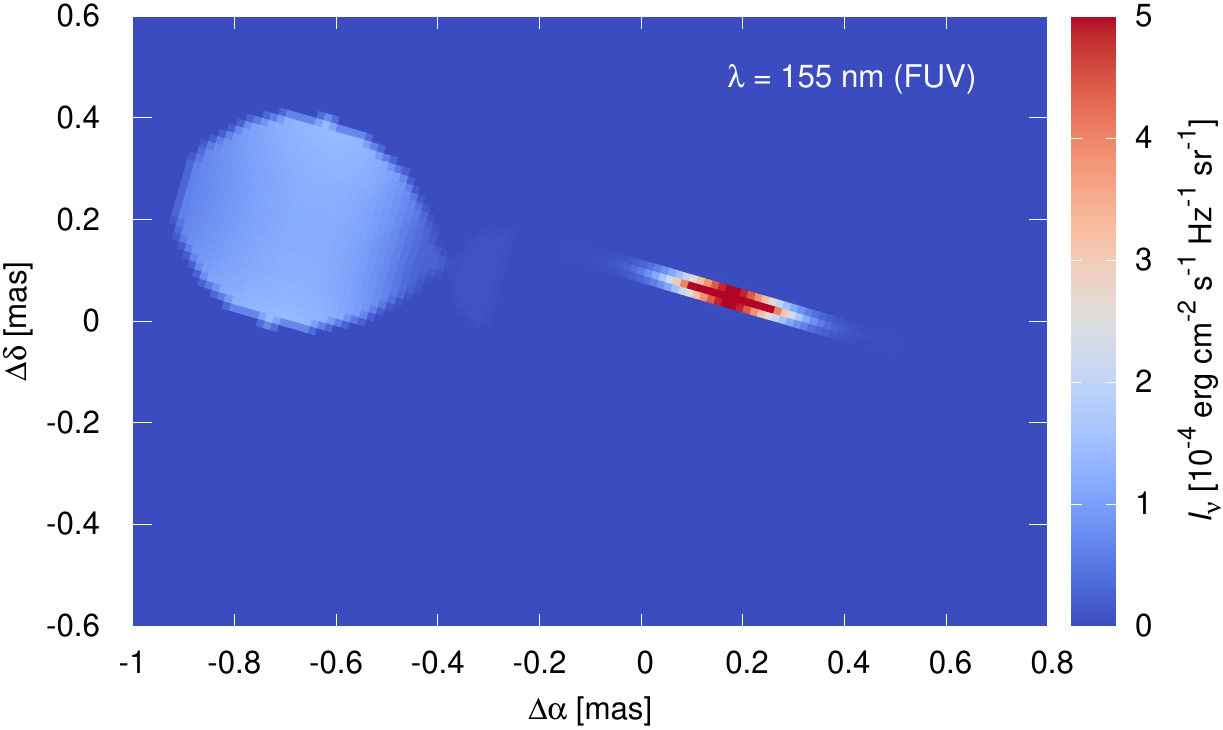}
\includegraphics[width=7.5cm]{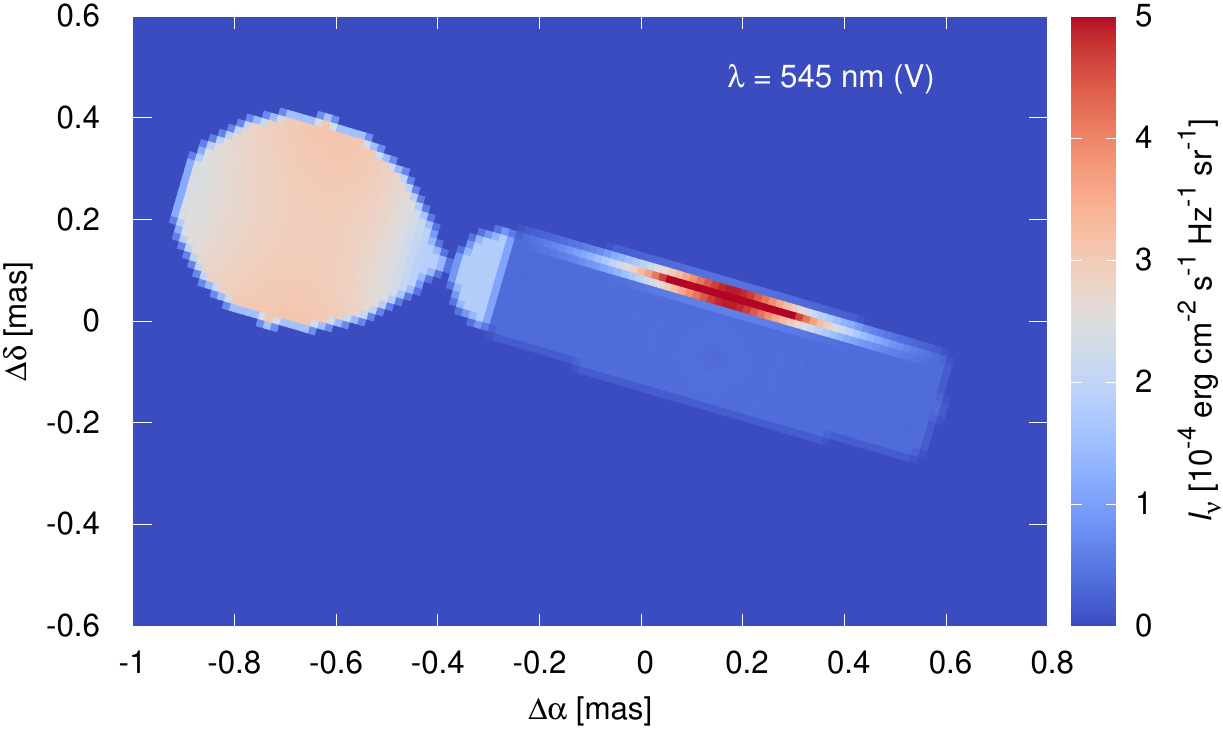}
\includegraphics[width=7.5cm]{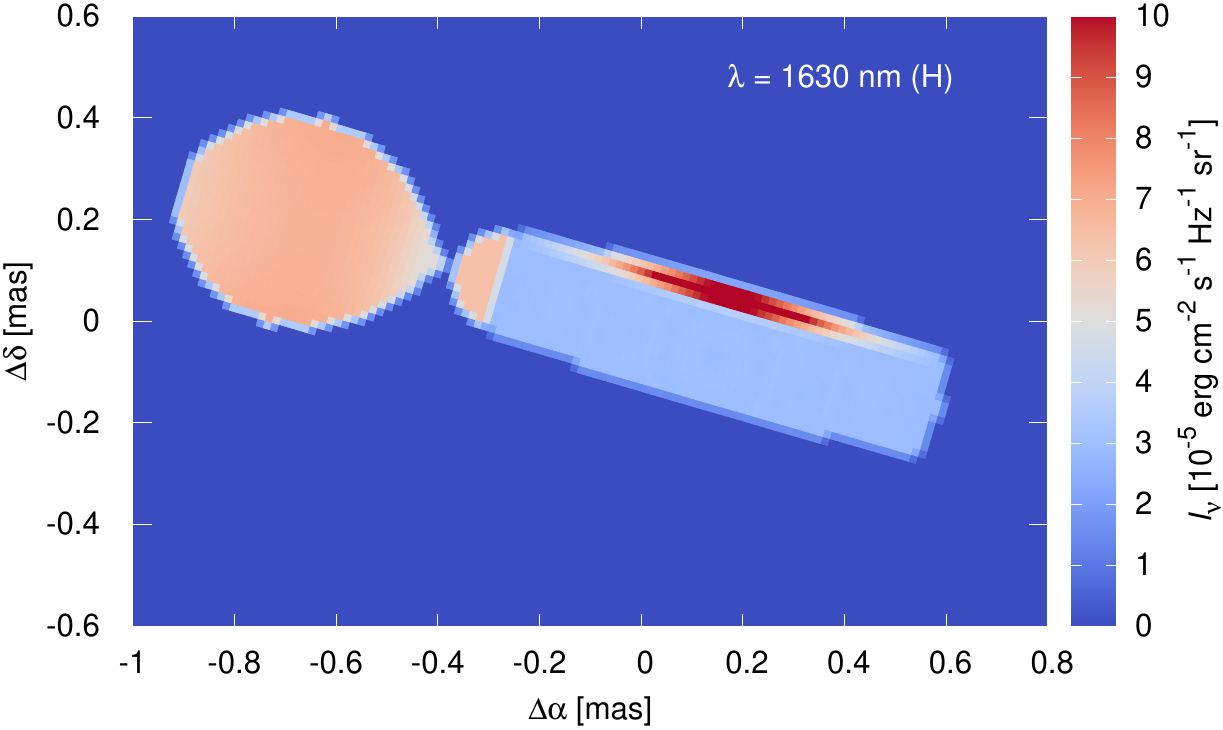}
\includegraphics[width=7.5cm]{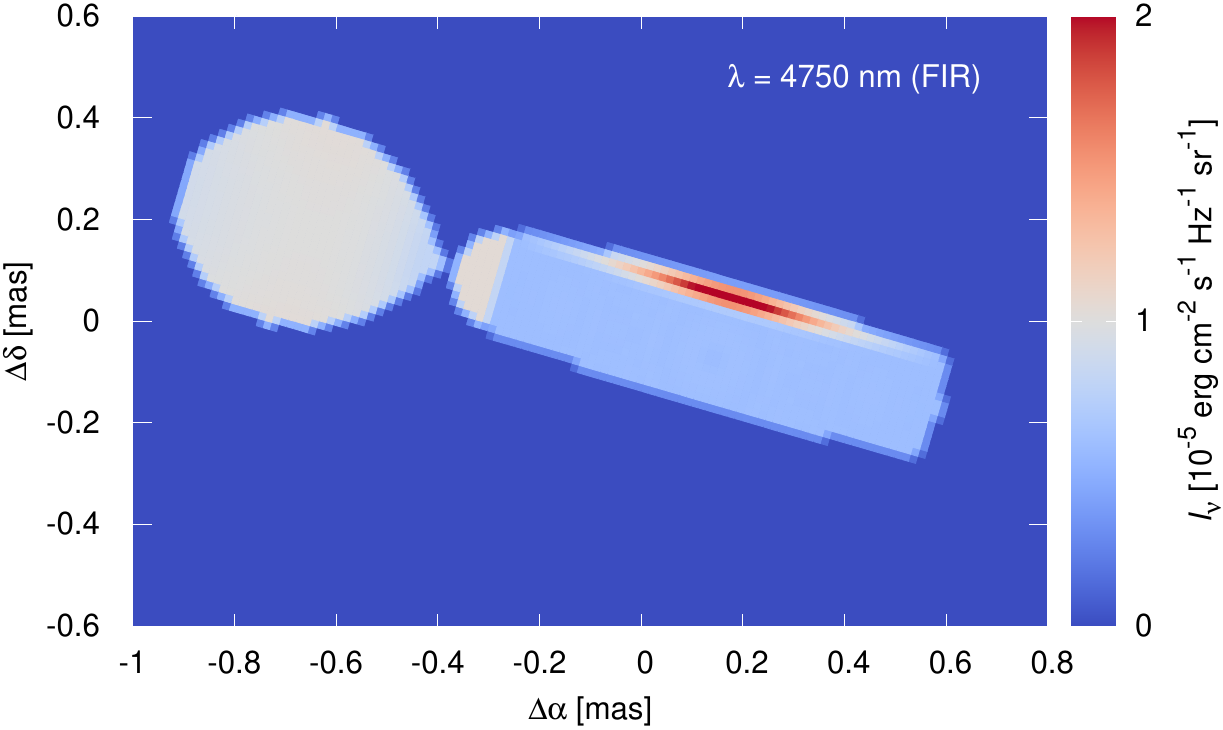}
\caption{Synthetic images of slab power-law model which was
further improved by a ``spot" (that is a~spherical structure
between the primary and secondary, partly hidden inside the slab).
The total (not reduced) $\chi^2$ value was decreased from $105\,202$
(without the spot) down to $102\,005$, which is a significant
improvement. Four wavelengths are shown:
$\lambda = 155\,{\rm nm}$ (FUV), 545\,nm (V~band), 1630\,nm (H), and 4750\,nm (M).}
\label{spot_img_oao2_1550}
\end{figure}

The existence of a~region heated by an interaction of the incoming flow and the
accretion disk is usually required by theoretical models
\citep[e.g.,][]{lubow1975}. For \bl \citet{lomax2012,mennickent2013} used
the hot spot to explain a presence of bumps (or irregularities) in the light
curve and polarized flux. We carried out an~attempt to confirm their
findings in continuum. A~spot represented by a~homogeneous sphere has been
added to the {\em slab power-law} model (see Fig.~\ref{spot_img_oao2_1550}).
Parameters defining its radial position
$r_\mathsf{s}$ within the accretion disk, the position
angle~$\theta_\mathsf{s}$ with respect to line joining centers of both binary
components, radius~$R_\mathsf{s}$, density~$\rho_\mathsf{s}$, and
temperature~$T_\mathsf{s}$ were optimized using the differential evolution
and simplex algorithms; the optimal values are listed in Table~\ref{tab:hotspot}.

First, only spot parameters were converged.
As the spot is a substantial non-axisymmetric feature,
the fit converged quickly to a location just between the primary and the disk.
The spot temperature~$T_\mathsf{s}$ about 10\,000\,K is logically
between those of the primary and the outer rim of the disk.
Second, all parameters were set free and converged again,
because the original parameters might have been affected by a systematic error of the model
(namely the missing spot).
The procedure helped to decrease the original (not reduced)
$\chi^2 = 105\,202$ down to $102\,005$,
which is a statistically significant improvement
and we thus may confirm the existence of the spot on the basis of continuum observations.
We verified that adding a spot to other models leads to an improvement of the same order.

As illustrated by Fig.~\ref{spot_img_oao2_1550}, the `spot' detected by our modeling is not
a tiny structure corresponding to the area of interaction
between the gas stream from the donor and the disk.
It likely represents an illuminated part of the disk rim,
where the reflection of donor light, and irradiation heating occur.
We note that such lateral temperature gradient has been observationally proven
for another object with an optically thick disk, $\varepsilon$~Aur
\citep[cf., e.g.,][]{hoard2012}.

In principle, it should be possible to add a second spot to the model,
and so on, and expect further improvements of the $\chi^2$, but we
prefer to keep a simple model as long as possible. Otherwise,
systematic uncertainties among different types of observations
(light curves, squared visibilities, closure phases, etc.) might
be hidden by a complex model. Moreover, there are techniques
(like spectro-interferometry in lines, or Doppler tomography)
better suited to pinpoint the orbital position of such non-axisymmetric features.

\begin{table}
\centering
\renewcommand{\arraystretch}{1.3}
\caption{Properties of the hot spot added to slab power-law model of $\beta$~Lyr~A.
\label{tab:hotspot}}
\begin{tabular}{lrllr@{$\,\pm\,$}l}
\hline\hline
Par. & Unit & Min & Max & \multicolumn{2}{c}{Value} \\
\hline
$r_\mathsf{s}$      & (\rs)          & 20         & $R_\mathsf{out}$ & 30.28 &$2.7 $\\
$\theta_\mathsf{s}$ & (deg)          & 0          & 360              & 0.6   &$11.4$\\
$R_\mathsf{s}$      & (\rs)          & 4          & 1.5$H$           & 6.48  &$2.21$\\
$\rho_\mathsf{s}$   & (g\,cm$^{-3}$) & $10^{-11}$ & $10^{-7}$        & 5.36  &$5.03 \times 10^{-8}$\\
$T_\mathsf{s}$      & (K)            & 6\,000     & 25\,000          & 9781  &$1893$\\
\hline
\end{tabular}
\tablefoot{The resulting total (not reduced) $\chi^2 = 102\,005$,
with individual contributions
$\chi^2_\mathsf{LC} = 8\,093$,
$\chi^2_\mathsf{V^2} = 54\,067$,
$\chi^2_\mathsf{CP} = 27\,290$, and
$\chi^2_\mathsf{T_3} = 12\,553$. For comparison with Table~\ref{tab:chi2}, the reduced $\chi^2$ is now $3.74$.}
\end{table}

%%%%%%%%%%%%%%%%%%%%%%%%%%%%%%%%%%%%%%%%%%%%%%%%%%%%%%%%%%%%%%%%%%%%%%%%

\subsection{Comparison of SED from models and observations}

\begin{figure}
\centering
\includegraphics[width=9cm]{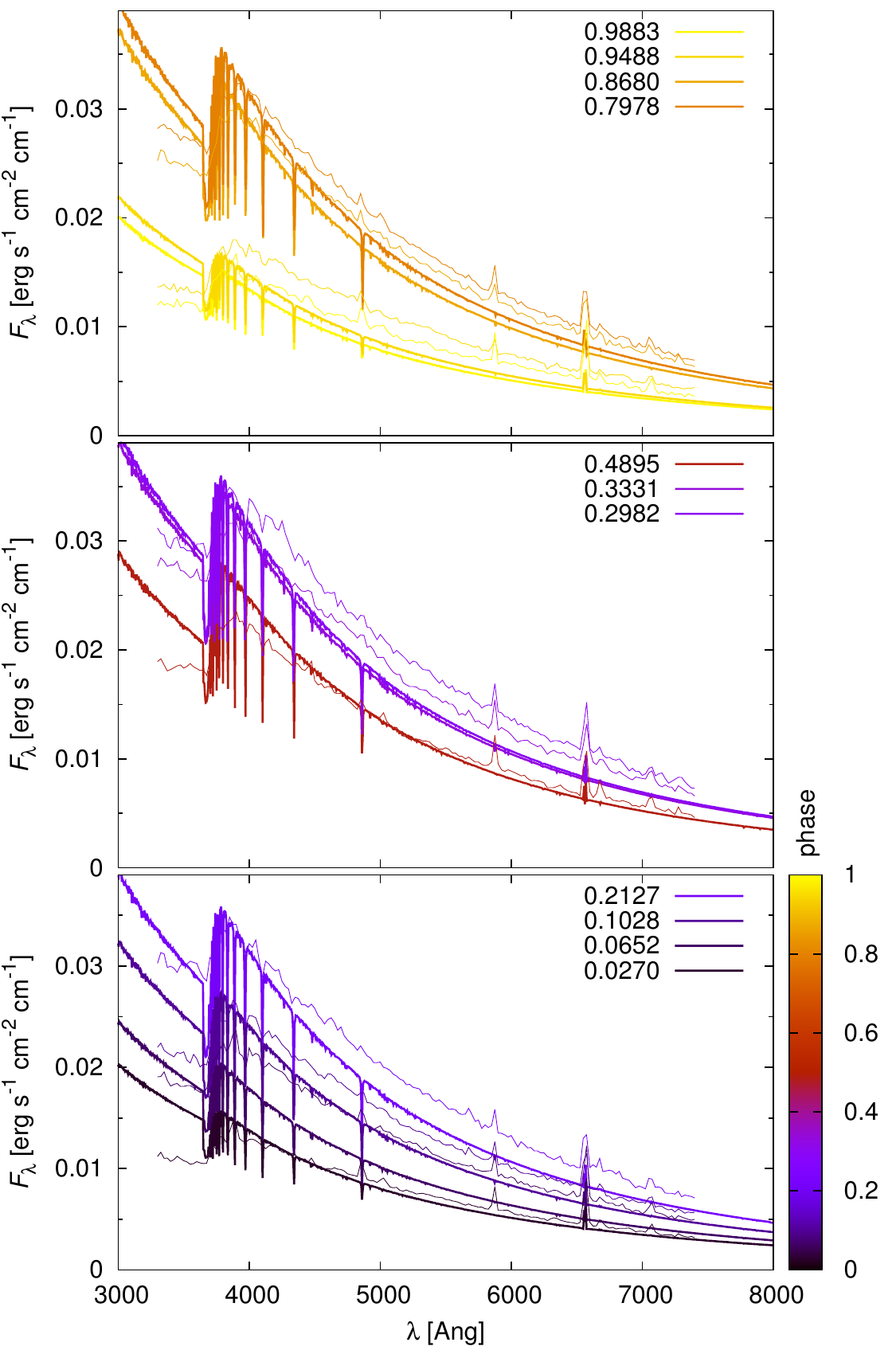}
\caption{Comparison of synthetic spectral--energy distributions
(thick lines) of $\beta$~Lyr~A and observed SED (thin lines) according to \cite{burnashev78}.
The monochromatic flux $F_\lambda$ (in ${\rm erg}\,{\rm s}^{-1}\,{\rm cm}^{-2}\,{\rm cm}^{-1}$)
was measured in the range of $\lambda = 330\hbox{ to }740\,{\rm nm}$.
The color scale correspond to the orbital phase.
Observation uncertainties are on the order of 5\,\%.
The model (corresponding to a 'nebula' with $t_{\rm inv} = 4.53$)
was {\em not\/} converged with respect to these observations
and the SEDs thus inevitably exhibit some systematic offsets.}
\label{nebula_powlaw_shellspectrum_BURNASEV}
\end{figure}

Although we did not attempt to fit the spectral energy distribution (SED)
of the system, it would be an important check of the model. That is why
we compare synthetic SEDs to the observations of \cite{burnashev78}
(see Figure~\ref{nebula_powlaw_shellspectrum_BURNASEV}).
It seems inevitable that SEDs exhibit some systematic offsets.
Also the resolution of synthetic spectra does not match the observations.
Moreover, we cannot expect that emission lines will be computed correctly,
because the model is mostly focused on opaque medium and continuum flux.
However, the absolute fluxes and their spread among primary minimum,
secondary minimum and out of the eclipses seem to at least roughly correct.
The expected systematic uncertainties of the observations (5\%)
might be almost of the same order, especially in UV, where the extinction
is strong and variable. We thus believe it should be possible to match
the observed SEDs with a future version of our model.

%%%%%%%%%%%%%%%%%%%%%%%%%%%%%%%%%%%%%%%%%%%%%%%%%%%%%%%%%%%%%%%%%%%%%%%%
%%%%%%%%%%%%%%%%%%%%%%%%%%%%%%%%%%%%%%%%%%%%%%%%%%%%%%%%%%%%%%%%%%%%%%%%

\section{Conclusion and outlook}

The properties of opaque bodies within \bla system were
studied. Our analysis was primarily targeted
on the properties of the accretion disk surrounding
the mass-gaining component of this close interacting binary.

For the description of interacting binary systems, we created a tool based on the
\shellspec\ code. It permitted us to significantly improve the modeling
of stellar systems and the computations of radiation transfer
in the co-moving circumstellar medium by~\citet{budaj2004}.
Apart from improvements suitable for an optically thick medium,
we can also compute interferometric observables
and proceed with both local- and global-optimisation methods.

We then constructed several disk models that differed in shape,
density and temperature profiles. These models
were fitted to series of spectro-interferometric and
photometric observations, both sampling the whole orbit.
We also compared our results to those obtained by earlier investigators
of the system \citep[especially][]{al2000,mennickent2013}
and to theoretical models of accretion disks.

Our results indicate that the opaque parts of the accretion disk have
the outer radius $R_\mathsf{out} = (30.0\pm1.0)$\,\rs,
the semi-thickness $H = (6.5\pm1.0)$\,\rs\ (for slab and wedge shapes),
or equivalently the scale-height multiplication factor $h_{\rm mul} = 4.3\pm0.3$
(for nebula model; see the overview in Fig.~\ref{fig:model:disk}). But the true location of the disk pseudo-photosphere slightly depends on the wavelength.
The minimum mass should be $10^{-4}$ to $10^{-3}$\,\ms.
Given the thickness, the disk clearly cannot be in a vertical hydrostatic equilibrium.
We have also determined
the orbital inclination $i = (93.5\pm1.0)$\,deg (as an average and range for admissible models),
the longitude of ascending node $\Omega = (253.7\pm1.0)$\,deg,
and the probable distance to the $\beta$~Lyr~A system $d = (319.7\pm2.7)$\,pc.

The power-law temperature profiles (and also the steady-disk for wedge)
seem compatible with the observations,
but the central values remain very unconstrained,
because the disk continuum is formed only a few solar
radii below the disk rim.

An addition of a~hot spot to our model improved the $\chi^2$,
so that we can consider the existence of the spot to be
confirmed in the continuum radiation, although it may be actually
a compensation of missing reflection from the disk,
or heating of the disk by the companion. Its position may also correspond
to a flow of material from the primary (donor).

The radiative and kinematic properties of neighboring \blb
have been determined too. Even though we were unable
to prove \bla and~B orbit each other,
they both likely originate from the same association.

\vskip\baselineskip
This study presents a~springboard to forthcoming analyses
of the optically {\em thin\/} circumstellar medium in \bla
--- it is crucial to know the properties of the opaque material too.
Using a~series of spectroscopic and
spectro-interferometric observations of strong
emission lines we intend to resolve and describe
the structure and kinematics of the optically thin
medium within this remarkable system.
Consequently, it should be possible to better determine
the radial profiles of the disk atmosphere.
At the same time, the mass of jets would provide an~accurate estimate
of the mass and angular momentum loss from
the system, which would offer an~invaluable test for
models of mass transfer in binary systems.

%%%%%%%%%%%%%%%%%%%%%%%%%%%%%%%%%%%%%%%%%%%%%%%%%%%%%%%%%%%%%%%%%%%%%%%%

\begin{acknowledgements}
The constructive criticism of an earlier version of this manuscript
by the anonymous referee is appreciated.
This research was supported by the grants P209/10/0715, GA15-2112S,
and~GA17-00871S of the Czech Science Foundation, by the grant no.~250015 of
the Grant Agency of the Charles University in Prague.\\
This work is based upon observations obtained with the
Georgia State University Center for High Angular
Resolution Astronomy Array at Mount Wilson Observatory.
The CHARA Array is supported by the National Science Foundation
under Grant No. AST-1211929, AST-1411654, AST-1636624, and AST-1715788.
Institutional support has been provided from the GSU
College of Arts and Sciences and the GSU Office of the
Vice President for Research and Economic Development.
We thank M.~Zhao for participating in the MIRC observations,
and B.~Kloppenborg for his support in the earlier attempts to use
the SIMTOI modeling tool. Two Reticon spectra of \blb were obtained by Dr.~P.~Hadrava.\\
HB acknowledges financial support from the Croatian Science Foundation
under the project 6212 ``Solar and Stellar Variability".\\
The work of JB was supported by the VEGA 2/0031/18 and APVV-15-0458 grants.\\
We acknowledge the use of the electronic database from the CDS, Strasbourg,
and the electronic bibliography maintained by the NASA/ADS system.\\
This work has made use of data from the European Space Agency (ESA)
mission \emph{Gaia} (\url{http://www.cosmos.esa.int/gaia}),
processed by the \emph{Gaia} Data Processing and Analysis Consortium
(DPAC, \url{http://www.cosmos.esa.int/web/gaia/dpac/consortium}).
Funding for the DPAC has been provided by national institutions,
in particular those participating in the \emph{Gaia} Multilateral Agreement.\\

In April 2012, our late friend and colleague Olivier Chesneau pointed out to some of us
the publication by \citet{lomax2012} and expressed these inspirational words: "...the hot spot...is quite extended and should
show up in the interferometric observables". This was the starting point of this study and our results are dedicated to him.

\end{acknowledgements}

%%%%%%%%%%%%%%%%%%%%%%%%%%%%%%%%%%%%%%%%%%%%%%%%%%%%%%%%%%%%%%%%%%%%%%%%

\bibliographystyle{aa}
\bibliography{bl22saga}

%%%%%%%%%%%%%%%%%%%%%%%%%%%%%%%%%%%%%%%%%%%%%%%%%%%%%%%%%%%%%%%%%%%%%%%%

\begin{appendix}

%%%%%%%%%%%%%%%%%%%%%%%%%%%%%%%%%%%%%%%%%%%%%%%%%%%%%%%%%%%%%%%%%%%%%%%%

\section{Details on the interferometric data reductions}
\label{sec:apa}
Additional details on the reduction process of interferometric observations are presented in this section. Detailed characteristics of all observations are presented in electronic Table~\ref{tab:if:logobs}. Individual interferometric observation are available in the {\tt OIFITS} format electronically through CDS. These files contain only {\em reduced} data, that is calibrated squared visibilities and closure phases. Raw data are made available upon request.
\begin{table}
	\centering
	\caption{Detailed journal of interferometric observations.
		\label{tab:if:logobs}}
	\begin{tabular}{ccccccc}
		\hline\hline\noalign{\smallskip}
		Date & RJD & $\phi_\mathsf{O}$ & Tel. & $\Delta\lambda$ & $N_\mathsf{CH}$ & Src. \\
		(yyyy-mm-dd) & (d) & & & (nm) & & \\
		\hline\noalign{\smallskip}
		\multicolumn{7}{c}{This table is available electronically through CDS.} \\
		\noalign{\smallskip}\hline
	\end{tabular}
	\tablefoot{``Date'' denotes the observation date,
		``RJD'' the mid-exposure epoch,
		$\phi_\mathsf{O}$ the full number of orbital cycles
		since the reference epoch by~\citet{ak2007},
		``Tel.'' the configuration of the source instrument,
		$\Delta\lambda$ the passband, and
		$N_\mathsf{CH}$ number of channels into which the
		passband $\Delta\lambda$ was sliced.
		Columns ``Src.'' denotes source instrument of the
		observations. They are the following:
		1. CHARA/VEGA,
		2. CHARA/MIRC, and
		3. NPOI.}
\end{table}
\begin{table*}
	\centering
	\caption{Journal of calibrator stars that were used to calibrate the interferometric observations of \be.\label{tab:vega:calibrators}}
	\begin{tabular}{rlcccc}
	\hline\hline\noalign{\smallskip} 	
	Parameter & Unit &	\multicolumn{4}{c}{Calibrator}	\\	
	& & HD\,176437 & HD\,192640 &	HD\,189849 & HD\,168914 \\
	\hline\noalign{\smallskip}
	Spectral type & & B9III &	A2V	& A4III	& A7V \\
	$T_\mathsf{eff}$ & (K) & 11\,226 & 8\,7774 & 7\,804	 & 7\,600	\\
	$\log g_\mathsf{[cgs]}$ & & 4.11 & 4.42 &	3.89 & 4.20	\\
	$V$ &(mag)&	3.250& 4.949$^{3}$ & 4.65$^{1}$	& 5.12$^{2}$\\
	$\theta_\mathsf{LD}$ & (mas) &0.755$\pm$0.019$^{\rm a}$&0.471$\pm$0.033$^{\rm b}$&0.517$\pm$0.036$^{\rm b}$&0.456$\pm$0.023$^{\rm c}$	\\
	$\theta_\mathsf{UD}(\mathsf{V})$ & (mas) &0.727$\pm$0.018 &0.449$\pm$0.031 &0.489$\pm$0.034 &0.432$\pm$0.022 \\
	$\theta_\mathsf{UD}(\mathsf{R})$ & (mas) &0.733$\pm$0.018 &0.454$\pm$0.032 &0.496$\pm$0.035 &0.437$\pm$0.022 \\
	$\theta_\mathsf{UD}(\mathsf{I})$ & (mas) &0.738$\pm$0.019 &0.458$\pm$0.032 &0.501$\pm$0.035 &0.441$\pm$0.022 \\
	$\theta_\mathsf{UD}(\mathsf{H})$ & (mas) &0.737$\pm$0.015 &0.465$\pm$0.033 &0.510$\pm$0.036 &0.463$\pm$0.033 \\
	\noalign{\smallskip}\hline
	\end{tabular}	
	\tablefoot{$V$ denotes apparent magnitude in the Johnson's V passband,
		$\theta_\mathsf{LD}$ limb-darkened-disk angular diameter,
		$\theta_\mathsf{UD}$ uniform-disk angular diameter, and
		V, R, I, H denote Johnson-series passbands.
		Apparent magnitudes and limb-darkened angular
		diameters were taken from the following sources:
		$^{1}$\citet{johnson1966},
		$^{2}$\citet{haggkvist1966},
		$^{3}$\citet{moffett1979}
		$^{\rm a}$\citet{monnier2012},
		$^{\rm b}$\citet{lafrasse2010},
		$^{\rm c}$\citet{touhami2013}
		}
\end{table*}

\subsection{Details on the reduction of CHARA/VEGA observations}\label{sec:apa:vega}
\begin{figure}
\centering
\includegraphics[width=7.5cm]{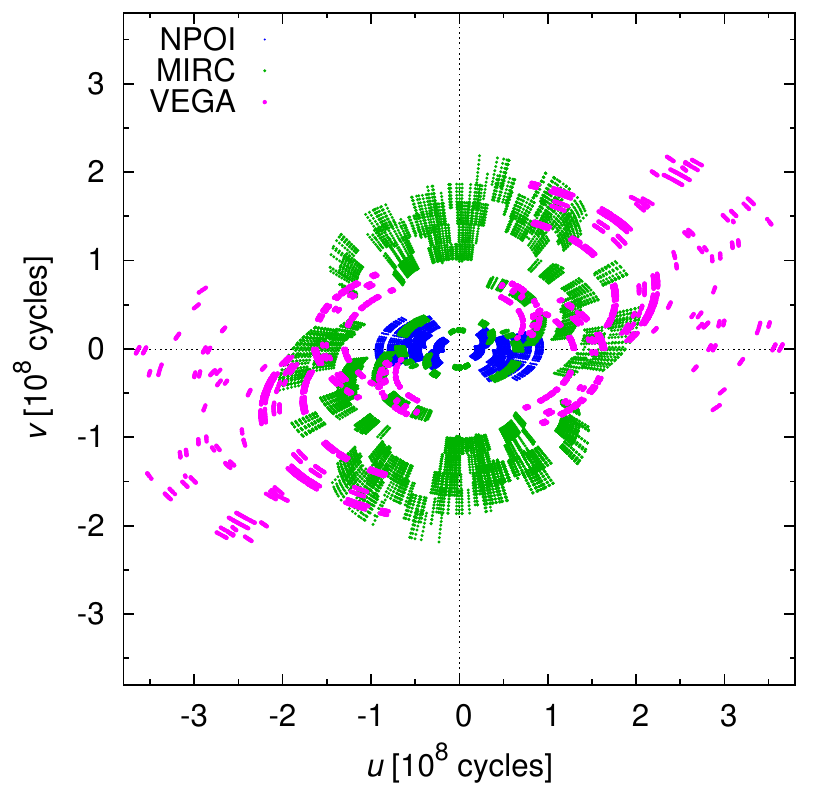}
\caption{$(u, v)$ coverage of all interferometric observations.
Colors correspond to three different instruments:
NPOI (blue),
CHARA/MIRC (green),
CHARA/VEGA (magenta).}
\label{fig:uv}
\end{figure}

The observations were carried out during a~11 days long campaign in 2013. Both MIRC and VEGA were observing at the same time, sharing the visible and infrared photons. The majority of observations was taken in the three-telescope (3T) mode. The only two exceptions were the first ($22^\mathsf{nd}$ Jun, 2013) and the last ($1^\mathsf{st}$ Jul, 2013) nights. Observations
from those two nights were taken in two-telescope mode. Only four CHARA telescopes (denoted E1, E2, W1, and W2) were used. The long baselines were in general east-west oriented and
the short baselines north-south oriented, because the projection of \bl orbit on the sky is roughly east-west oriented (see Fig.~\ref{fig:uv}). The length of projected baselines ranged from $\simeq 65$\,m to $\simeq 250$\,m.

The observations were carried out with two cameras in four passbands that were centered at following wavelengths $\lambda_\mathsf{C} \in \{535, 656, 706.5, 815\}$\,nm. Resolution of the recorded spectra was $R = 5\,000$. Individual frames were recorded with frequency $100\,$Hz, and were grouped into blocks containing 2500 frames. An~observation typically contained $\simeq 20$ blocks. Within these blocks the frames were coherently summed and raw squared visibility was determined for each block. The whole passband was not used, but two narrow channels were selected in each passband. The following channels were selected:
$\Delta_\mathsf{CH} \in
\{
	525-535,
	535-545,
	643-653,
	658-667,
	687-702,
	709-724,
	810-825,
	825-840
\}$\,nm. The channels avoid major spectral lines in these regions. The only exceptions are regions $643-653$\,nm, and $658-667$\,nm, which are partially affected by~wings of H$\alpha$ line, and $825-840$\,nm, which is affected by water-vapor lines. Narrow channels were chosen because of the limited coherence of the waves due to the atmospheric turbulence.

Four calibrators were observed in order to calibrate the instrumental visibilities. Due to variability of atmospheric conditions during night, a~calibrator was observed before and after each observation of \be. A~list of calibrators and their properties are listed in Table~\ref{tab:vega:calibrators}. The calibrators were chosen with help of tool SearchCal\footnote{The tool is available at \url{http://www.jmmc.fr/searchcal}.} developed by~\citet{bonneau2006}. The uniform-disk diameters of calibrators were taken from the JMMC catalog of stellar
diameters by~\citet{lafrasse2010}.

In order to avoid modeling of highly inaccurate observations, all blocks that have instrumental visibility with $\mathrm{S/N} < 2$ were removed. Also we performed simple filtering based on residual optical path delay (OPD). Correct blocks of visibility measurements have very similar OPD. Hence if OPD of one (or more) blocks deviate significantly from a~mean OPD based on all blocks acquired within one measurement, this block is very likely wrong. Therefore if a~block OPD differed from the mean by more than two standard deviations, it was removed.

\subsection{Details on the reduction of CHARA/MIRC observations}\label{sec:apa:mirc}

The instrument CHARA/MIRC was used to measure squared visibilities and closure phases. MIRC performs real-time group delay tracking in the \emph{H}-band. The observations come from two observational runs: (i)~2006 -- 2007 campaign that was already analyzed by~\citet{zhao2008} and whose description can be found there, and (ii)~2013 campaign, together with VEGA, whose description follows. All observations used here were taken across the near-IR H-band.

Since 2011, MIRC can combine light from six telescopes, so it is able to record $15$ squared visibilities and $20$ closure phase observations.  The H-band is split into eight channels with absolute wavelength accuracy $\pm 0.25\%$. A~more thorough description of the instrument is in studies by~\citet{monnier2004,monnier2010}, and~by~\citet{che2010,che2012}. Using Fourier transform techniques, the visibilities are measured, averaged, and corrected for biases. The bispectrum is formed using the phases and amplitudes of three baselines that form a closed triangle \citep{monnier2007}. Amplitude calibration was performed using real-time flux estimates derived through use of a beam splitter following spatial filtering
for improved performance. Lastly, observations of reference calibrators (see Table~\ref{tab:vega:calibrators}) throughout the night allowed for correction of time-variable factors such as atmospheric coherence time, vibrations, differential dispersion, and birefringence in the beam train. The uncertainties of closure phase measurements were adjusted (see Sect.~\ref{sec:shellspec:datamod}). MIRC and VEGA shared the same calibrators for this observing campaign.

\subsection{Details on the reduction of NPOI observations}\label{sec:apa:npoi}
All NPOI observations were taken with the six-beam combiner. Visibilities, complex triple amplitudes, and closure phases were recorded in $16$ narrow-band channels between $5500$\,\r{A}
and $8500$\,\r{A}.
The calibrators are taken from a list of single stars maintained at NPOI with diameters estimated from $V$ and $(V-K)$ using the surface brightness relation by \citet{mozurkewich2003} and \citet{vanbelle2009} (see Table~\ref{tab:vega:calibrators}). Values of $E(B-V)$ were derived from comparison of the observed and theoretical colors as a function of spectral type by Schmidt-Kaler in \citet{aller1982}. Values for the extinction derived from $E(B-V)$ were compared to estimates based on the maps by \citet{drimmel2003}, and used to correct $V$ if
they agreed within $0.5$\,mag. Even though the surface brightness relation based on $(V-K)$ colors is  to first order  independent of the reddening, we included this small correction. For these observations, only one calibrator was used: HD\,176437.

NPOI data and their reductions followed the procedure described by \citet{npoi} and \citet{hummel2003}. A~pipeline written in GDL\footnote{\url{http://gnudatalanguage.sourceforge.net}} was used for the OYSTER\footnote{\url{http://www.eso.org/~chummel/oyster}} NPOI data reduction package. The pipeline automatically edits the one-second averages produced by another pipeline directly from the raw frames, based on expected performance such as the variance of fringe tracker delay, photon count rates, and narrow-angle tracker offsets. Visibility bias corrections are derived as usual from the data recorded away from the stellar fringe packet. After averaging the data over the full length of an observation, the closure phases of the calibrators were  automatically unwrapped so that their variation with time, as well as that of the visibility amplitude, could be interpolated for the observations of \be. For the calibration of the visibilities, the pipeline used all calibrator stars observed during a~night to obtain smooth averages of the amplitude and phase-transfer functions using a~Gaussian kernel of 80 minutes
in length. The residual scatter of the calibrator visibilities and phases around the average set the level of the calibration uncertainty and was added in quadrature to the intrinsic data errors. The amplitude calibration error of typically a~few percent in the red channels up to 15\% in the blue channels was added in quadrature to the intrinsic error of the visibilities. The phase calibration was good to about a~couple of degrees.

%%%%%%%%%%%%%%%%%%%%%%%%%%%%%%%%%%%%%%%%%%%%%%%%%%%%%%%%%%%%%%%%%%%%%%%%

\section{Details on the photometric data reductions}\label{sec:apb}
Hvar \ubvr\ observations are differential observations,
relative to $\gamma$~Lyr (HD~176437), for which
the following mean Hvar all-sky values from excellent
nights were adopted:
\begin{eqnarray}
	V & = & 3.253\,\mathrm{mag}, \nonumber \\
	B-V & = & -0.0644\,\mathrm{mag}, \nonumber \\
	U-B & = & -0.0308\,\mathrm{mag}, \nonumber \\
	V-R & = & -0.0024\,\mathrm{mag}. \nonumber
\end{eqnarray}
The check star $\nu^2$~Lyr (HD~174602)
\begin{eqnarray}
	V & = & 5.243\,\mathrm{mag}, \nonumber \\
	B-V & = & 0.0980\,\mathrm{mag}, \nonumber \\
	U-B & = & 0.1038\,\mathrm{mag}, \nonumber \\
	V-R & = & 0.1000\,\mathrm{mag}, \nonumber
\end{eqnarray}
was observed as frequently as the variable and \blb was also
observed as another check star on a number of nights.
All Hvar \ubvr\ observations were transformed to
standard system through non-linear transformation
formul\ae\ using the {\tt HEC22} reduction program
\citep[see][for the observational strategy and
data reduction]{hhj94, hechor98}.
\footnote{The whole program suite with a detailed manual,
examples of data, auxiliary data files, and results is available at
\url{http://astro.troja.mff.cuni.cz/ftp/hec/PHOT}.}
All observations were reduced with the latest
{\tt HEC22 rel.18.2} version of the program, which allows the time variation
of linear extinction coefficients to be modeled in the course of
observing nights. The \ubvr\ observations were reduced to Johnson bright
standards, for which we derived robust mean \ubvr\ values from individual
observations published by \citet{johnson1966}.
The uncertainties of these observations estimated from
measurements of the check star are
$\sigma_\mathsf{U,1} = 0.010$\,mag,
$\sigma_\mathsf{B,1} = 0.014$\,mag,
$\sigma_\mathsf{V,1} = 0.007$\,mag, and
$\sigma_\mathsf{R,1} = 0.013$\,mag.
We underline that our $R$ magnitudes were reduced
to Johnson, not Cousins $R$ values.

Differential Johnson-Cousins \bvr\ observations were acquired
at private observatory of Mr.~Svoboda in Brno, Czech~Republic
using SBIG~ST-7XME CCD~camera mounted at 34\,mm
refractor. \bl and comparison
star $\gamma$\,Lyr were observed simultaneously.
The atmospheric extinction was assumed constant
over the image. Seasonal transformation of the instrumental
magnitudes into standard system was carried
out using linear formul\ae~similar to Eq.~(4) in
\citet{hec1977}. Upper limit on the uncertainties
of these observations are
$\sigma_\mathsf{B,2} = 0.013$\,mag,
$\sigma_\mathsf{V,2} = 0.013$\,mag, and
$\sigma_\mathsf{R,2} = 0.010$\,mag.

All calibrated \ubvr\ observations acquired at Hvar observatory,
and differential \bvr\ observations acquired by PS
are listed in Table~\ref{tab:lc:data}. This table
is available electronically at CDS.

\begin{table}
	\centering
	\caption{List of calibrated \bl photometric measurement acquired
		at Hvar observatory.
		\label{tab:lc:data}}
	\begin{tabular}{cccccr}
		\hline\hline\noalign{\smallskip}
		HJD & $U$ & $B$ & $V$ & $R$  & Source\\
		(d) & \phantom{00}(mag) & \phantom{00}(mag) & \phantom{00}(mag) & \phantom{00}(mag) \\
		\hline\noalign{\smallskip}
		\multicolumn{6}{c}{This table is available electronically through CDS.} \\
		\noalign{\smallskip}\hline
	\end{tabular}
	\tablefoot{HJD denotes heliocentric Julian date of mid-exposure,
		$U$, $B$, $V$, and $R$ are Johnson \ubvr\ apparent magnitudes.
		Column ``Source'':
		1. Hvar observatory,
		2. differential \bvr\ photometry acquired by PS.
		The uncertainties of the Hvar estimated from
		measurements of the check star ($\nu^2$~Lyr) are
		$\sigma_\mathsf{U,1} = 0.010$\,mag,
		$\sigma_\mathsf{B,1} = 0.014$\,mag,
		$\sigma_\mathsf{V,1} = 0.007$\,mag, and
		$\sigma_\mathsf{R,1} = 0.013$\,mag.
		The upper limit on uncertainties of differential
		\bvr\ measurements collected by~PS are
		$\sigma_\mathsf{B,2} = 0.013$\,mag,
		$\sigma_\mathsf{V,2} = 0.013$\,mag, and
		$\sigma_\mathsf{R,2} = 0.010$\,mag.
		}
\end{table}
%

%%%%%%%%%%%%%%%%%%%%%%%%%%%%%%%%%%%%%%%%%%%%%%%%%%%%%%%%%%%%%%%%%%%%%%%%

\section{Supplementary material to analysis of \ble}\label{sec:apc}
Supplementary material to analysis of \blb
(see Sect.~\ref{sec:blb}) is presented here.

\subsection{Details on the spectroscopic observations}
All $13$ electronic spectrograms were obtained
in the coud\'e focus of the $2$\,m reflector and have
linear dispersion of $17.2$\,\Am and two-pixel resolution
$12\,600$ ($11-12$\ks per pixel). The first $7$ spectra (until
$\mathrm{RJD} \simeq 50\,235$)  were taken with a~Reticon 1872RF
linear detector and cover a~spectral region from
$6300$ to $6730$\,\r{A}. Complete reductions (bias subtraction,
flat-fielding, extraction of 1-d spectrum,
wavelength calibration, normalization) of
these spectrograms were carried out by PH with the program
\spefoe. The remaining spectra were secured with
a~SITe-$5\,800\times2\,000$ CCD detector and cover
wavelength interval from $6260$ to $6760$\,\r{A}.
Their initial reductions (bias subtraction, flat-fielding,
extraction of 1-d spectrum, and wavelength calibration) were carried
out by M\v{S} in \texttt{IRAF} \footnote{\texttt{IRAF} is distributed by the~National Optical Astronomy Observatories,
  operated by the~Association of Universities for Research in Astronomy,
  Inc., under contract to the~National Science Foundation of the~United States.} and their normalization
by PH in \spefoe. In both cases the stellar continuum
was approximated by Hermite polynomials that were fitted
through several (suitably chosen) continuum points.

Photometric observations of \blb were obtained
at Hvar observatory and their reduction
procedure is described in Appendix~\ref{sec:apb}.

\subsection{Kinematic and radiative properties of \ble}
\begin{figure*}
	\centering
	\includegraphics[width=\textwidth]{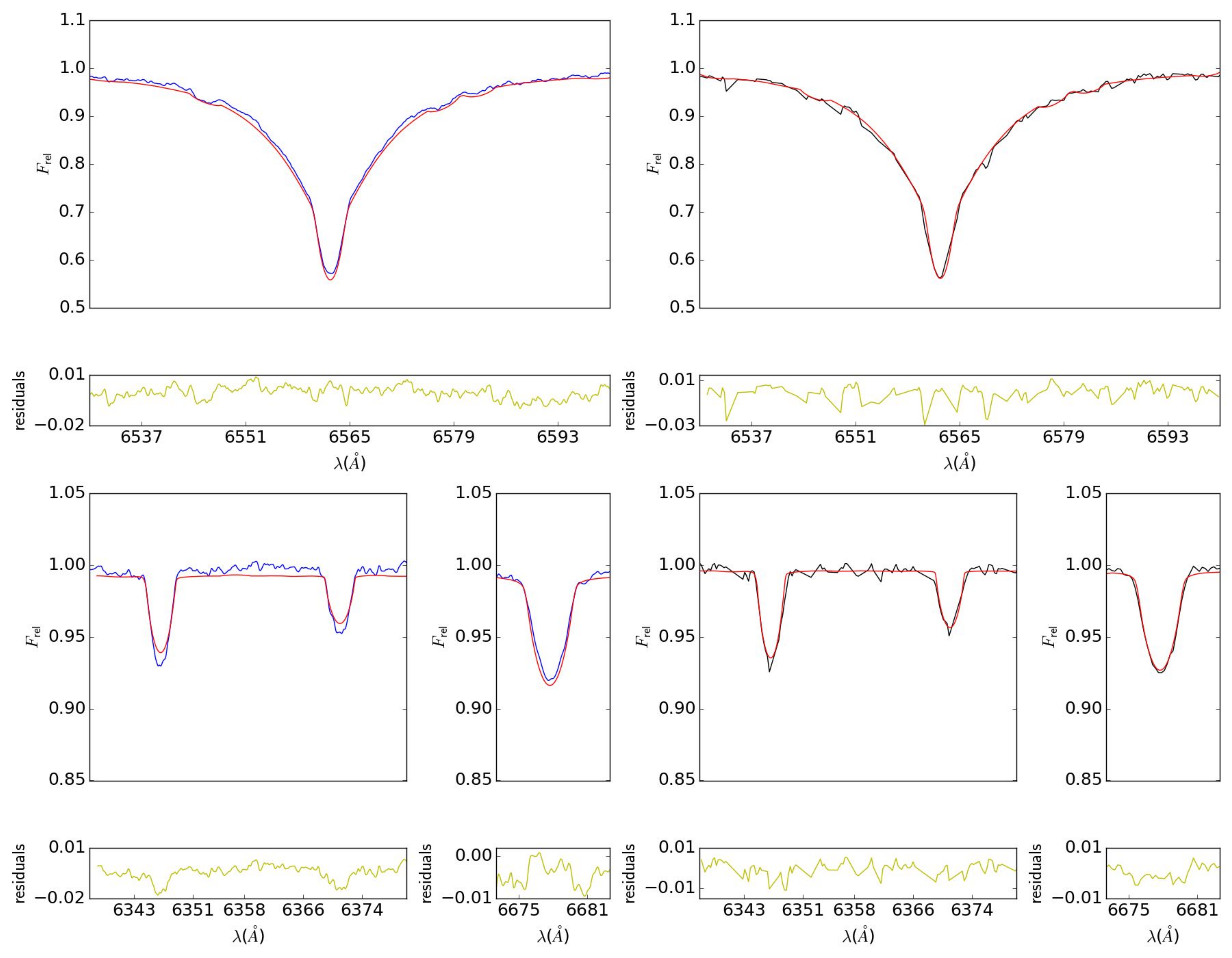}
	\caption{Comparison of an~observed spectrum of \blb obtained on $\mathrm{RJD} =
		49\,866.4348$ and disentangled spectra of~\blb with the best
		fitting spectra corresponding to their respective solutions
		listed in Table~\ref{tab:blb:spectrafit}, that is synthetic
		spectra compared to observed spectrum correspond to slightly
		different parameters than those compared to disentangled
		spectra.
		The blue line denotes disentangled spectrum,
		the black line the observed spectrum,
		the red line best-fitting synthetic spectrum, and
		the yellow line fit residuals.
		Only principal spectral lines
		(\ion{Si}{ii}{\,$6347$}\,\r{A},
		\ion{Si}{ii}{\,$6371$}\,\r{A},
		\hae,
		\ion{He}{i}{\,$6678$}\,\r{A})
		that are present within the studied spectral region
		$\Delta\lambda \simeq 6200 - 6800$\,\r{A} and their
		surroundings are plotted.
		\label{fig:blb:spectrafit}}
\end{figure*}

Additional details on the measuring of RVs and
the modeling of observed spectra with synthetic ones
follow:
\begin{itemize}
\item The RV measurements of \blb obtained manually
with \spefoe, and through comparison with synthetic
spectra using~\pyt
are listed in Table~\ref{tab:blb:rv}. Four spectral
lines were measured with the manual method on each
spectrum. Instead of individual measurements,
their mean and corresponding standard deviation
are listed, because measurements on each spectral line did
not differ systematically from RVs measured on
the remaining spectral lines. The manually measured
RVs give an~impression that they slowly vary,
but similar trend is not present in automatic
measurements or RVs measured on~photographic plates.

\item Uncertainties of kinematic and radiative
properties of \blb obtained through modeling
of its observed (or disentangled) spectra
with synthetic spectra (solutions~1 and~2 in
Table~\ref{tab:blb:spectrafit}) were obtained through
Markov chain Monte Carlo simulation implemented
within {\tt emcee}\footnote{The library is
available through GitHub~\url{https://github.com/dfm/emcee.git}
and its thorough description is
at~\url{http://dan.iel.fm/emcee/current/}.}
Python library by~\citet{foreman2013}. The
posterior probability distribution of
for each individual optimized was fitted
with a~Gaussian function. Standard deviation
of the function was taken for uncertainty
of the optimized parameter. Only statistical
part of the total uncertainty was pinpointed
by this approach.
The wavelength ranges fitted for observed spectra were
$\Delta\lambda=\{6337-6410, 6530-6600, 6660-6690\}$\,\r{A},
and for disentangled spectra
$\Delta\lambda=\{6338-6605, 6673-6724\}$\,\r{A}.
The region $\Delta\lambda = 6337-6410$\,\r{A} was not
modeled for observed spectra, because it contains only
few weak stellar lines and is densely polluted by
telluric lines.

\item Agreement between the observed (or disentangled)
spectra and their best-fitting synthetic spectra
(given by solutions~1 and~2 in Table~\ref{tab:blb:spectrafit})
is demonstrated by Fig.~\ref{fig:blb:spectrafit}.

Only the major spectral lines and their vicinity
are plotted. Only fit of one observed spectrum
with high S/N is shown. Jags in the observed
spectrum are remnants of telluric lines. Also
we note that \pyt does not require that the modeled
spectra are equidistant. Hence it was not
necessary to fill the gaps in spectra that emerged after
the removal of telluric lines.
\end{itemize}

\begin{table}
	\caption{RVs of \ble.
	\label{tab:blb:rv}}
	\begin{tabular}{ccccccl}
	\hline\hline\noalign{\smallskip}
	RJD      & $RV_\mathsf{SPEFO}$& $RV_\mathsf{PYTERPOL}$&Instrument\\
	\noalign{\smallskip}\hline\noalign{\smallskip}
	49\,866.4348&$-19.9\pm2.1$&$-20.15\pm0.61$&1\\
	49\,899.5024&$-16.4\pm2.1$&$-15.22\pm1.09$&1\\
	49\,907.5209&$-18.8\pm1.6$&$-18.02\pm1.33$&1\\
	49\,919.4182&$-22.1\pm1.8$&$-19.51\pm0.40$&1\\
	49\,931.4558&$-23.1\pm1.7$&$-18.18\pm0.94$&1\\
	49\,941.4280&$-18.8\pm0.5$&$-15.40\pm0.36$&1\\
	50\,235.4949&$-16.7\pm1.9$&$-15.94\pm0.64$&1\\
	52\,856.4091&$-17.9\pm0.5$&$-18.13\pm0.47$&2\\
	57\,332.2830&$-17.2\pm2.7$&$-18.72\pm0.49$&2\\
	57\,349.2456&$-10.6\pm2.6$&$-14.97\pm0.77$&2\\
	57\,417.6691&$-14.0\pm2.0$&$-16.23\pm0.68$&2\\
	57\,445.6286&$-19.2\pm0.5$&$-20.39\pm0.56$&2\\
	57\,464.5450&$-18.9\pm1.0$&$-18.61\pm0.40$&2\\
	\noalign{\smallskip}\hline
	\end{tabular}\\
	\tablefoot{Two different ways to measure RVs were used.
	$RV_\mathsf{SPEFO}$ denotes average RVs based on manual
	measurements of four spectral lines in \spefoe, and
	$RV_\mathsf{PYTERPOL}$ RVs measured through automatic
	comparison of observed and synthetic spectra.  	
	Instruments:
	1. Reticon 1872RF detector,
	2. CCD detector.}
\end{table}

\subsection{Proper motion of \bla and~B}\label{sec:apc:pm}
Proper motions of \bla and~B were downloaded
from the Vizier portal. Each 2-d vector had
a~component along the declination $\mu_\delta$
and in the perpendicular direction along the
right ascension $\mu_\alpha$. The latter coordinate
was corrected for declination of both systems.
The following coordinates were used:
\begin{eqnarray}
	\label{eq:pm0}
	X_\mu &=& \mu_\delta, \\
	\label{eq:pm1}
	Y_\mu &=& \mu_\alpha\cos\delta,
\end{eqnarray}
where $\delta$ denotes declination.
For \bla $\delta_\mathsf{A} = 33.362667$\,deg, and
for \blb $\delta_\mathsf{B} = 33.351856$\,deg were adopted.
All studied records in coordinates given
by~Eqs.~(\ref{eq:pm0}), and (\ref{eq:pm1}) are
listed in Table~\ref{tab:blb:pm}.
\begin{table}
	\centering
	\caption{Proper motion measurements of \ble.
	\label{tab:blb:pm}}
	\begin{tabular}{ccccccc}
		\hline\hline
		\noalign{\smallskip}
		Comp.&$Y_\mu$&$\Delta Y_\mu$&$X_\mu$&$\Delta X_\mu$& Source\\
		\hline
		A&0.92&0.44&-4.46&0.51&1\\
		A&6.60&3.80&-5.60&3.10&2\\
		B&2.84&2.30&-1.90&2.60&2\\
		A&0.84&0.40&-4.10&0.50&3\\
		B&-0.50&1.20&-0.80&1.30&3\\
		A&2.79&1.38&-5.24&1.18&4\\
		A&1.59&0.12&-3.53&0.20&5\\
		A&2.11&0.40&-3.51&0.50&6\\
		B&1.07&1.82&-0.64&1.86&6\\
		A&1.59&1.00&-3.50&1.00&7\\
		B&-0.08&1.10&-6.50&0.90&7\\
		B&4.31&0.26&-2.22&0.26&8\\
        B&4.37&0.09&-0.98&0.10&9\\
		\hline
	\end{tabular}
	\tablefoot{$X_\mu$ and $Y_\mu$ are proper motions given
		by Eqs.~(\ref{eq:pm0}) and~(\ref{eq:pm1}). $\Delta X_\mu$
		($\Delta Y_\mu$) denotes uncertainty of the corresponding
		quantity. The unit of all listed quantities is
		mas\,yr$^{-1}$. ``Comp.'' denotes component of the
		\be~visual system.
		Column ``Source'':
		1. the Hipparcos and Tycho Catalogs \citep{esa97},
		2. the Tycho Reference Catalog \citep{Hog1998},
		3. the Tycho 2 Catalog \citep{Hog2000},
		4. Astrometric position and proper motion of 19 radio stars \citep{Boboltz2003},
		5. Hipparcos the new reduction \citep{leeuw2007b,leeuw2007a},
		6. All-sky Compiled Catalog of $2.5$ million stars \citep{ascc},
		7. the Four US Naval Observatory CCD Astrograph Catalog \citep{ucac4},
		8. the First Data Release of \emph{Gaia} mission (DR1) \citep{gaiadr1}.
        9. the second Data Release of \emph{Gaia} mission (DR2)
	}
\end{table}

%%%%%%%%%%%%%%%%%%%%%%%%%%%%%%%%%%%%%%%%%%%%%%%%%%%%%%%%%%%%%%%%%%%%%%%%

\section{A note on disk model with a vertical temperature jump}\label{sec:apd}

\begin{figure}
\centering
\includegraphics[width=8cm]{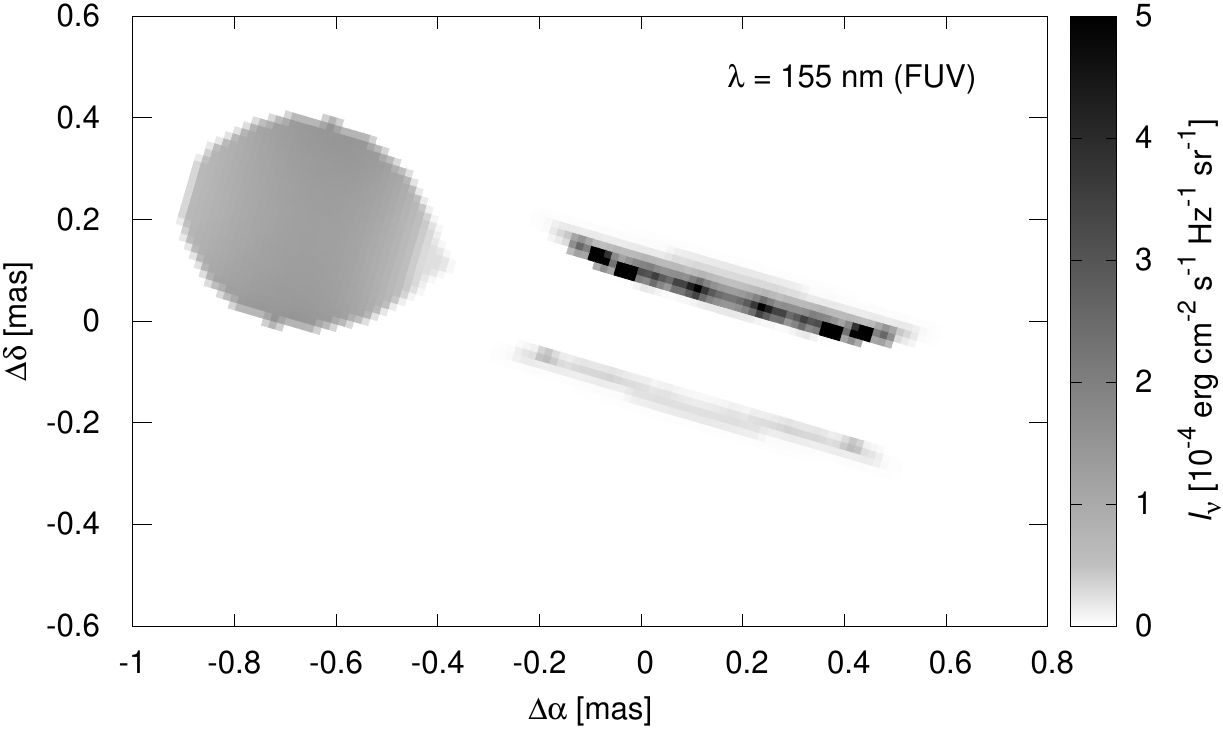}
\caption{Synthetic image of $\beta$~Lyr~A
for $\lambda = 155\,{\rm nm}$ (OAO2 band).
This model is based on a ``nebula`` with a vertical temperature jump,
$T(z)$ profile is determined by its scale height~$H$
and multiplication factor $h_{\rm mul} = 5.27$,
together with temperature inversion scale $h_{\rm inv} = 5.19$
and factor $t_{\rm inv} = 4.87$.
The resulting total $\chi^2 = 99\,430$, with individual contributions as low as
$\chi^2_\mathsf{LC} = 4\,932$,
$\chi^2_\mathsf{V^2} = 53\,534$,
$\chi^2_\mathsf{CP} = 28\,282$,
$\chi^2_\mathsf{T_3} = 12\,684$,
which is significantly better than the nominal nebula model presented in Table~\ref{tab:shellspec:result}.
However, the artifacts (bright spots close to the outer rim; black in this color scale) are clearly visible.}
\label{nebula_img_oao2_1550_BW}
\end{figure}

The previous version of \shellspec\ contained a ``nebula`` disk model
with an exponential vertical density profile~$\rho(z)$,
and a possible jump in the corresponding temperature profile~$T(z)$
to mimic a hotter disk atmosphere irradiated by the star.
For this model, we realized there are bright spots in FUV close to the outer rim
(see Figure~\ref{nebula_img_oao2_1550_BW}).
They actually helped to decrease the $\chi^2_\mathsf{LC}$ contribution.

This is a well-known problem caused by a simple linear interpolation
of both $\rho$ and $T$ quantities within the critical step
of the optical depth, where the medium changes from thick to thin,
and there is a large source~$S_\nu$ (and~contribution) function in the middle.
When the resolution of the model is increased twice,
or four times, these spots subsequently disappear
and $\chi^2_\mathsf{LC}$ increases again.
Users should be aware of these discretization artifacts,
because they sometimes appear in the course of convergence
(e.g., when the orbital inclination changes).
The same would be true for models with overlapping optically thick
and optically thin objects, like slab\,+\,slab, or wedge\,+\,flow.
Nevertheless, this problem is an indication for us that FUV radiation
and corresponding light curves should be described by a more complete
model of the disk atmosphere.

%%%%%%%%%%%%%%%%%%%%%%%%%%%%%%%%%%%%%%%%%%%%%%%%%%%%%%%%%%%%%%%%%%%%%%%%

\clearpage

\end{appendix}
\end{document}